\documentclass[aps,prd,twocolumn,superscriptaddress,floatfix,nofootinbib]{revtex4-2}

\usepackage{graphicx}
\usepackage{adjustbox}
\usepackage{dcolumn}
\usepackage{tensor}
\usepackage{bm}
\usepackage{float}
\usepackage{color}
\usepackage[usenames,dvipsnames]{xcolor}
\usepackage{listings}
\usepackage{fancyvrb}
\usepackage{slashed}
\usepackage{amsmath}
\usepackage{amssymb}
\usepackage{amsfonts}
\usepackage{enumitem}
\usepackage{listings}
\usepackage{mathtools}
\usepackage{slashed}
\usepackage{verbatim}
\usepackage{hyperref}
\usepackage[margin=1.0in]{geometry}
\usepackage{slashed}
\usepackage{braket}
\usepackage[normalem]{ulem}
\usepackage{cancel}
\usepackage{makecell}
\usepackage{tablefootnote}
\usepackage[symbol]{footmisc}
\newcommand{\cev}[1]{\reflectbox{\ensuremath{\vec{\reflectbox{\ensuremath{#1}}}}}}

\usepackage{accents}
\usepackage[normalem]{ulem} 
\makeatletter
\DeclareRobustCommand{\cev}[1]{%
  \mathpalette\do@cev{#1}%
}
\newcommand{\do@cev}[2]{%
  \fix@cev{#1}{+}%
  \reflectbox{$\m@th#1\vec{\reflectbox{$\fix@cev{#1}{-}\m@th#1#2\fix@cev{#1}{+}$}}$}%
  \fix@cev{#1}{-}%
}
\newcommand{\fix@cev}[2]{%
  \ifx#1\displaystyle
    \mkern#23mu
  \else
    \ifx#1\textstyle
      \mkern#23mu
    \else
      \ifx#1\scriptstyle
        \mkern#22mu
      \else
        \mkern#22mu
      \fi
    \fi
  \fi
}

\makeatother

\def\ee{\end{eqnarray}}

\def\nn{\nonumber}

\newcommand{\be}{\begin{eqnarray}}
\newcommand{\en}{\end{eqnarray}}
\newcommand{\bea}[1]{\left(\begin{array}{#1}}
\newcommand{\ena}{\end{array}\right)}
\newcommand{\ba}{\begin{eqnarray}}
\newcommand{\ea}{\end{eqnarray}}
\newcommand{\CL}{\mathcal{L}}
\newcommand{\CM}{\mathcal{M}}
\newcommand{\CO}{\mathcal{O}}

\newcommand\lrnab{\raise .8ex\hbox{$^\leftrightarrow$} \hspace{-8.8pt}
\nabla}
\newcommand\lnab{\raise .8ex\hbox{$^\leftarrow$} \hspace{-9.8pt}
\nabla}
\newcommand\rnab{\raise .8ex\hbox{$^\rightarrow$} \hspace{-9.8pt}
\nabla}

\begin{document}


\title{Nuclear-level Effective Theory of $\mu \rightarrow e$ Conversion:\\
 Formalism and Applications}

\author{W. C. Haxton}
\affiliation{Department of Physics, University of California, Berkeley, California 94720, USA}
\affiliation{Lawrence Berkeley National Laboratory, Berkeley, California 94720, USA}
\author{Evan Rule}
\affiliation{Department of Physics, University of California, Berkeley, California 94720, USA}
\author{Ken McElvain}
\affiliation{Department of Physics, University of California, Berkeley, California 94720, USA}
\author{Michael J. Ramsey-Musolf\,}
\affiliation{Tsung-Dao Lee Institute and School of Physics and Astronomy, Shanghai Jiao Tong University, 800 Dongchuan Road, Shanghai 200240, China}
\affiliation{Amherst Center for Fundamental Interactions, Department of Physics, University of Massachusetts, Amherst, Massachusetts 01003, USA}

\date{\today}

\begin{abstract}
Over the next 
decade new $\mu \rightarrow e$ conversion searches at Fermilab (Mu2e) and J-PARC (COMET, DeeMe) are expected to advance 
limits on charged lepton flavor violation (CLFV) by more than four orders of magnitude.  By considering the
consequence of $P$ and $CP$ on elastic $\mu \rightarrow e$ conversion and the structure of possible charge and current densities, we show that rates are governed by six
nuclear responses and a single scale, $q/m_N$, where $q \approx m_\mu$ is the momentum transferred from the leptons to the nucleus.  To relate this result to microscopic
formulations of CLFV, we construct in nonrelativistic effective theory (NRET) the CLFV nucleon-level interaction, pointing out the relevance of the dimensionless scales $y=({q b \over 2})^2 > |\vec{v}_N| >  |\vec{v}_\mu| > |\vec{v}_T|$,
where $b$ is the nuclear size, $\vec{v}_N$
and $\vec{v}_\mu$ are the nucleon and muon intrinsic velocities, and $\vec{v}_T$ is the target recoil velocity.  We discuss previous work, noting the lack of a systematic treatment of the various small parameters.
Because the parameter $y$ is not small, a proper calculation of $\mu \rightarrow e$ conversion requires a 
full multipole expansion of the nuclear response functions, an 
apparently daunting task with Coulomb-distorted electron partial waves.  We demonstrate that the multipole expansion can be carried out to high precision by 
introducing a simplifying local momentum $q_\mathrm{eff}$ for the electron.   Previous work has been limited to simple charge or spin interactions,
thereby treating the nucleus effectively as a point particle.   We show that such formulations are not compatible with the general form of the 
$\mu \rightarrow e$ conversion rate, failing to generate three of the six allowed nuclear response functions.   The inclusion of the nucleon velocity $\vec{v}_N$ 
yields an NRET with 16 operators and a rate of the general form.   Consequently, in
the current discovery era for CLFV, it provides the most sensible starting point for experimental analysis, defining what can and cannot
be determined about CLFV from the highly exclusive process of $\mu \rightarrow e$ conversion.   Finally, we expand the NRET operator basis to account for the effects of
$\vec{v}_\mu$,  associated with the muon's lower component, generating corrections to the CLFV coefficients of the point-nucleus response functions.
Using advanced shell-model methods, we  compute $\mu \rightarrow e$ conversion rates for a series
of experimental targets, deriving bounds on the coefficients of the CLFV operators.  These calculations are the first to include a general basis of CLFV operators, full
evaluation of the associated nuclear response functions, and an accurate treatment of electron and muon Coulomb effects.
We discuss  target selection as an experimental ``knob" that can be turned to probe the
microscopic origins of CLFV.  We describe two types of coherence that enhance certain CLFV operators
and selection rules that blind elastic $\mu \rightarrow e$ conversion to others. 
We discuss the matching
of the NRET onto higher level effective field theories, such as those constructed at the light quark level, noting opportunities to build on existing work in direct detection of dark matter.
 We discuss the relation of $\mu \rightarrow e$ conversion to $\mu\rightarrow e+\gamma$ and $\mu\rightarrow 3e$,
showing how MEG II and Mu3e results will complement those of Mu2e and COMET.  Finally we describe a accompanying
script -- in Mathematica and Python versions -- that can be used to compute $\mu \rightarrow e$
conversion rates in various nuclear targets for the full set of NRET operators.
\end{abstract}


\maketitle

\section{Introduction\label{sec:sec1}} 
Muon-to-electron conversion, in which a muon bound to a nucleus converts to a mono-energetic outgoing electron, 
occurs at an observable level only if there are new sources of flavor violation, beyond those responsible for neutrino mixing \cite{Baldini,Kuno,Marciano,DeGouvea}.  
It has long been recognized  that $\mu \rightarrow e$ conversion and other charged lepton flavor violating (CLFV) processes  (e.g., $\mu^+ \rightarrow e^+ + \gamma$ and
$\mu^+ \rightarrow e^+e^-e^+$) are among our most sensitive tests of new physics beyond the standard model \cite{Barbieri,Cirigliano2009,Calibbi}.  
This has motivated a series of experimental advances that, in sum, have improved limits on
$\mu \rightarrow e$ conversion rates by $\approx$12 orders of magnitude over
the past 75 years \cite{Bernstein}.

The experimental attributes of $\mu \rightarrow e$ conversion are quite attractive.  Intense muon beams exist,
with rates on target of $\approx 10^{11}$/s expected in the experiments now under construction.
Muons can be readily stopped in thick targets, where they quickly
cascade into $1s$ Coulomb orbits around target nuclei.  Conversion occurring without nuclear excitation
produces a monoenergetic electron carrying almost all of the 
energy released ($\approx m_\mu$), generating a signal that can be distinguished from the continuous spectrum
of background electrons coming from the free decay of the muon, $\mu^- \rightarrow e^- + \nu_\mu + \bar{\nu}_e$.  
If the process is mediated by photon exchange or otherwise involves an operator that couples to nuclear charge, 
the rate can be enhanced by the coherent elastic 
response  \cite{Bernstein}.

Bounds on $\mu \rightarrow e$ conversion are typically expressed as the branching ratio with respect to muon capture
in the same nucleus.  The best current limits on $\mu \rightarrow e$ conversion correspond to branching ratios 
$\approx 10^{-12}-10^{-13}$.  Depending on the CLFV operator (including its isospin, which influences the coherence),
existing branching ratio limits are testing CLFV up to energy scales of $\approx 10^3$ TeV.  Within the next few years
Fermilab's  Mu2e \cite{Mu2Eover,Abusalma}  experiment and  
J-PARC's COherent Muon to Electron Transition (COMET) \cite{COMETI,COMETII} and DeeMe \cite{DeeMeover} experiments 
should begin taking data.  Mu2E and COMET are designed to reach branching ratios of $\approx 10^{-17}$, probing CLFV scales up to $\approx 10^4$ TeV.

The nuclear physics of $\mu \rightarrow e$ conversion is somewhat unusual.  As experimentalists select out the
elastic channel in order to suppress backgrounds due to free muon decay, contributing CLFV operators are constrained by both $P$ and $CP$ selection rules.  The energy transfer to the nucleus is negligible, while the three-momentum
scale $ q \approx m_\mu$ is comparable to the inverse nuclear size, guaranteeing sensitivity to the internal structure of the nucleus.
These same conditions arise in another process of considerable interest, WIMP dark matter (DM) direct detection, where the low WIMP velocity restricts the
scattering to the elastic channel in almost all nuclei, but where the momentum transfer typically
peaks at $q \approx 100-150$ MeV.   These conditions lead to unusual relations between nucleon-level 
effective theories of DM direct detection and the nuclear responses that matter to experiment, enhancing some operators and suppressing others \cite{Liam1,Liam2}.

This paper is the second of two (see \cite{RulePRL}) in which the nonrelativistic effective theory (NRET) formulation of $\mu \rightarrow e$ conversion is developed at the nucleon level,
and then embedded in the nucleus.  As was found in the DM studies, this embedding has significant consequences for the CLFV physics of elastic $\mu \rightarrow e$ conversion, blinding experiment
to certain interactions, while enhancing sensitivity to others.   In \cite{RulePRL} we demonstrated that the
rate for elastic $\mu \rightarrow e$ conversion depends on six symmetry-allowed nuclear response functions $W_i$, evaluated at a momentum scale $q \approx m_\mu$  that probes the structure of the nucleus:
this result was obtained through inclusion of the nucleon velocity operator $\vec{v}_N$.  In fact, this
general form of the elastic $\mu \rightarrow e$ conversion rate can be deduced without specifying the microscopic forms of the nucleon charges and currents -- all that is needed are the parity and
time-reversal properties of the available densities.  The derivation is presented in Sec. \ref{sec:sec3A}.  
The leptonic coefficients of these response functions, denoted as $R_i$, represent the CLFV  physics that can be ``mined" in experiments that employ appropriately selected targets.
Work prior to \cite{RulePRL} concentrated on just two of the response functions, associated with the simplest charge and spin interactions.
As we will illustrate in this paper, such interactions may not be present, and even if they are, do not always dominate the conversion rate.

The Galilean-invariant nucleon-level microscopic description of CLFV we develop allows one to relate CLFV limits
obtained from different targets and also to connect limits obtained at the nuclear level to higher-energy formulations.  A microscopic operator expansion can be used with the impulse approximation (or its generalizations)
to make these comparisons and connections.  The expansion, carried out to a given order in the
 available small parameters, produces a nucleon-level nonrelativistic effective theory (NRET) 
with a complete set of operators in that order.

In an NRET, all possible interactions are constructed consistent with the applicable symmetries and the available nucleon
and lepton operators, which include the nucleon and lepton charges, their spins, the direction of the relativistic electron $\hat{q}$, the intrinsic 
nucleon $\vec{v}_N$ and muon  $\vec{v}_\mu$ velocities, and the target recoil velocity $\vec{v }_T$.  Together with the momentum transferred from the leptons $q$, these operators
generate a hierarchy of dimensionless scales, $y \equiv ({q b \over 2})^2 > |\vec{v}_N| >|\vec{v}_\mu|> |\vec{v}_T|$, that should
be respected in NRETs.  Here $b$,  the harmonic oscillator parameter employed in building Slater determinants for the nucleus,
represents the nuclear size.
In this paper we construct three levels of NRET corresponding to successively incorporating the physics associated
with $y$, $\vec{v}_N$, and $\vec{v}_\mu$.

We discuss past work and the approximations employed, which in general has involved a very limited set of operators
and expansions that do not respect the above hierarchy.  In particular, the large value of $y$ --
which ranges from
0.20 (C) to 0.37 (Cu) for the targets we explore here, and reaches 0.54 for the heavy target tungsten -- implies
significant angular momentum transfer between the leptons and nucleus.  That is, several multipoles can contribute
to the nuclear response functions that govern elastic $\mu \rightarrow e$ conversion.

Previous investigators, perhaps
daunted by the task of handling the required number of electron distorted waves,
have generally avoided calculating the full nuclear response functions.  This is an uncontrolled approximation, governed by terms of higher order in $y$
and sensitive to details of the underlying nuclear microphysics; we show, among similar targets, errors can range from negligible ($\approx$ 5\%)
to $o(1)$.
Yet, as summarized in Table \ref{tab:pastwork} of Sec. \ref{sec:sec2}, only a few calculations
have gone beyond the the lowest multipole, and those employed electron plane waves and the simplest operators, spin and charge.   
All work employing Dirac solutions
limited the electron waves to those with $j=\textstyle{1 \over 2}$
(so Dirac $| \kappa|$ =1), so that (at most) only the lowest multipole operator contributes.  The error induced by this Dirac truncation
generally exceeds the size of the Coulomb corrections being evaluated.

In Sec. \ref{sec:sec2} we introduce a simple trick that, later in the paper, enables us to properly evaluate the response functions:
the effects of the
Coulomb interaction on the electron -- its shortened wavelength in the nuclear interior and its enhanced amplitude -- can be 
incorporated through the introduction of an effective electron momentum $q_\mathrm{eff}$.  We demonstrate this 
approximation is very accurate for both low- and high-$Z$ targets and for the range of relevant Dirac partial waves.  
The introduction of $q_\mathrm{eff}$ allows one
to exploit the full power of standard spherical vector harmonic Bessel expansions to generate rather elegant expressions
for the needed response functions that include the Dirac physics of the electron and bound muon.

Section \ref{sec:sec2} also includes a discussion of the upper and lower $1s$ muon radial wave functions $g(r)$ and $f(r)$ that 
appear in the nuclear transition densities of the full rate formula. While we later present such a formula, we note
that in light- and medium-mass nuclei the muon is weakly bound, with a Bohr radius significantly larger than the nuclear size.
Consequently, its Dirac wave function varies slowly over the nucleus.  Thus there are opportunities to streamline rate formulas by taking 
advantage of the relative simplicity of the muon physics.  By replacing the muon wave function
with a suitably determined average value, rate formulas can be obtained in which the nuclear and particle physics are more clearly factorized
and the roles of various small parameters more apparent.  Noting that past averaging has followed procedures more appropriate to muon
capture, we suggest an improved procedure that reproduces leading multipoles exactly but remains quite accurate for others.
We quantify errors associated with our procedure for averaging, showing that they are quite small.

Section \ref{sec:sec3} begins with a derivation of the general form of the elastic $\mu \rightarrow e$ conversion rate, using
only symmetry arguments and an inventory of available charges and currents.  We then
develop the nucleon-level NRET Hamiltonian, forming the most general interaction from the available nucleon and lepton operators.
The NRET describes the nonrelativistic intrinsic Hamiltonian, the physics that remains after the phase-space integration removes 
the relative motion of the relativistic electron with respect to the daughter nucleus.
In addition to the electron direction $\hat{q}$ and the electron
and nucleon spin and charge operators, the various velocity operators come into play: here one can make use of the
 hierarchy, $|\vec{v}_N| > |\vec{v}_\mu| > |\vec{v}_T|$ to develop various levels of NRET.
 We show that this NRET operator basis needs to include at least terms linear in $\vec{v}_N$ to be complete, in the sense of
 reproducing the six nuclear response functions allowed by symmetry.  
 
For completeness, we extend the NRET to include the muon velocity operator $\vec{v}_\mu$ and thus the
contribution from the muon's lower component. This addition introduces no new
response function physics, but instead generates corrections to the coefficients $R_i$ proportional to $\langle f \rangle/\langle g \rangle$,
the ratio of the muon's lower and upper components.  We stress that this distinguishes the effects of $\vec{v}_\mu$ -- 
corrections to response functions that influence rates at the level of 5\% for $^{27}$Al -- from those of $\vec{v}_N$, which generates new
response functions, sensitivity to new aspects of CLFV, and contributions to rates that can be $o(1)$, depending on the
source of the CLFV.  This reflects the fact that the leptonic vertex in elastic $\mu \rightarrow e$ is inclusive, with the sum over
partial waves guaranteeing that muon lower-component contributions are always accompanied by upper-component
contributions.  In contrast, the nucleon vertex is exclusive, with $\vec{v}_N$ changing the parity and $CP$ of operators it generates,
thus altering the physics.

Until the present work, no formulation of $\mu \rightarrow e$ conversion has been available that combined the desired elements:
a complete basis of operators, the full form of the nuclear response functions $W_i$, 
and an accurate treatment of electron and muon Coulomb effects and lepton and nucleon velocities.   In fact, the effects
of $\vec{v}_N$ appear to have been ignored universally.  

Because the NRET basis is complete, any UV
theory should reduce at low energy to a form compatible with the NRET.   More importantly, as experiments are done with
largely nonrelativistic nuclei, the NRET provides one with a complete but minimal basis for analyzing and comparing limits
or results obtained from a range of nuclear targets.  
Consequently, information extracted
from $\mu \rightarrow e$ conversion limits, encoded as constraints on the $R_i$'s, can be ``ported upward" to constrain
theories formulated at higher energy scales, which necessarily have more degrees of freedom.
Thus a ``top-down" reduction of a favored UV theory to make contact with experiment would only need to be carried out to
the nucleon-level NRET: there would be no need to repeat the nuclear 
physics performed here.  

At the end of Sec. \ref{sec:sec3} we present a simple example of such a reduction: we construct the 20 nucleon-level relativistic operators that can arise
for semi-leptonic CLFV interactions mediated by scalar or vector mediators.  We show that these 
interactions reduce to a subset of the NRET operators we introduced (reflecting the specialization to scalar and vector exchanges), under a non-relativistic reduction
that treats the velocities of bound nucleons and the muon to first order.  We also discuss at the end of this section some of the nontrivial aspects
of such matching, including the interpretation of NRET operator coefficients derived from nuclei as effective.

This strategy follows one that has been used very effectively in studies of DM direct detection,
an elastic process with very similar kinematics.
The NRET DM formulation \cite{Liam1}  has become a very popular bridge between the nuclear scale and UV theories or, more commonly, effective field theories (EFTs) formulated at higher energy scales.
For example, open-source tools like DirectDM \cite{Bishara:2017nnn}, based on operator formulations at the light-quark and gluon level, were designed 
to work with the analogous NRET tool DMFormFactor  \cite{Liam2}.  This formalism that has been adopted by most of the major experimental groups \cite{DMexp1,DMexp2,DMexp3,DMexp4,DMexp5},
some of whom have developed specialized cuts helpful in isolating specific NRET interactions.  
We describe in Appendix \ref{AppendixC} a script 
for $\mu \rightarrow e$ conversion quite analogous to DMFormFactor.  By making the code publicly available, we hope to encourage similar
developments for $\mu \rightarrow e$ conversion.

In Sec. \ref{sec:sec4} (and in Appendix \ref{AppendixB}) we present details of the standard multipole formalism for embedding the nucleon-level interaction in a nucleus,
leading to the elastic $\mu \rightarrow e$ conversion rate formula.   We present rates for a succession of NRETs, according to the hierarchy of small parameters:
the extreme nonrelativistic limit, the inclusion of effects linear in $\vec{v}_N$, and the inclusion of additional effects linear in $\vec{v}_\mu$.

Rates are expressed as sums over
CLFV coefficients $R_i$ multiplied by nuclear response functions $W_i$, with the latter expressed as a sum over contributing multipoles.
Each $R_i$ is a bilinear function of the NRET operator coefficients.  
This factorization can be helpful in experimental analysis, if CLFV is discovered.
It allows one to view the $W_i$'s as experimental ``knobs" that can be dialed to isolate the CLFV 
coefficients $R_i$: the nuclear response functions depend on properties of the ground state, such as
its angular momentum, isospin, spin-orbit properties of valence nucleons, and the spin-orbit structure of the core.
Separating the various contributions will be challenging, of course.   Yet by 
selecting nuclear targets with specific properties, it should be possible to learn more about the operator source of the CLFV
found in $\mu \rightarrow e$ conversion than would be possible from results on a single nuclear target.

In Sec. \ref{sec:sec5} we discuss the nuclear physics of elastic $\mu \rightarrow e$, taking the long-wavelength
limit to isolate the leading nuclear operator for each response function.  We point out two distinct forms of
coherence, one of which is quite novel and especially relevant to targets like $^{27}$Al due to the
spin-orbit structure of the ground state.  We stress that the naive counting of operators at the nucleon NRET
level no longer holds at the nuclear level: some operators are enhanced by the two coherence mechanisms,
others are suppressed by nucleus-imposed selection rules.  We note that some
of the nuclear response functions $W_i$ are closely related to standard-model responses,
and thus potentially could be measured in iso-elastic muon capture or elastic electron scattering, while
others are not and thus necessarily require microscopic evaluation.    We present graphics showing 
the wide variation in target responses to CLFV operators; target attributes experimentalists may be able to exploit.

Section \ref{sec:sec5} also describes the large-basis configuration-interaction shell-model calculations we performed
to compute nuclear responses, using the best available effective interactions.  The needed one-body density matrices are obtained for a 
variety of nuclear targets that either have been used in past experiments or might be suitable choices for future ones.  
Calculations were performed for 11 targets ranging from C to Cu.    As
the included spaces we use are separable and the calculations are done without any basis truncations, we can  
project spurious center-of-mass motion, for consistency with our Galilean-invariant NRETs.

In Sec. \ref{sec:sec6} we derive constraints on the operator coefficients -- the low-energy constants or LECs -- 
from existing limits on $\mu \rightarrow e$ conversion and estimate how these limits will be sharpened when Mu2e and COMET results become available.
The constraints take the form of limits on bilinear combinations of the LECs, thereby showing what can and cannot be learned about the
CLFV from elastic $\mu \rightarrow e$ conversion.
We compare these constraints to those that can be obtained  from $\mu \rightarrow e+\gamma$ and $\mu\rightarrow 3e$.

In the concluding Sec. \ref{sec:sec7}  we summarize this paper and describe some future goals, including extending the current formalism to allow for nuclear excitation.  Such
an extension can yield additional physics:  because of the absence of $CP$ constraints on inelastic transitions, four LECs that cannot be probed in elastic
$\mu \rightarrow e$ conversion then play a role.

\begin{figure}[ht]   
\centering
\includegraphics[scale=0.40 ]{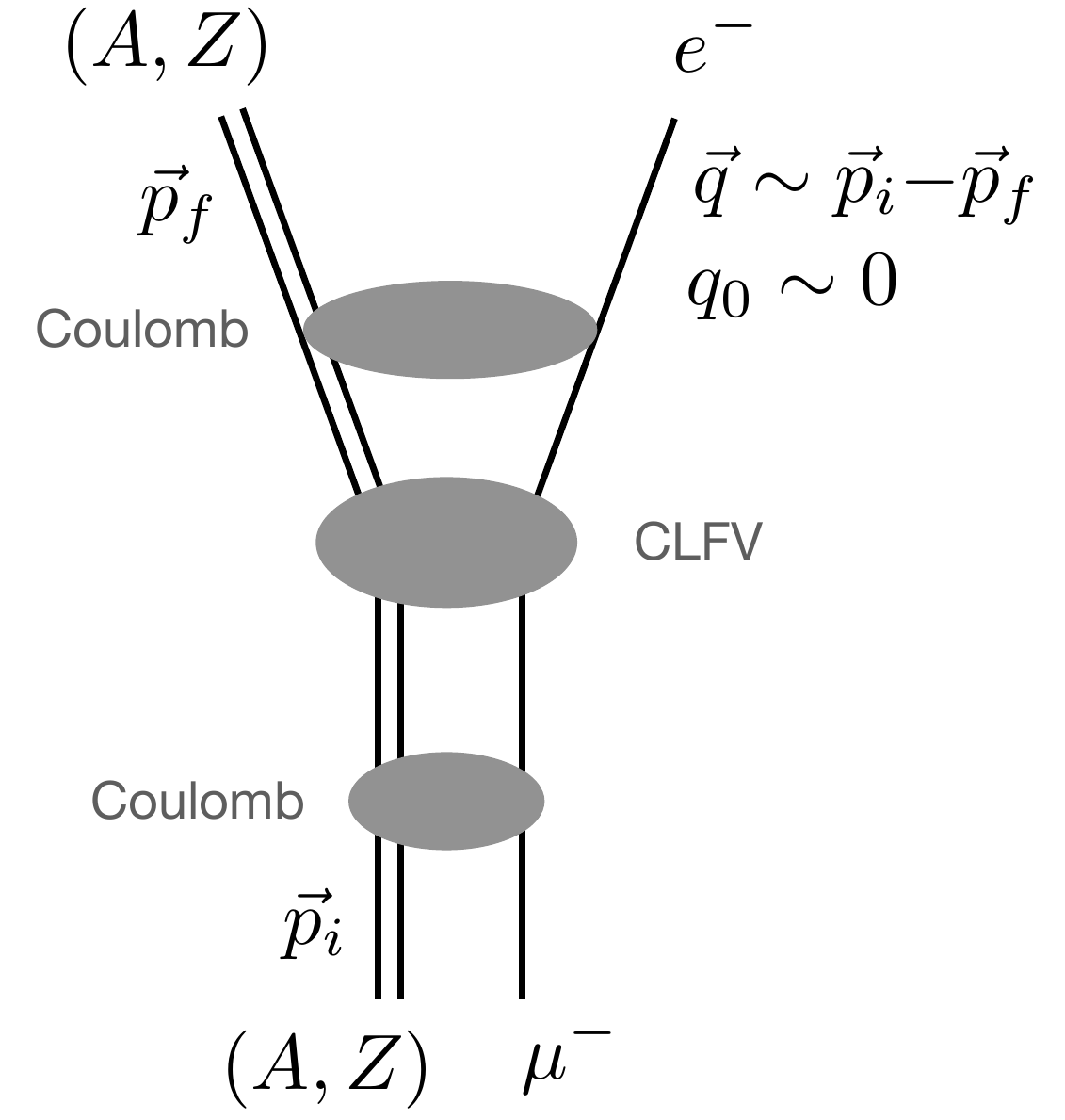}
\caption{Depiction of elastic $\mu \rightarrow e$ conversion.  The nuclear Coulomb potential binds
the $1s$ initial-state muon and distorts the outgoing electron wave function.   Neglecting nuclear recoil,
the electron's energy is the muon mass minus its Coulomb binding.} 
\label{fig:Decay}
\end{figure}

\section{Leptonic Interactions}
\label{sec:sec2}
Figure 1 depicts the conversion of a muon, bound in a $1s$ atomic orbital by the Coulomb field of the nucleus,  into an outgoing 
electron of fixed energy $E_e \approx m_\mu$.  If this CLFV process occurs without nuclear
excitation -- the nucleus remains in its ground state -- one finds
\be
E_e =m_\mu-E^\mathrm{bind}_\mu -{\vec{q}^{\,2} \over 2 M_T},
\label{eq:kin1}
\ee
where $\vec{q}$ is the three-momentum transferred from the nucleus to the electron, $m_\mu$ and $M_T$ are respectively
the muon and nuclear masses, and $E^\mathrm{bind}_\mu$ is the muon's binding energy, defined here as a positive quantity. 
Working to first order in $m_\mu/M_T$ and ignoring all smaller quantities in $1/M_T$ one finds
\be
\vec{q}^{\,2} ={M_T \over m_\mu + M_T} \left[ (m_\mu-E^\mathrm{bind}_\mu)^2 - m_e^2 \right],
\label{eq:kin2}
\ee
which can be substituted into Eq. (\ref{eq:kin1}) to determine the nuclear recoil energy.

The CLFV Hamiltonian consists of a series of terms of the form
\be
\mathcal{\hat{H}}_i &=&c_i ~\int  d\vec{r} ~ \bar{\Phi}_e(\vec{r}) \mathcal{O}_L\Phi_\mu(\vec{r})~ \Phi^*_N(\vec{r}) \mathcal{O}_N \Phi_N(\vec{r}) .~~~
\label{eq:LAG}
\ee
The four-fermion form is quite general: up to small nuclear recoil effects, the momentum transfer $\vec{q}$ is fixed, and consequently, any propagator effects can be absorbed
into the coefficients $c_i$.   In general, the leptonic and 
nucleon operators  $\CO_L$ and $ \CO_N$ (and the corresponding currents) can be scalar or vector (in which case they must couple to an overall scalar).
The choices of the $\CO_L$ and $\CO_N$, and consequently the number of coefficients $c_i$, are limited by the available
charge and current densities that one can form.  Elastic matrix elements of the interaction can be further
limited by parity and $CP$ selection rules.  In this paper, we will construct a general operator basis for
this process, derive rates, and compare those rates to existing and anticipated limits of $\mu \rightarrow e$ conversion.
A first step is the derivation of the Coulomb-distorted leptonic wave functions that appear in the transition density of Eq. (\ref{eq:LAG}).

\subsection{Coulomb effects}
\label{sec:2A}
Taking the matrix element of the interaction in Eq. (\ref{eq:LAG}) between lepton and nuclear states yields
\be
c_i \langle \mathcal{O}_L \rangle \int  d\vec{r} \langle f |  {\Psi}^*_e(\vec{r}) \Psi_\mu(\vec{r})~ \sum_{i=1}^A  \hat{O}_N(i) \delta (\vec{r}-\vec{r}_i) |i \rangle~~~~
\label{eq:convolution}
\ee
where $|i \rangle$ and $|f \rangle$ are the initial and final nuclear states and $\langle \mathcal{O}_L \rangle$ is the matrix
element of the leptonic operator between spin states.  We consider here a nuclear operator that is one-body,
which we write somewhat schematically 
in first quantization.   The nuclear transition density thus involves a convolution of the electron and muon wave functions
with the nuclear density.  The outgoing highly relativistic electron is distorted and enhanced in the region of the nucleus by the Coulomb
potential, which is sourced by the nucleus's extended charge distribution.  
The wave function and bound-state energy of a muon in the $1s$ orbital are an eigensolution of the Dirac equation
for the same Coulomb field.  

Two of the terms in the integrand of Eq. (\ref{eq:convolution}),
the nuclear density and the electron wave function, vary over a length sale typical of the nuclear radius:  the electron carries a momentum $q \approx m_\mu \approx 1/1.86$ fm, while
the shell-model oscillator parameter $b$ for $^{27}$Al is 1.85 fm.  The third term, the $1s$ muon wave function, varies more slowly for
the targets considered in this paper, which range from C to Cu.  The muon's Bohr radius in $^{27}$Al is $a_0^\mu \approx$ 20 fm.

As noted in the Introduction, the neglect of higher multipole operators in computing elastic $\mu \rightarrow e$ rates will in general lead to a response 
function that is correct only through $o(y)$.  Such multipoles will contribute to elastic $\mu \rightarrow e$ conversion if the ground-state angular momentum 
$j_N \ge 1$.  In $^{27}$Al, where $j_N=\textstyle{5 \over 2}$, the symmetry-allowed charge multipoles are $M_0$, $M_2$,
and $M_4$, while the allowed transverse (longitudinal) spin operators are $\Sigma_1^\prime$, $\Sigma_3^\prime$, and $\Sigma_5^\prime$ 
($\Sigma_1^{\prime \prime}$, $\Sigma_3^{\prime \prime}$,
and $\Sigma_5^{\prime \prime}$).  (The specific forms of these operators will be given later in this paper.)
The error one induces by truncating the multipole expansion is not simple to estimate: the numerical impact of $o(y^2)$ corrections will depend on both the
NRET operator under study and the specific nuclear physics of the target.  For example, the $M_2$ operator is related to the quadrupole moment,
which can be quite enhanced in mid-shell nuclei due to collective deformation.

Most past work has retained only the leading multipole contribution, while also making other simplifications, such as neglecting the Coulomb distortion
of the outgoing electron partial waves and treating only the simplest nuclear operators. A relatively
complete compilation of past work is given in Table \ref{tab:pastwork}.   In the few cases in Table \ref{tab:pastwork} where the full response function  
was computed \cite{Siiskonen2000,Kosmas2001,Cirigliano2018,Davidson2018}, electron plane waves were
employed and only
the charge and/or spin nuclear operators were considered.   Others used distorted electron waves -- solutions of the Coulomb Dirac equation -- 
but retained only those waves with Dirac $|\kappa| =1$ ($j=\textstyle{1 \over 2})$.  This
guarantees that (at most) one multipole operator is generated; an approximation that one expects to become increasingly problematic
with increasing $A$ and $y$.  While the shell-model rate calculations reported in this paper only extend up to Cu, nearly 40\%
of the spin response probability in $^{63}$Cu is carried by the $J=3$ multipole.

In Sec. \ref{sec:2B} we show that a full multipole expansion using distorted electron waves can be carried out to high accuracy for any of our NRET
operators.  This is accomplished by employing an approximation familiar from relativistic electron scattering:
the effects of the Coulomb field in shortening the wavelength of the electron wave function and 
enhancing its overlap with the nucleus can be incorporated very accurately
through the introduction of an effective momentum $q_\mathrm{eff}$.  We demonstrate the accuracy of the  $q_\mathrm{eff}$ approximation
as the Dirac quantum number $\kappa$ and target charge $Z$ are varied.
This  substitution allows us to complete the full multipole expansion using standard Bessel vector spherical harmonic expansions.

In Sec. \ref{sec:2C} we discuss the bound muon wave function.  The muon is largely nonrelativistic: in  $^{27}$Al the
 ratio of lower-to-upper components at the peak of the transition density is $\approx 0.027$.
Consequently, $\mu \rightarrow e$ conversion rates are dominated by the muon's upper component.
While in our master rate formula $g(r)$ and $f(r)$ are retained in the transition densities, we also simplify that result
by replacing these functions with average values.  This is a common practice in muon capture studies \cite{Walecka}, as well as in $\mu \rightarrow e$ conversion,
as can be seen in Table \ref{tab:pastwork}.   Here the purpose of the averaging is to simplify the rate formula to bring out
the underlying physics more clearly.  We also do the averaging in a way that exactly preserves the values of leading multipoles
and minimizes errors for other multipoles.  We demonstrate that the errors induced by using averages
$\langle g \rangle$ and $\langle f \rangle$ are small.

In our NRET the muon's lower component is generated from the upper component through the action of the muon's velocity
operator.  As the NRET forces one to think about all of the theory's velocity operators in a consistent way, it brings out some inconsistencies in past
work.  Several of the papers listed in Table \ref{tab:pastwork} included the lower component, but none prior to our work in \cite{RulePRL} treated the nucleon velocity operator,
whose magnitude for aluminum is about four times larger.  The two velocities have a common origin: the NRET is the Galilean invariant intrinsic Hamiltonian
that describes the relative motion of the nuclear constituents.  $A-1$ of the relative (Jacobi) coordinates describe the relative positions of the $A$ nucleons, 
while the $A$th Jacobi coordinate gives the muon's position relative to the 
center-of-mass of the $A$ nucleons.  NRET velocity operators are conjugate to these relative coordinates,
with $|\vec{v}_N| > |\vec{v}_\mu|$ for all targets considered here.   We discuss the muon's
lower component in Sec. \ref{sec:2D} and address its role in the NRET later in the paper.

\begin{table*}
\caption{An incomplete survey of elastic $\mu\rightarrow e$ conversion studies including the nuclear targets considered, the nuclear multipole operators evaluated, and the form of the lepton wave functions employed. $\mathcal{O}_{J;\tau}$ means that both isospin structures and all allowed $J$ were included. For the Dirac electron, all of the references surveyed restrict attention to the lowest partial waves $\kappa=\pm 1$. Besides the Dirac solution, the remaining forms of the muon wave function are all constant approximations: $\braket{|\psi_{\mu}|^2}_{\rho}$ is obtained by averaging the probability of the Dirac solution over the nuclear density, $|\phi^Z_\mathrm{1s}(\vec{0})|^2$ is the probability of the point-like Schrodinger solution evaluated at the origin, $|G(R_N)|^2$ is the upper component of the muon's Dirac wave function evaluated at the nuclear radius. Superscript $\dag$ indicates that the reference considers the inelastic process as well, although the information in the table reflects only the treatment of the elastic process.}
\centering
\begin{tabular}{llllll}
\hline
~~\\[-.25cm]
Author(s) & Year [Ref] & Target & Operators & $\psi_e$ & $\psi_{\mu}$\\[.1cm]
\hline
Weinberg and Feinberg$^{\dag}$& 1959 \cite{Weinberg1959} & multiple & $M_{0;p}$ & Plane wave & $\braket{|\psi_{\mu}|^2}_\rho$ \\
Marciano and Sanda & 1977 \cite{Marciano1977} & multiple &$M_{0;\tau}$ & Plane wave & $\braket{|\psi_{\mu}|^2}_\rho$ \\
Shanker & 1979 \cite{Shanker1979} & multiple & $M_{0;\tau}$ & Dirac, $|\kappa|=1~~$ & Dirac \\
 Kosmas and Vergados$^{\dag}$ & 1990 \cite{Kosmas1990} & multiple & $M_{0;\tau}$ & Plane wave & $\braket{|\psi_{\mu}|^2}_\rho$ \\
Chiang et al.$^{\dag}$ & 1993 \cite{Chiang1993} & multiple & $M_{0;\tau}$ & Plane wave & $\braket{|\psi_{\mu}|^2}_\rho$ \\
Kosmas et al.$^{\dag}$ & 1994 \cite{Kosmas1994} & $^{48}$Ti & $M_{0;\tau}$ & Plane wave & $\braket{|\psi_{\mu}|^2}_\rho$ \\
 Czarnecki, Marciano, and Melnikov & 1998 \cite{Czarnecki1998} & $^{27}$Al, $^{48}$Ti, $^{208}$Pb & $M_{0;\tau}$ & Dirac, $|\kappa|=1$ & Dirac \\
Siiskonen, Suhonen, and Kosmas$^{\dag}$ & 2000 \cite{Siiskonen2000} & $^{27}$Al, $^{48}$Ti & $M_{J;\tau}\;\Sigma'_{J;\tau}\;\Sigma''_{J;\tau}$ & Plane wave & $\braket{|\psi_{\mu}|^2}_\rho$ \\
Kosmas$^{\dag}$ & 2001 \cite{Kosmas2001} & $^{48}$Ti, $^{208}$Pb & $M_{J;\tau}\;\Sigma'_{J;\tau}\;\Sigma''_{J;\tau}$ & Plane wave & $\braket{|\psi_{\mu}|^2}_\rho$ \\
Kitano, Koike, and Okada & 2002 \cite{Kitano2002} & multiple& $M_{0;\tau}$ & Dirac, $|\kappa|=1$ & Dirac \\
Kosmas & 2003 \cite{Kosmas2003} & multiple & $M_{0;\tau}$ & Plane wave & Dirac \\
Cirigliano et al. & 2009 \cite{Cirigliano2009} & multiple & $M_{0;\tau}$ & Dirac, $|\kappa|=1$ & Dirac \\
Crivellin et al. & 2017 \cite{Crivellin2017} & $^{27}$Al, $^{197}$Au & $M_{0;\tau}$ & Dirac, $|\kappa|=1$ & Dirac \\
Bartolotta and Ramsey-Musolf & 2018 \cite{Bartolotta2018} & $^{27}$Al & $M_{0;\tau}$ & Dirac, $|\kappa|=1$ & Dirac \\
Cirigliano, Davidson, and Kuno & 2018 \cite{Cirigliano2018} & $^{27}$Al & $M_{0;\tau}\;\Sigma'_{J;\tau}\;\Sigma''_{J;\tau}$ & Plane wave & $|\phi_{1s}^{Z}(\vec{0})|^2$ \\
Davidson, Kuno, and Saporta & 2018 \cite{Davidson2018} & $^{27}$Al, Ti & $M_{0;\tau}\;\Sigma'_{J;\tau}\;\Sigma''_{J;\tau}$ & Plane wave & $|\phi_{1s}^{Z}(\vec{0})|^2$\\
Civitarese and Tarutina$^{\dag}$ & 2019 \cite{Civitarese2019} & $^{208}$Pb & $M_{0;\tau}$ & Plane wave & $|G(R_N)|^2$ \\
Rule, Haxton, and McElvain & 2021 \cite{RulePRL} & $^{27}$Al, Ti & $M_{J;\tau}\;\Sigma'_{J;\tau}\;\Sigma''_{J;\tau}$ & Full Dirac ($q_\mathrm{eff}$) & $|\phi_{1s}^{Z_\mathrm{eff}}(\vec{0})|^2$ \\
 & & & $\Delta_{J;\tau}\;\tilde{\Phi}'_{J;\tau}\;\Phi''_{J;\tau}$ & & \\
Heeck, Szafron, and Uesaka & 2022 \cite{Heeck2022} & multiple & $M_{0;\tau}$ & Dirac, $|\kappa|=1$ & Dirac \\
Cirigliano et al. & 2022 \cite{Cirigliano2022} & $^{27}$Al,$^{48}$Ti,$^{197}$Au,$^{208}$Pb~~ & $M_{0;\tau}$& Dirac, $|\kappa|=1$ & Dirac \\
Hoferichter, Men\'endez, and No\"el & 2022 \cite{Hoferichter:2022mna} & $^{27}$Al,Ti~~ & $M_{J;\tau}\;\Sigma'_{J;\tau}\;\Sigma''_{J;\tau}$ & Plane wave & $\braket{|\psi_{\mu}|^2}_{\rho}$ \\
 & & & $\Phi''_{J;\tau}$ & & \\
\hline
\end{tabular}
\label{tab:pastwork}
\end{table*}

\subsection{The electron's distorted partial waves}\label{sec:2B}
Somewhat schematically, in the plane-wave limit of Eq. (\ref{eq:convolution}) one obtains a nuclear transition density of the form
\[ \sum_{i=1}^A  e^{-i \vec{q} \cdot \vec{r}_i} \hat{\mathcal{O}}_N(i) \]
where for simplicity we have omitted the s-wave contribution of the muon.  Here $\hat{\mathcal{O}}_N(i)$ would be one of the NRET operators 
that we will construct from nucleon charges, spins, and velocities and their longitudinal and transverse projections.  As $\vec{r}_i$ operates on
nucleon $i$, one must expand the exponential, regrouping powers of $\vec{r}_i$ with $\hat{\mathcal{O}}_N(i)$ to form nuclear operators.
Because $|\vec{q} \cdot \vec{r}_i| \approx o(1)$ for $^{27}$Al and other targets of interest, a full expansion of the exponential should be made.
The elegant way to do this is by an expansion in spherical Bessel functions, spherical harmonics, and vector
spherical harmonics (depending on whether $\hat{\mathcal{O}}_N(i)$ is a charge or current operator).  From this construction emerge
nuclear multipole operators of good angular momentum, $P$, and $CP$.  Such techniques are particularly powerful for elastic processes like
$\mu \rightarrow e$ conversion because the contributing multipolarities $J$ then are restricted by the ground-state angular momentum $j_N$,
($J \le 2 j_N$), as well as by good $P$ and $CP$ of the ground state.  

However, the generalization of this construction for distorted electron partial waves is nontrivial, especially when a large NRET operator
basis is used.  The paper of Kitano, Koike, and Okada \cite{Kitano2002} listed in Table \ref{tab:pastwork} begins by discussing
an operator basis somewhat analogous to our NRET.
The authors also took the electron partial waves from solutions of the Coulomb Dirac equation, but, in estimating rates, a series
of simplifications were made, including the retention of just the $|\kappa|$=1 partial waves.  In the end,
the pseudoscalar, axial vector, tensor, and space-like vector operators of their
Lagrangian were eliminated from consideration.  Only the lowest multipole of the charge operator $M_0$ was retained, generated by the nucleon-level scalar interaction and the charge component of the vector interaction.  

This kind of procedure is not compatible with our ``bottom up'' NRET approach, which requires that we construct and evaluate all possible NRET operators,
limited only by basic symmetries: in this approach one makes
no apriori assumptions about the UV source of the CLFV, or, consequently, how it will be manifested through specific low-energy operators
in elastic $\mu \rightarrow e$ conversion.  Thus we need to include the full set of NRET operators and, given our remarks about the nuclear response functions,
all contributing electron Coulomb-distorted partial waves.
To accomplish this, we show that the distorted-wave problem can be greatly
simplified by recognizing that almost all of the distortion can be absorbed through a simple scaling of the plane-wave solution.  This allows us to
complete the full multipole expansion for the entire set of NRET operators, and when combined with muon averaging, analytically evaluate all
of the one-body matrix elements of these operators.  This is an enormous simplification with very little sacrifice of accuracy, as we will show.

The Dirac solutions for the continuum electron and bound-state muon can be expanded in partial waves.  Assuming spherical symmetry,
\be
\psi_\kappa (\vec{r}) =  \left( \begin{array}{c}  i{g_{\ell j}}  \left[ \begin{array}{r} \langle\textstyle{ {1 \over 2}, {1 \over 2}} |(\ell {1 \over 2}) j m \rangle \\ [.2cm] \langle \textstyle{{1 \over 2}, \mbox{-}{1 \over 2}}  |(\ell {1 \over 2}) j m \rangle \end{array} \right] \\~~\\
-{f_{\ell j} }  \left[ \begin{array}{r} \langle {1 \over 2}, {1 \over 2}  |(\ell \pm1 \, {1 \over 2}) j m \rangle \\[.2 cm] \langle {1 \over 2}, \mbox{-}{1 \over 2}  |(\ell \pm 1 \, {1 \over 2}) j m \rangle \end{array} \right]  \end{array} \right)
\label{eq:Dirac1}
\ee
Solutions are indexed by the parameter
$\kappa = \dots, -3,-2,-1,1,2,3, \dots$ where $j=|\kappa| - \textstyle{1 \over 2}$ and
\be
\kappa = \left\{ \begin{array}{rr} -(\ell+1) & ~~\kappa<0 \\ \ell &~~ \kappa>0 \end{array} \right.
\ee
The angular-spin wave functions are
\be
\langle \theta \phi  | (\ell {1 \over 2}) j m \rangle = \sum_{m_\ell m_s}  \langle \ell m_\ell \textstyle{1 \over 2} m_s | j m \rangle Y_{\ell m_\ell}(\theta,\phi) \xi_{m_s} ~~~~
\ee
where $\xi_{m_s}$ is the Pauli spinor. 

In the plane-wave limit the radial solutions are spherical Bessel functions
\begin{eqnarray}
{g_{\ell j} (r) \over q} \equiv {G_{\ell j}  \over qr} &=&  j_\ell( qr)  \nonumber \\
{f_{\ell j}(r) \over q} \equiv{F_{\ell j} \over qr} &=& \left[ {E-m \over E+m} \right]^{1 \over 2}  \left\{ \begin{array}{cc}~~ j_{\ell-1}(q r) & \kappa>0 \\ -j_{\ell+1}(q r) & \kappa<0 \end{array} \right.~~
\end{eqnarray}
In the presence of a Coulomb potential, the radial solutions satisfy
\begin{eqnarray}
(E-\mu-V_c(r)) G_{\ell j} &=& -{d \over dr} F_{\ell j} +{\kappa \over r} F_{\ell j} \nonumber \\
(E+\mu-V_c(r)) F_{\ell j} &=& {d \over dr} G_{\ell j} +{\kappa \over r} G_{\ell j}
\label{eq:Dirac2}
\end{eqnarray}
where $\mu$ is the reduced mass of the lepton
\[ \mu = {m M_T \over m+ M_T} \]
with $m$ the lepton mass and $M_T$ the nuclear mass.
$V_c(r)$ is  the Coulomb potential, which we
compute using an extended nuclear charge distribution of the form
\be
 \rho(r) = {n_0 \over 1+\mathrm{exp}[{r-c \over \beta}]}~~\mathrm{with}~~\int_0^\infty dr\; r^2 \rho(r) =Z ~~~~~ 
 \label{eq:density}
 \ee
where the values of $c$ and $\beta$, typically determined from elastic electron scattering, are taken from Ref. \cite{density} (see Table \ref{tab:input}).  
$Z$ is the nuclear charge.
The normalization $n_0$ can be evaluated analytically in terms of the polylogarithm function $\mathrm{Li}_n(z)$,
\[ n_0= -{Z \over 2 \beta^3 \mathrm{Li}_3(-\mathrm{exp}[c/\beta])} \]
as can the associated potential.


A procedure familiar from electron scattering studies 
is the replacement of the Coulomb solution in the vicinity of the nucleus by a plane wave solution with a shifted momentum,  $\vec{q} \rightarrow \vec{q}_\mathrm{eff}$:
\be
\displaystyle{e^{i \vec{q} \cdot \vec{r}}  \rightarrow {q_\mathrm{eff} \over q}  e^{i \vec{q}_\mathrm{eff} \cdot \vec{r}} ~~~~ r \lesssim R}
 \label{eq:electronspinor}
 \ee
where $\vec{q}_\mathrm{eff}$ is the local momentum in the Coulomb well and $R$ the nuclear radius.  This is a local replacement, not affecting 
the flux at infinity.  
For derivations and discussion see \cite{Dirac12,Dirac13,Dirac15}.  The use of $\vec{q}_\mathrm{eff}$ in the exponential
accounts for the more rapid oscillation of the electron wave function in the attractive Coulomb well created by the nucleus, while $q_\mathrm{eff}/q$ 
accounts for the enhancement of the amplitude due to that attraction.
Here we determine $q_\mathrm{eff}$ using a constant potential whose depth is equated to the average of the Coulomb potential over the nuclear charge distribution
\begin{eqnarray}
\bar{V}_c &=&{  \int_0^\infty dr\;r^2 \rho(r) V_c(r) \over  \int_0^\infty dr\;r^2 \rho(r) } \nonumber \\
\vec{q}^{\, 2}_\mathrm{eff} &=& {M_T \over m_\mu +M_T} \left[ (m_\mu-E_{\mu}^\mathrm{bind} -\bar{V}_c)^2 - m_e^2 \right]~~~~~~
\end{eqnarray}
The values we obtained for a series of possible nuclear targets are given in Table \ref{tab:input}.  

\begin{figure*}[ht]   
\centering
\includegraphics[scale=0.47 ]{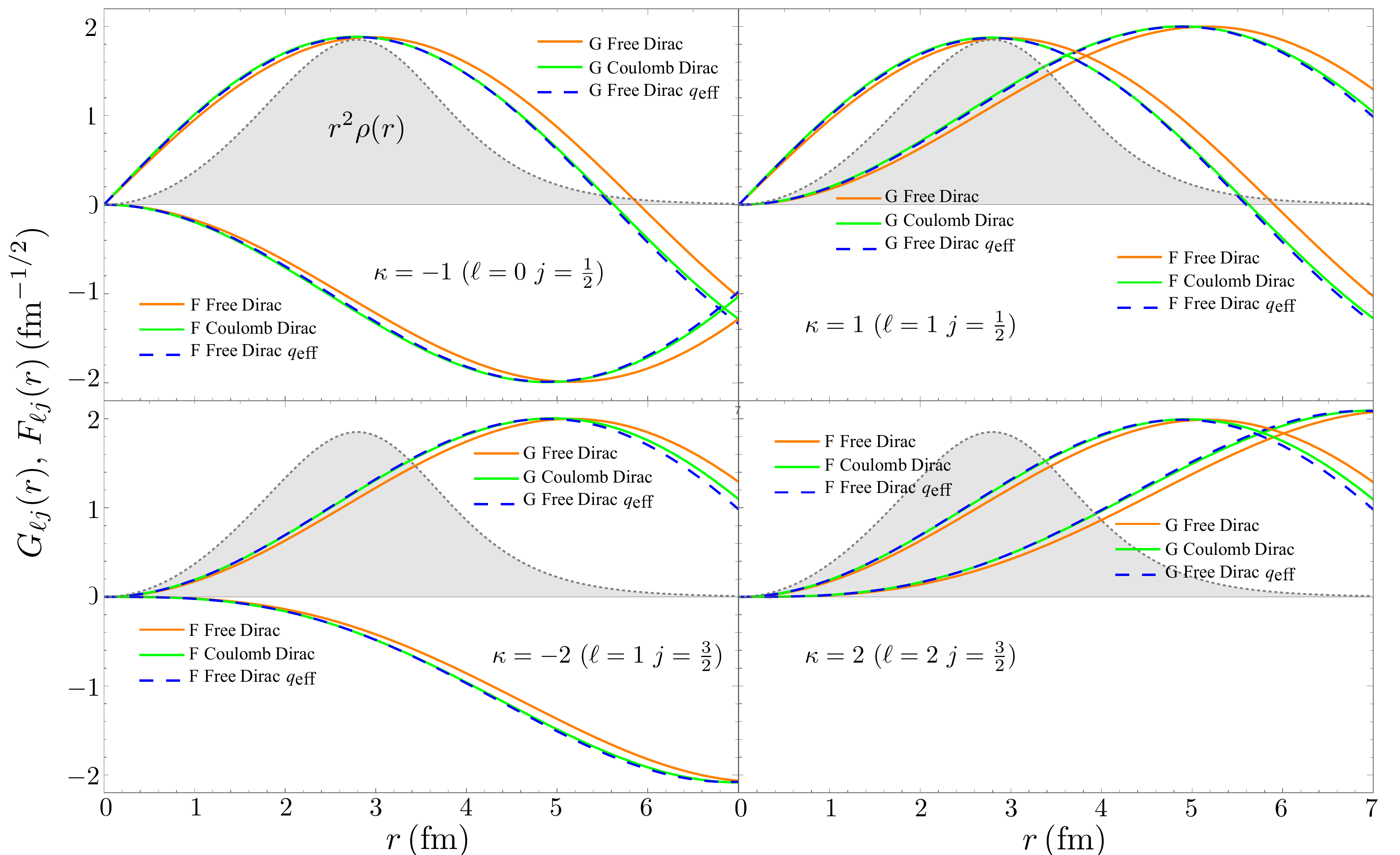}
\caption{The Dirac Coulomb solutions $G(r)$ and $F(r)$ for the highly relativistic outgoing electron produced in $\mu \rightarrow e$
conversion in $^{27}$Al (green line) are compared to the free solution (orange) and to the free solution evaluated with ${q}_\mathrm{eff}$ (blue
dashed), for low partial waves.  The nuclear charge distribution is shown by the shading (arbitrary normalization).  The agreement between
the Coulomb and free solutions evaluated with $q_\mathrm{eff}$ is quite good.  }
\label{fig:AAe1}
\end{figure*}


\begin{figure*}[ht]   
\centering
\includegraphics[scale=0.47 ]{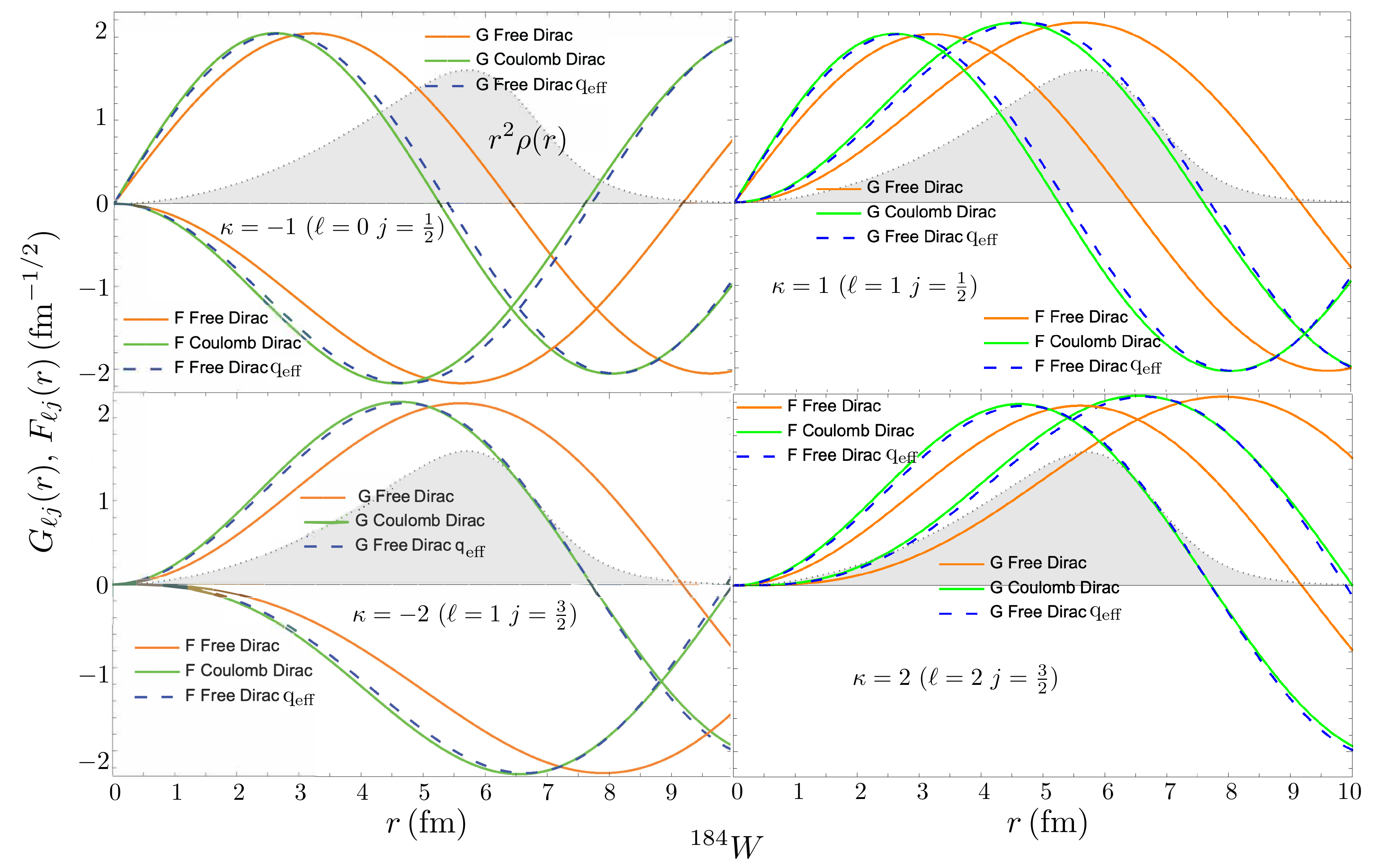}
\caption{As in Fig. \ref{fig:AAe1}  but for $^{184}$W.}
\label{fig:AAe3}
\end{figure*}

Consequently, for an ultra-relativistic electron, we identify
\begin{eqnarray}
{G^\mathrm{Coulomb}_{\ell j}  \over qr} &\leftrightarrow& {q_\mathrm{eff} \over q} j_\ell( q_\mathrm{eff}r)  \nonumber \\
{F^\mathrm{Coulomb}_{\ell j} \over qr} &\leftrightarrow&  {q_\mathrm{eff} \over q} \left\{ \begin{array}{cc}~~ j_{\ell-1}(q_\mathrm{eff} r) & \kappa>0 \\ -j_{\ell+1}(q_\mathrm{eff} r) & \kappa<0 \end{array} \right.
\label{eq:qeffC}
\end{eqnarray} 
This approximate equivalence was exploited in electron scattering studies using the targets $^{40}$Ca and  $^{208}$Pb \cite{Aste2005}.
In early work, the constant potential used in calculating $q_\mathrm{eff}$ was set to the Coulomb 
potential at the nuclear origin, which overestimates the attraction.   A physically more reasonable choice,
as calculations later demonstrated, is to fix the potential to its nuclear-density-weighted average, as is done here.  In Figs. \ref{fig:AAe1} and \ref{fig:AAe3}
the shifted plane waves of Eq. (\ref{eq:qeffC}) are compared to the numerically computed Dirac equation solutions.  These solutions and $\bar{V}_c$ 
are both generated from the nuclear charge distribution of Eq. (\ref{eq:density}).  Results are shown for $^{27}$Al and $^{184}$W
for the partial waves $|\kappa| \le 2$ to show the quality of the $q_\mathrm{eff}$ approximation in various partial waves 
and for low- and high-$Z$ targets.
The computed  $\bar{V}_c$'s for $^{27}$Al and $^{184}$W are  -5.84 and -18.4 MeV, respectively.

The correspondence between the $q_\mathrm{eff}$ approximation and the exact Dirac solutions is very good.   
In $^{27}$Al, deviations are only discernible on the far tail of the nuclear density, where contributions to the convolution are very small.
The agreement is also excellent for $^{184}$W, where the Coulomb distortion is very large.  This agreement is achieved
without tuning:  $q_\mathrm{eff}$ is derived from $\bar{V}_c$, which is computed from the same nuclear density used in the Dirac equation.
$q_\mathrm{eff}$ was not determined by fitting the Dirac solution.


We employ in this paper the free-electron Dirac spinor  \cite{Bjorken}
\be
U (q,s) &=&\sqrt{{E_e+m_e \over 2m_e}} \left( \begin{array}{cc}  \xi _s\\[0.2cm] \displaystyle\frac{\vec{\sigma}\cdot \vec{q}}{E_e+m_e}  \xi_s \end{array} \right)
\ee
with $\xi_s$ the Pauli spinor.  For our highly relativistic electron,
the net effect of the Coulomb interaction with the nucleus is the replacement
\be
U(q,s) e^{i \vec{q} \cdot \vec{r}}  \rightarrow {q_\mathrm{eff} \over q}  \sqrt{{E_e \over 2 m_e}} \left( \begin{array}{c} \xi_s \\ \vec{\sigma} \cdot \hat{q} \xi_s \end{array} \right) e^{i \vec{q}_\mathrm{eff} \cdot \vec{r}} ~~
\label{eq:Delectron}
\ee
where $\hat{q}$ is the unit vector in the direction of the electron.  

With $q_\mathrm{eff}$  one can accurately account for Coulomb distortions of the electron while not abandoning the elegant
multipole formalism of plane waves.
Table \ref{tab:pastwork} includes
four calculations in which full response functions have been calculated \cite{Siiskonen2000,Kosmas2001,Cirigliano2018,Davidson2018},
but those calculations employed undistorted plane waves.  In all other cases, only the leading multipole has been retained.   The associated error
depends on the detailed nuclear physics of the ignored multipoles and can be large.   For example, among the 11 targets we will consider in this paper, there are
three -- Na, Al, Cu -- with only odd isotopes, an unpaired proton, and a ground-state angular momentum of at least $\textstyle{3 \over 2}$,
so that more than one multipole contributes.  We picked one familiar NRET interaction, $\vec{\sigma}_L \cdot \vec{\sigma}_N$, and set
the isospin to be $(1+\tau_3)/2$, so that the operator couples to the unpaired valence proton.  Evaluating rates first with the leading multipole and then with all contributing multipoles, we find rate 
increases of 22.4\% (Na), 4.7\% (Al), and 65.4\% (Cu).   In the case of Al, ignoring higher multipoles produces a modest error, but in Cu the error is nearly $o(1)$.  All three isotopes have
strong spin responses.   One concludes that truncating the multipole expansion to the leading operator is not in any sense a
controlled approximation.   The associated error fluctuates considerably from nucleus to nucleus.  Consequently,
there is no shortcut: to assess the importance of neglected multipoles, one must calculate them.  The use of 
$q_\mathrm{eff}$ makes this practical even if one's operator basis is quite large, as it is here.

The $q_\mathrm{eff}$ substitution also helps one to see connections between $\mu \rightarrow e$ conversion and
standard-model processes.
Experimental $\mu \rightarrow$ e conversion rates are  expressed as a branching ratio with respect to the
muon capture rate.  In treatments like that presented here, the use of $q_\mathrm{eff}$ will allow one to express the numerator and denominator
in quite similar forms (polynomials in $y$), potentially making assessments of nuclear structure uncertainties more transparent.
In some cases it may be possible to do more.  For example, isovector responses contributing to 
elastic $\mu \rightarrow e$ conversion in $^{27}$Al can be related directly to
the iso-elastic process of the mirror $\beta$ decay of $^{27}$Si to the ground state of $^{27}$Al.  Such
relationships could be
important if CLFV is discovered, and a more quantitative assessment of nuclear structure uncertainties becomes
a priority. 

\begin{table}
 \caption{ \label{tab:fg} The ratio of the lower to upper components of the muon wave function evaluated at the radius $r_0$ where the monopole charge transition
 density peaks.}
\begin{tabular}{|c|c|c|c|c|c|}
\hline
& & & & &    \\[-.3cm]
 Target &~ $\left| \textstyle{f(r_0)\over g(r_{0}) } \right|$ ~&  Target  &  ~ $\left| \textstyle{f(r_0) \over g(r_0) } \right|$ ~&  Target  &   ~$\left| \textstyle{f(r_0) \over g(r_0) } \right|$ ~ \\
\hline
& & & & & \\[-.2cm]
$^{12}$C & 0.014 & $^{27}$Al & 0.027 & $^{48}$Ti & 0.043  \\
$^{16}$O & 0.017 & $^{28}$Si & 0.030 &  $^{56}$Fe  & 0.049\\
$^{19}$F & 0.019 &  $^{32}$S & 0.036 &   $^{63}$Cu  &  0.061 \\
$^{23}$Na & 0.024 &  $^{40}$Ca  & 0.040  &  $^{184}$W & 0.132 \\
 \hline
 \end{tabular}
\end{table}

\subsection{The $1s$ muon's upper component}\label{sec:2C}
We have computed bound muon Dirac wave functions for
the 11 targets listed in Table \ref{tab:input}.   These targets are both interesting
experimentally and tractable for the shell model:  high quality nuclear effective interactions are
available, allowing us to complete full-space diagonalizations with exact projection of the center-of-mass,
as we describe later.
We also treat $^{184}$W, as a
high-$Z$ comparison target.

\begin{figure*}[ht]   
\centering
\includegraphics[scale=0.47 ]{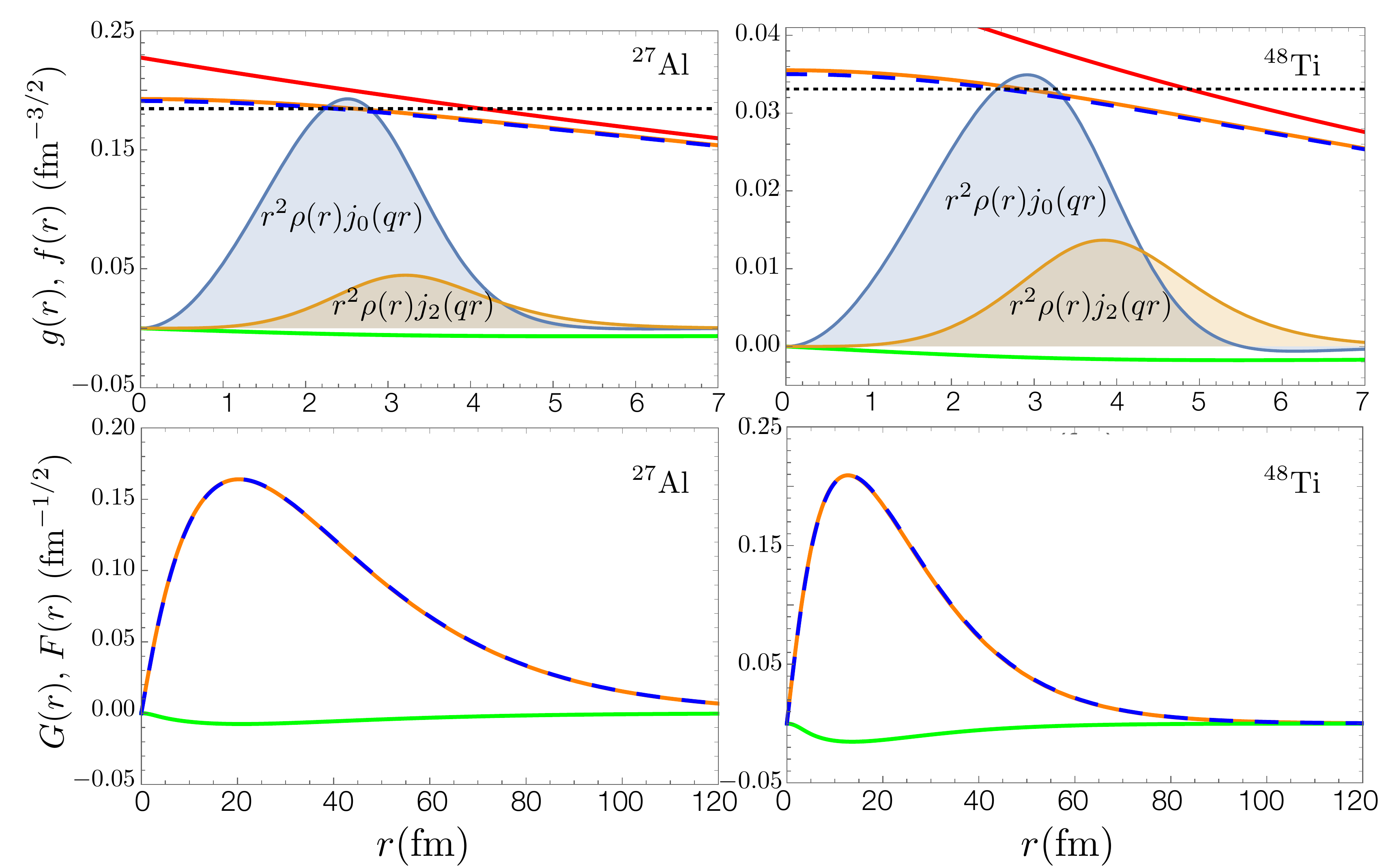}
\caption{Lower panels: muon $\kappa=-1$ bound state Dirac solutions $G(r)$ (orange line) and $F(r)$ (green) are shown for $^{27}$Al (left) and $^{48}$Ti (right), along with the Schr\"odinger
solution (blue dashed).   These solutions are computed for extended nuclear charges, using the parametrizations of
Table \ref{tab:input}, and are properly normalized.  Upper panels: $g(r)$ (orange), $f(r)$ (green), and the Schr\"odinger solutions for extended (blue dashed) and point (red) charge distributions.  
Also shown are the volume-weighted charge distributions $r^2 \rho(r) j_0(q r)$ and $r^2 \rho(r) j_2(q r)$ (shaded).
The overall normalization (but not the relative normalization)
of the two densities is arbitrary.  As 
the muon wave function varies slowly over the nucleus, it is appropriate to use an average value: the black dotted line is the value obtained by averaging over $r^2 \rho(r) j_0(q r)$ (see text) .}
\label{fig:muon}
\end{figure*}

Solutions were obtained by numerically integrating from the origin outward.
The eigenvalue is determined by requiring the wave function to vanish at infinity.  
Results for the $^{27}$Al and $^{48}$Ti $1s$ muon states ($\kappa=-1$) are shown in Fig. \ref{fig:muon}.
The corresponding Schr\"{o}dinger solutions are also plotted: they are
almost indistinguishable from the Dirac radial solutions $g(r)$.  
The Dirac binding energies are nearly identical to the Schr\"{o}dinger values, exceeding the latter by just
0.18\% and 0.37\%, respectively.

Figure \ref{fig:muon} shows that the muon is nearly nonrelativistic, with $g(r)$ and the Schr\"{o}dinger solution
effectively indistinguishable.   For the first 11 entries of Table \ref{tab:fg}, the ratio $|f(r)/g(r)|$ evaluated at
the peak of charge transition density $ r^2 j_0(q_\mathrm{eff} r) \rho(r)$ ranges from 0.014 (C) to 0.061 (Cu).  
The $^{27}$Al value is 0.027, while that for the high-Z target $^{184}$W is 0.132.

The Dirac bound-state normalization
\be  
\int_0^\infty dr\;\left[G^2_{\ell j}(r) + F^2_{\ell j}(r)\right] =1
\label{eq:FGsq}
\ee  
provides an alternative measure of the relativity:
the lower-component contributions to this normalization integral are 0.2\% and 0.5\%, respectively, for $^{27}$Al and $^{48}$Ti.
For $^{184}$W the result is 2.2\%.\\
~~\\
\noindent
One concludes that, for lighter nuclear targets, retention of just the upper component of the muon wave function -- or nearly equivalently,
use of the Schr\"{o}dinger solution --  can be a sensible procedure.  There is a long history of muon-capture studies in light- and medium-mass
nuclei that take advantage of the muon's low velocity in the initial state \cite{Walecka}.
However, the three papers listed in Table \ref{tab:pastwork} that have made this approximation 
either did so for a high-Z nucleus ($^{208}$Pb) \cite{Civitarese2019} or equated the Schr\"{o}dinger density
to its point-nucleus value \cite{Cirigliano2018,Davidson2018}, which, as we will later discuss in this section, can substantially overestimate the muon density even for a relatively light nucleus like $^{27}$Al.

\subsection{The $1s$ muon's  lower component}\label{sec:2D}
Despite the remarks made above, in most of the past work summarized in Table \ref{tab:pastwork} the muon's lower component was
included (either explicitly in the transition density or as an average value).  
In those cases where the lower component was treated explicitly,  the electron partial-wave expansion was truncated to 
$|\kappa|=1$, and only the charge
operator was considered: the authors thus depend on the coherent enhancement of the monopole charge operator to dilute 
contributions from higher multipoles.  As we have noted, this choice of approximations would not in general be successful for
operators other than the charge operator -- so clearly would not be appropriate in an NRET -- and in fact would be a poor choice
even for the charge operator, were the coupling isovector and thus not coherent.  In cases where an average value was used,
the electron was treated as a plane wave.   Thus it is difficult to see a consistent pattern of approximations in past work,
even if we limit our attention to simple operators like the charge.
%

All papers in Table \ref{tab:pastwork} that treated the muon's lower component omitted internucleon velocities.
Not only is this parametrically difficult to justify, but nucleon velocities generate new operators and a new source of coherent enhancement,
as we will later see.  This leads to new nuclear response contributions to the rate that in principle might be new observables, if they can be  isolated by comparing
rates in a variety of targets.

In contrast, $\vec{v}_\mu$ plays a relatively minor role in $\mu \rightarrow e$ conversion.  Numerically, $|\vec{v}_N| \gtrsim |\vec{v}_\mu|$ for the nuclear targets considered here; for example, 
$\sqrt{\langle \vec{v}^{\,2}_\mu \rangle}  \approx 0.05$ for $^{27}$Al while for an unpaired $1d_{5/2}$ valence proton $\sqrt{\langle \vec{v}_N^{\,2} \rangle} \approx 0.21$.  
But there are more important reasons than magnitude for $\vec{v}_\mu$'s relative lack of impact.

\begin{figure}[ht]   
\centering
\includegraphics[scale=0.30 ]{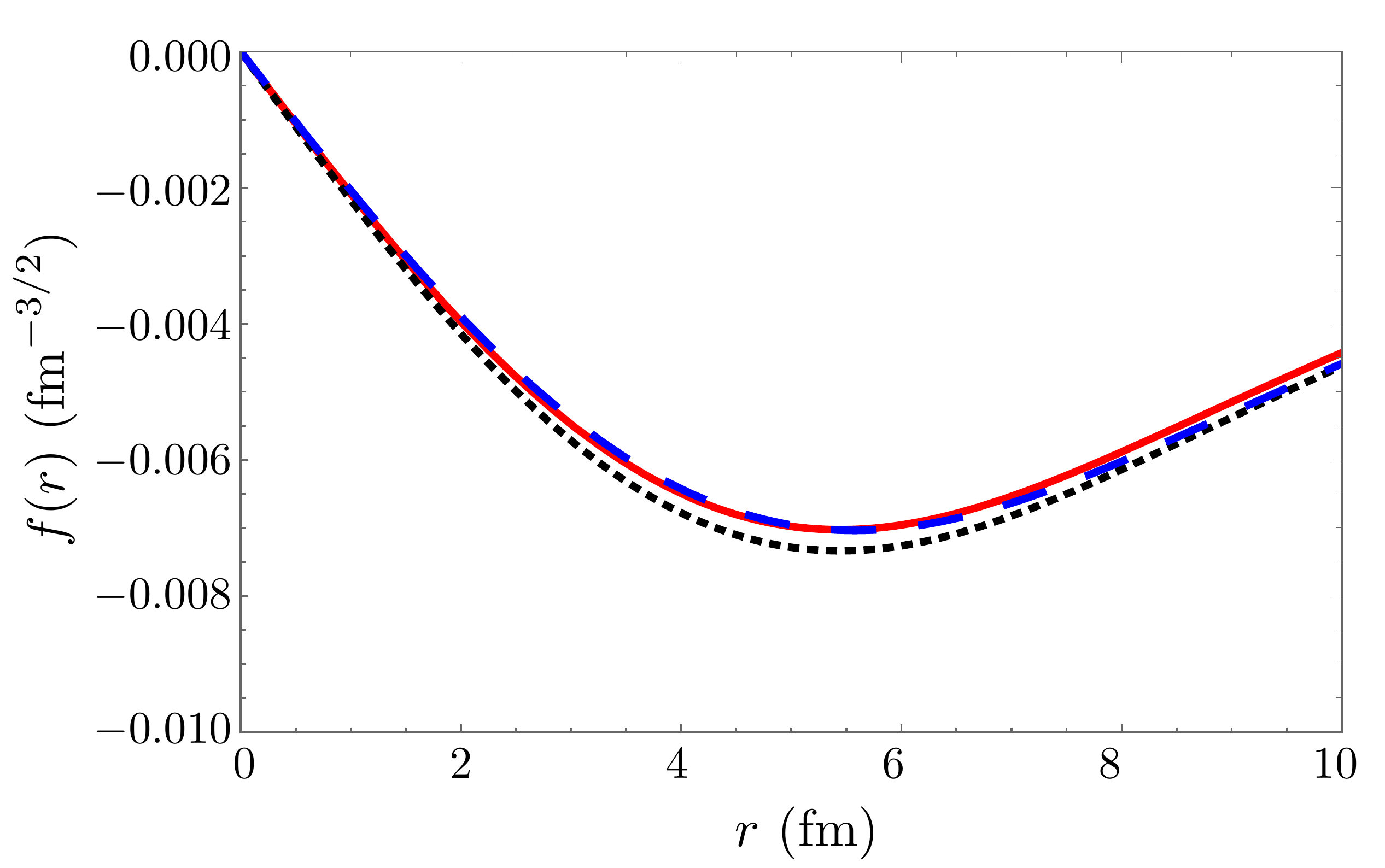}
\caption{The muon's lower-component wave function $f(r)$ computed for $^{184}$W from the Dirac equation (solid  line),
and approximated as a derivative of $g(r)$ using either $\mu$ (dotted line) or $\mu^*$ (dashed line).}
\label{fig:lower}
\end{figure}

The muon's velocity is the operator that generates the lower component Dirac solution $f(r)$ from the upper component $g(r)$,
\begin{eqnarray}
 \psi_{\kappa=-1}^\mu(\vec{r}) &=& \left[ \begin{array}{c}  \xi_s \\  { \vec{\sigma} \cdot \vec{p}_\mu \over 2 \mu^*}  \xi_s \end{array}  \right] i g(r) Y_{00}(\Omega) \nonumber \\ 
 &\approx& \left[ \begin{array}{c}  \xi_s \\  { \vec{\sigma} \cdot \vec{v}_\mu \over 2 }  \xi_s\end{array}  \right]{  i g(r) \over \sqrt{4 \pi}}.
 \label{eq:lowerCC}
\end{eqnarray}
as
\be
 f(r) \approx {1 \over 2 \mu^*} {d g(r) \over d r} \approx {1 \over 2 \mu} {dg(r) \over d r} .
 \label{eq:muCC}
 \ee
 Here  $\mu$ is the muon's reduced mass and $\mu^*$ is the effective reduced mass.  If the Coulomb potential in the vicinity of the nucleus is replaced by an average value $\bar{V}_c$, 
then $\mu^* \equiv  \mu - (|E_{\mu}^\mathrm{bind}|+\bar{V}_c)/2$.
(In $^{27}$Al the difference between $\mu$ and $\mu^*$ is 2.6\%.)  The three functions in Eq. (\ref{eq:muCC}) 
are compared for the high-$Z$ target $^{184}$W in 
Fig. \ref{fig:lower}; differences are small.

The leptonic amplitude involves a product of this wave function with that for the electron, given by Eq. (\ref{eq:Delectron}),
whose upper and lower components are both $o(1)$.  
Consequently, whether
the operator connecting the muon to the electron is even or odd in the Foldy-Wouthuysen sense, $\vec{v}_\mu$ always 
appears in combination with the much larger electron velocity.  
Furthermore, the leptonic amplitude is inclusive, a sum over partial waves.    As we will show by explicit calculation, these properties limit the
role of $\vec{v}_\mu$ to that of a nuclear form factor Coulomb correction, typically at the level of $\approx$ 5\% for the light
and medium mass targets considered here.  It plays no other role in the CLFV physics of $\mu \rightarrow e$ conversion.

\begin{table}
  \caption{ \label{tab:average1} Exact calculations of transition probabilities involving the muon's upper component $g(r)$ have been calculated
  for the charge $M$, transverse spin $\Sigma^\prime$, and longitudinal spin $\Sigma^{\prime \prime}$ operators, evaluated using
  both leading multipoles (e.g., $W_{M_0M_0}$) or full response functions (e.g., $W_{MM}$), and employing
  isoscalar (00), isovector (11), and proton (pp) couplings.  From each calculation we extract a value of $R$ (see text; the top line for each
  target), the ratio of the exact result to that obtained when
  $g(r)$ is replaced by the point Schr\"{o}dinger analog.  If we take $R$ from $W^{00}_{M_0M_0}$ as a baseline, 
  percentage variations in $R$, representing the error introduced if the baseline is used globally, are given in the second line.} 
 \begin{adjustbox}{width=\columnwidth,center}
 \begin{tabular}{|c|c|c|c|c|c|c|}
 \hline
 \rule{0cm}{0.4cm}
 &  \multicolumn{5}{c}{~~~~~~~$R$ and \% variation}  &\\[0.15cm]
\hline
\rule{0cm}{0.5cm}
 Target & $W_{M_0M_0}^{00}$ & $W_{MM}^{00}$ & $W_{M_0M_0}^{pp}$ & $W_{MM}^{pp}$ & $W_{M_0M_0}^{11}$ & $W_{MM}^{11}$ \\[0.15cm]
 \hline
 \rule{0cm}{0.4cm}
 $^{27}$Al & 0.6566 & 0.6565 & 0.6570 & 0.6569  & 0.6434 & 0.6417  \\
  & 0.00\% & $-0.01\%$ & 0.07\% & 0.06\% &  $-2.01\%$ & $-2.26\%$ \\
 \rule{0cm}{0.4cm}
  $^{63}$Cu & 0.3204 & 0.3204 & 0.3200 & 0.3199 & 0.3287 & 0.3281  \\
  & 0.00\% & $-0.01\%$ & $-0.15\%$ & $-0.16\%$   & $2.58\%$ & $2.38\%$ \\ [.15cm]
 \hline
\rule{0cm}{0.5cm}
 Target & $W_{\Sigma^{\prime }_1\Sigma^\prime_1}^{00}$ & $W_{\Sigma^\prime\Sigma^\prime}^{00}$ & $W_{\Sigma^\prime_1\Sigma^\prime_1}^{pp}$ & $W_{\Sigma^\prime\Sigma^\prime}^{pp}$ &  $W_{\Sigma^\prime_1\Sigma^\prime_1}^{11}$ & $W_{\Sigma^\prime\Sigma^\prime}^{11}$ \\[0.15cm]
 \hline
 \rule{0cm}{0.4cm}
 $^{27}$Al & 0.6475 & 0.6463 & 0.6513 & 0.6495 & 0.6561 & 0.6533 \\
  & $-1.38\%$ & $-1.56\%$ & $-0.81\%$ & $-1.08\%$ &  $-0.07\%$ & $-0.49\%$ \\
 \rule{0cm}{0.4cm}
  $^{63}$Cu & 0.3689 & 0.3141 & 0.3673 & 0.3085 &  0.3655 &  0.3026 \\
  & $15.11\%$ & $-1.99\%$ & $14.62\%$ & $-3.72\%$ & $14.05\%$ &  $-5.58\%$ \\ [.15cm]
  \hline
\rule{0cm}{0.5cm}
 Target & $W_{\Sigma^{\prime \prime}_1\Sigma^{\prime \prime}_1}^{00}$ & $W_{\Sigma^{\prime \prime}\Sigma^{\prime \prime}}^{00}$ & $W_{\Sigma^{\prime \prime}_1\Sigma^{\prime \prime}_1}^{pp}$ & $W_{\Sigma^{\prime \prime}\Sigma^{\prime \prime}}^{pp}$ & $W_{\Sigma^{\prime \prime}_1\Sigma^{\prime \prime}_1}^{11}$ & $W_{\Sigma^{\prime \prime}\Sigma^{\prime \prime}}^{11}$ \\[0.15cm]
 \hline
 \rule{0cm}{0.4cm}
 $^{27}$Al & 0.6345 & 0.6339 & 0.6338 & 0.6332 &  0.6331 &  0.6323 \\
  & $-3.36\%$ & $-3.45\%$ & $-3.46\%$ & $-3.56\%$ & $-3.58\%$ & $-3.69\%$ \\
 \rule{0cm}{0.4cm}
  $^{63}$Cu & 0.3275 & 0.2951 & 0.3228 & 0.2921 & 0.3182 & 0.2891  \\
  & $2.20\%$ & $-7.92\%$ & $0.74\%$ & $-8.85\%$ & $-0.71\%$ & $-9.77\%$ \\ [.15cm]
  \hline
 \end{tabular}
 \end{adjustbox}
\end{table}

\begin{table}
  \caption{\label{tab:average2} Errors incurred when $4 \pi W_{O^gO^{(i)f}}^{\tau\tau'}$ is replaced by $\braket{g}\braket{f}W_{OO^{(i)}}^{\tau\tau'}$, where $\braket{g}$ is obtained by averaging over the isoscalar monopole operator $M_{00;0}(q_\mathrm{eff}r)$ and $\braket{f}$ is obtained by averaging over the corresponding lower-component operator $M^{(2)}_{00;0}(q_\mathrm{eff}r)$.}
\centering
 \begin{tabular}{|c|c|c|c|c|}
 \hline
 \rule{0cm}{0.4cm}
 &  \multicolumn{3}{c}{~~~~~~~\% variation in $\langle g \rangle \langle f \rangle$}  &\\[0.1cm]
\hline\rule{0cm}{0.4cm}
  Target& $W_{MM^{(2)}}^{00}$ & $W_{MM^{(2)}}^{11}$ & $W_{MM^{(2)}}^{10}$ & $W_{MM^{(2)}}^{01}$ \\[0.15cm]
  \hline
   \rule{0cm}{0.4cm}
 $^{27}$Al & 0.00\% & $-5.19\%$ & $0.98\%$ & $-5.91\%$ \\
  $^{63}$Cu & 0.00\% & $-4.05\%$ & $-1.27\%$ & $-2.83\%$ \\ [0.15cm]
  \hline\rule{0cm}{0.4cm}
 Target & $W_{\Sigma'\Sigma'^{(0)}}^{00}$ & $W_{\Sigma'\Sigma'^{(0)}}^{11}$ & $W_{\Sigma'\Sigma'^{(0)}}^{10}$ & $W_{\Sigma'\Sigma'^{(0)}}^{01}$ \\[0.15cm]
 \hline
  \rule{0cm}{0.4cm}
 $^{27}$Al & $-5.58\%$ & $-6.78\%$ & $-6.20\%$ & $-6.17\%$ \\
  $^{63}$Cu & $-3.26\%$ & $-4.16\%$ & $-2.69\%$ & $-4.76\%$ \\ [0.15cm]
  \hline\rule{0cm}{0.4cm}
  Target & $W_{\Sigma'\Sigma'^{(2)}}^{00}$ & $W_{\Sigma'\Sigma'^{(2)}}^{11}$ & $W_{\Sigma'\Sigma'^{(2)}}^{10}$ & $W_{\Sigma'\Sigma'^{(2)}}^{01}$ \\[0.15cm]
   \hline
    \rule{0cm}{0.4cm}
 $^{27}$Al & $-5.61\%$ & $-6.47\%$ & $-6.38\%$ & $-5.68\%$ \\
  $^{63}$Cu & $-10.29\%$ & $-13.03\%$ & $-11.04\%$ & $-12.27\%$ \\ [0.15cm]
  \hline\rule{0cm}{0.4cm}
  Target  & $W_{\Sigma''\Sigma''^{(0)}}^{00}$ & $W_{\Sigma''\Sigma''^{(0)}}^{11}$ & $W_{\Sigma''\Sigma''^{(0)}}^{10}$ & $W_{\Sigma''\Sigma''^{(0)}}^{01}$ \\[0.15cm]
   \hline
    \rule{0cm}{0.4cm}
 $^{27}$Al & $-2.70\%$ & $-22.12\%$ & $-2.59\%$ & $-22.11\%$ \\
  $^{63}$Cu & $3.57\%$ & $3.27\%$ & $4.58\%$ & $2.37\%$ \\ [0.15cm]
  \hline\rule{0cm}{0.4cm}
  Target  & $W_{\Sigma''\Sigma''^{(2)}}^{00}$ & $W_{\Sigma''\Sigma''^{(2)}}^{11}$ & $W_{\Sigma''\Sigma''^{(2)}}^{10}$ & $W_{\Sigma''\Sigma''^{(2)}}^{01}$ \\[0.15cm]
   \hline
    \rule{0cm}{0.4cm}
 $^{27}$Al & $-5.33\%$ & $-3.79\%$ & $-5.29\%$ & $-3.82\%$ \\
  $^{63}$Cu & $-5.78\%$ & $-7.27\%$ & $-4.15\%$ & $-8.86\%$ \\ 
  \hline
 \end{tabular}
\end{table}

~~\\
\noindent
{\it Averaging the muon wave function:}  In most of the papers
listed in Table \ref{tab:pastwork}, the gentle variation of the muon's wave function was exploited to remove it from the 
transition density convolution integral, in favor
of an average value.   Rates are reasonably insensitive to how this averaging is done, though we point out below
that past averaging has employed procedures more appropriate for the inclusive process of muon capture.  In contrast,
the averaging we employ here is designed for elastic
$\mu \rightarrow e$ conversion and reproduces the exact results for the leading upper- and lower-component multipoles.

In this paper, as previously mentioned, we will use averaging to simplify our full rate expression.  For example, the effects of $\vec{v}_\mu$
are thereby reflected in one parameter $\langle f \rangle/\langle g \rangle$, which can be factored from all response functions.
There is a second advantage gained from averaging:  it significantly simplifies the nuclear physics. After averaging, we can exploit properties of the harmonic oscillator to analytically evaluate all charge and current one-body matrix elements
and, consequently, derive closed-form expressions for each nuclear response function.  This can be very helpful in understanding the
nuclear physics and how it effects CLFV sensitivities in targets of interest.  In addition, averaging produces $\mu \rightarrow e$ conversion
multipole operators identical to those employed in standard-model semileptonic weak interactions \cite{Donnelly,Walecka}.   The ability to use electron
scattering or beta decay measurements to determine CLFV transition densities could prove helpful in reducing nuclear structure 
uncertainties.

 %

The point-nucleus $1s$ muon  Schr\"{o}dinger solution provides a convenient normalization for the average
density
\be
\phi_{1s}^Z(\vec{r}=0) \equiv \phi_{1s}^Z(\vec{0}) = {1 \over \sqrt{\pi}} \left[ {Z \alpha \mu c \over \hbar} \right]^{3/2},
\ee
where $\mu$ is the reduced mass of the muon.  As the point density exceeds that obtained
by averaging over the nuclear volume,
a common practice is to express the volume-averaged result in the same functional form, but with $Z$ replaced by an effective charge $Z_\mathrm{eff} < Z$,
\[ \bar{\phi}^\mu_{1s}  \equiv \phi_{1s}^{Z_\mathrm{eff}}(\vec{0}). \]
The difference between the physical and effective charges $Z-Z_\mathrm{eff}$ will increase as Z increases.    Alternatively \cite{Walecka} one can express
the probability density $|\bar{\phi}^\mu_{1s}|^2$ as
  \be
\left|    \bar{\phi}^\mu_{1s}  \right|^2  \equiv R  \left| \phi^Z_{1s}(\vec{0})\right|^2,
 \ee
 with $R<1$ decreasing as $Z$ increases.  
 Values of $Z_\mathrm{eff}$ and $R$ for various nuclear
 targets are given in Table \ref{tab:input}, using the averaging procedure we describe below.  Both of these
 ways of expressing  $\bar{\phi}^\mu_{1s}$ in terms of the point Schr\"{o}dinger density have been
 used extensively in muon capture studies.

 The transition density for elastic $\mu \rightarrow e$ conversion is a convolution of a slowly
 varying muon wave function $\phi_{1s}^\mu(r)={1 \over \sqrt{4 \pi}} g(r) $ with a more rapidly varying multipole density whose shape is frequently
 nearly Gaussian (see Fig. \ref{fig:muon}).   The standard procedure for handling such convolution integrals is a Taylor expansion around
 the integrand's peak at $r_0$, generating errors that depends only on the second derivative of the slowly varying function.

When an isoscalar coupling to charge exists, the resulting coherence usually leads to a rate dominated by the isoscalar monopole
operator. Here we slightly modify the Gaussian method described above, effectively
choosing $r_0$ so that $g(r_0) \equiv \langle g \rangle$ yields the exact integrated monopole strength
\begin{eqnarray}
&&  \left|    {\phi}^{Z_\mathrm{eff}}_{1s}(\vec{0})  \right| =  {1 \over \sqrt{\pi} } \left[ { Z_\mathrm{eff} \alpha \mu c \over \hbar} \right]^{3 \over 2} \equiv  \left|    \bar{\phi}^\mu_{1s}  \right|  \equiv {1 \over \sqrt{4 \pi}}\langle g \rangle \nonumber \\
&& ~~~  ={|\langle g.s. | ~ \sum_{i} j_0(q_\mathrm{eff} r_i) Y_{00}(\Omega_i)  {1 \over \sqrt{4 \pi}} g(r_i) ~ |g.s. \rangle| 
 \over |\langle g.s. | \sum_i j_0(q_\mathrm{eff} r_i) Y_{00}(\Omega_i)  ~ |g.s. \rangle |} ~~~~~\nonumber \\
&&~~~~~~~~~~~={\int dr~ r^2 \rho(r) j_0(q_\mathrm{eff} r) {1 \over \sqrt{4 \pi}} \displaystyle{g(r)}
 \over \int dr~ r^2 \rho(r)  ~j_0(q_\mathrm{eff} r) }~~~~~~
 \label{eq:weight}
 \end{eqnarray}
Here $\rho(r)$ is the ground-state expectation value of the isoscalar density.  We make this choice not only because the monopole
 operator is often dominant, but also because the monopole charge density is the best understood property of nuclei, tightly constrained by
 elastic electron scattering.  Thus a precise treatment of
 the Coulomb physics for this multipole is warranted.
 
 The approximation induced by our averaging comes in using the same $\langle g \rangle$ in other multipoles.  
 Defining the muon-wave-function weighted and unweighted multipole operators by
 \be
  \hat{O}_{JM}^g(q)  &\equiv& \sum_{i=1}^A  {1 \over \sqrt{4 \pi}}g(x_i ) \hat{O}_{JM}(q x_i) 
  \ee
 (and similarly for $f(r)$), we have for the response function generated by $\hat{O}_{JM}^g(q)$
 \begin{eqnarray} 
 W_{O^g O^{\prime g}}^{\tau \tau^\prime}\equiv {4 \pi \over 2 j_N+1} \sum_J \langle j_N || O^g_{J;\tau}||j_N \rangle  \langle j_N || O_{J;\tau^\prime}^{\prime g} ||j_N \rangle \nonumber \\
 \rightarrow  {\langle g \rangle^2 \over 2 j_N+1}  \sum_J \langle j_N || O_{J;\tau}||j_N \rangle  \langle j_N || O_{J;\tau^\prime}^{\prime } ||j_N \rangle~~\nonumber \\
  =  {4 \pi |\phi_{1s}^{Z_\mathrm{eff}} (\vec{0}) |^2  \over 2 j_N+1}  \sum_J \langle j_N || O_{J;\tau}||j_N \rangle  \langle j_N || O_{J;\tau^\prime}^{\prime } ||j_N \rangle~~\nonumber \\
 \end{eqnarray}
To test the validity of this replacement, we
have computed the exact $R$ (equivalent to $Z_\mathrm{eff}^3$) for other multipole choices.  
For the charge operator, we vary the
isospin, and we repeat the averaging using the full response function (not just the monopole), thereby including higher multipoles.  We have also repeated these calculations for the transverse electric spin operator $\Sigma_J^\prime$ and longitudinal spin operator $\Sigma_J^{\prime \prime}$,
which we will introduce later in this paper.  The results for $^{27}$Al and $^{63}$Cu are presented in Table \ref{tab:average1}.  
Also shown in the table is the percent error one would make in using $R_0$, the reduction factor obtained for the isoscalar monopole 
charge operator, in all other cases.  For $^{27}$Al the average value of the variation in $R$ is 0.018 and the worse case is a deviation of 0.037.
As $R$ represents a probability, this means that use of $R_0$ globally would affect limits derived on CLFV LECs at the typical
level of $\lesssim 1$\% and in the worse case 2\%.   For $^{63}$Cu, the corresponding typical uncertainty
is $\approx 2.7$\% with a worse-case deviation of 7.6\%.  Consequently, one should associate uncertainties of this magnitude with
the LEC bilinears we later extract from experimental bounds.  In cases where the isoscalar monopole dominates due to its coherence,
the error will be much smaller, as this multipole is exactly reproduced by our averaging procedure.

As we will show in detail later, the lower component $f(r)$ alters the monopole transition density through an interference term
that similarly can be used to define an optimal $\langle f \rangle$:
\be
{1 \over \sqrt{4 \pi}} {\langle f \rangle} \equiv  {\int dr~ r^2 \rho(r) j_1(q_\mathrm{eff} r)  {1 \over \sqrt{4 \pi}} \displaystyle{f(r)}
 \over \int dr~ r^2 \rho(r) j_1(q_\mathrm{eff} r)   }.
 \label{eq:fgmono}
 \ee
As above, we can then test the validity of using this monopole value for $\langle f \rangle$ in other upper-lower component multipole
interference terms (defined explicitly later in this paper),
 \begin{eqnarray} 
 W_{O^g O^{(i) f}}^{\tau \tau^\prime}\equiv {4 \pi \over 2 j_N+1} \sum_J \langle j_N || O^g_{J;\tau}||j_N \rangle  \langle j_N || O_{J;\tau^\prime}^{(i) f} ||j_N \rangle \nonumber \\
 \rightarrow  { \langle g \rangle \langle f \rangle \over 2 j_N+1} \sum_J \langle j_N || O_{J;\tau}||j_N \rangle  \langle j_N || O_{J;\tau^\prime}^{(i)} ||j_N \rangle \nonumber \\
 =  {4 \pi |\phi_{1s}^{Z_\mathrm{eff}} (\vec{0}) |^2  \over 2 j_N+1} {\langle f \rangle \over \langle g \rangle } \sum_J \langle j_N || O_{J;\tau}||j_N \rangle  \langle j_N || O_{J;\tau^\prime}^{(i)} ||j_N \rangle \nonumber \\
 ~~
 \end{eqnarray}
 varying the
 operator and isospin. The small parameter that represents the effects of $\vec{v}_\mu$ is thus $\langle f \rangle/\langle g \rangle$, 
 for which results are shown in Table \ref{tab:average2} for $^{27}$Al and Cu.
 The average deviation is 6\%.  As the magnitude of the ratio $|\langle f \rangle/\langle g \rangle|  \approx 0.027$ for $^{27}$Al this implies a typical rate error
 of $\approx 0.1-0.2\%$; perhaps less, as the leading interference term is reproduced exactly. \\
~\\
{\it Previous averaging procedures.} Many of the references listed in Table \ref{tab:pastwork}
employed averaging.  But typically the authors followed the very early work \cite{Weinberg1959}, in which methods more appropriate to muon capture
were applied to elastic $\mu \rightarrow e$ conversion.  For example, in \cite{Chiang1993} 
\begin{eqnarray}
&&\langle \phi_{1s}\rangle^2 \equiv{\int d^3r\; \rho(r) |\phi_{1s}^\mu(\vec{r}\,)|^2  \over \int d^3r\; \rho(r) } ~ \nonumber \\
&&~~~~~\Rightarrow\langle \phi_{1s} \rangle^2 \equiv {\alpha^3 m_\mu^3 \over \pi} {Z^4_\mathrm{eff} \over Z}. ~~~~~~
\label{eq:Primakoff}
\end{eqnarray}
A nuclear-density-weighted average of the probability arose in Primakoff's early work
on the inclusive process of muon capture where
\begin{eqnarray}
&&\sum_f F(E_f) | \langle i | \Omega \,\phi_{1s}^\mu(\vec{r}\,) | f \rangle |^2  \nonumber \\
&&\approx F(\bar{E}_f) \sum_f  | \langle i | \Omega \,\phi_{1s}^\mu(\vec{r}\,) | f \rangle |^2 \nonumber \\
&&\approx F(\bar{E}_f) \langle i | \Omega \Omega^\dagger  |\phi_{1s}^\mu(\vec{r}\,)|^2  | i \rangle \nonumber \\
&&\approx { \int d^3r \;\rho(r) |\phi_{1s}^\mu(\vec{r}\,)|^2 \over \int d^3r \;\rho(r) } ~F(\bar{E}_f) \langle i | \Omega \Omega^\dagger   | i \rangle 
\end{eqnarray}
Here $\Omega$ is a nuclear operator.  After replacing the final-state phase space $F(E_f)$ by an average value, the sum over
final states was completed by closure.  The closure approximation relates a transition probability to the ground-state expectation
value of the muon probability.  However, the exclusive process of $\mu \rightarrow e$ conversion depends on a simpler ground-state
expectation value of an amplitude containing $g(r)$, and the average over this amplitude can be computed without 
approximations, producing an exact result.

Similarly, the appearance of $Z^4_\mathrm{eff}$ in Eq. (\ref{eq:Primakoff}) is another relic of early muon capture studies, derived in phenomenological fits to rates.
The muon probability contributes a factor of $Z^3$.  Muon capture rates are dominated by transitions to the giant
resonances, which are governed by the Thomas-Reiche-Kuhn sum rule, which predicts a scaling of $NZ/A \approx Z/2$.  The $Z^4_\mathrm{eff}$
that arises in muon capture thus reflects a rather complicated convolution of Coulomb and nuclear structure approximations.  The $Z_\mathrm{eff}$
from muon capture is not simply related
to the $Z_\mathrm{eff}$ of elastic $\mu \rightarrow e$ conversion, which arises solely from
the impact of the finite nuclear size in softening the Coulomb potential at small $r$.

\section{Effective Theory Formulation}
\label{sec:sec3}

\subsection{General construction of the response functions for $\mu \rightarrow e$ conversion}
\label{sec:sec3A}
One can construct the general form of the elastic $\mu \rightarrow e$ conversion rate by considering
the constraints imposed on current and charge matrix elements by the good parity and $CP$ (or $T$) of the
nuclear ground state.  We decompose the currents $\vec{\mathcal{J}}$ and charges $\rho$ into vector and axial-vector
components labeled by their transformation properties under $T$:
\begin{eqnarray}
\vec{\mathcal{J}}(\vec{x}) &\equiv& \vec{J}(\vec{x}) + \vec{J}_5(\vec{x}), ~~~~~ \mathcal{J}_0(\vec{x}) \equiv \rho(\vec{x}) + \rho_5(\vec{x}), \nonumber \\
\vec{J}(\vec{x})&=& \vec{J}^\mathrm{T}(\vec{x})+\vec{J}^{\slashed{\mathrm{T}}}(\vec{x}),~~~~\rho(\vec{x}) = \rho^\mathrm{T}(\vec{x}) +\rho^\slashed{\mathrm{T}}(\vec{x}),~~~ \nonumber \\
\vec{J}_5(\vec{x})&=& \vec{J}_5^\mathrm{T}(\vec{x})+\vec{J}_5^{\slashed{\mathrm{T}}}(\vec{x}),~~~\rho_5(\vec{x}) = \rho_5^\mathrm{T}(\vec{x}) +\rho_5^\slashed{\mathrm{T}}(\vec{x}),~~~ \nonumber \\
\end{eqnarray}
where T indicates time-reversal even and $\slashed{\mathrm{T}}$ time-reversal odd.  No further assumptions are made about
the currents and charges: they can be one-body or multi-body, etc. 

A current $\vec{\mathcal{J}}$ can be decomposed into its longitudinal, transverse electric, and transverse magnetic components.  Each multipole
contributing to an elastic matrix element can be evaluated according to its transformation properties under $P$ and $T$, with those
quantities odd in either quantity eliminated due to the good $P$ and $T$ of the nuclear ground state \cite{deForest}.  We can treat
the charge multipoles similarly.  As the transformation properties depend on whether the multipolarity $J$ is even or odd, we treat these separately.
A straightforward exercise, summarized in Tables \ref{tab:current1} and \ref{tab:current2}, identifies the allowed elastic multipoles.

\begin{table}
  \caption{ \label{tab:current1} The constraints imposed by parity and time reversal on candidate multipole operators mediating 
  elastic $\mu \rightarrow e$ conversion, assuming nuclear ground states of good parity and $CP$.  Slashes and backslashes indicate
  symmetry-forbidden multipoles. }
 \begin{tabular}{|c|c|c|c|c|c|c|}
 \hline
 \rule{0cm}{0.4cm}
 &  \multicolumn{6}{c|}{~ Symmetries:  Parity $\cancel{\mathrm{O}}$~Time $\bcancel{\mathrm{O}}$ }  \\[0.15cm]
\hline
\rule{0cm}{0.5cm}
 Current & J even & J odd~ & J even & J odd~ & J even & J odd~ \\[0.15cm]
 \hline
 \rule{0cm}{0.4cm}
$ \vec{J}^\mathrm{T}$ & $\mathcal{L}_J$  & $\bcancel{\cancel{\mathcal{L}_J}}$ & $\mathcal{T}^{el}_J$  & $\bcancel{\cancel{\mathcal{T}^{el}_J}}$  & $\cancel{\mathcal{T}^{mag}_J}$ & $\bcancel{\mathcal{T}^{mag}_J}$    \\
 \rule{0cm}{0.4cm}
$ \vec{J}^\slashed{\mathrm{T}}$ & $\bcancel{\mathcal{L}_J}$  & $\cancel{\mathcal{L}_J}$ & $\bcancel{\mathcal{T}^{el}_J}$  & $\cancel{\mathcal{T}^{el}_J}$  & $\bcancel{\cancel{\mathcal{T}^{mag}_J}}$ & $\mathcal{T}^{mag}_J$    \\
\rule{0cm}{0.5cm}
$ \vec{J}_5^\mathrm{T}$ & $\cancel{\mathcal{L}^5_J}$  & $\bcancel{\mathcal{L}^5_J}$ & $\cancel{\mathcal{T}^{5 \, el}_J}$  & $\bcancel{\mathcal{T}^{5 \, el}_J}$  & $\mathcal{T}^{5 \, mag}_J$ & $\bcancel{\cancel{\mathcal{T}^{5 \, mag}_J}}$    \\
 \rule{0cm}{0.4cm}
 $ \vec{J}_5^\slashed{\mathrm{T}}$ & $\bcancel{\cancel{\mathcal{L}^5_J}}$  & $\mathcal{L}^5_J$ & $\bcancel{\cancel{\mathcal{T}^{5 \, el}_J}}$  & $\mathcal{T}^{5 \,el}_J$  & $\bcancel{\mathcal{T}^{5 \, mag}_J}$ & $\cancel{\mathcal{T}^{5 \, mag}_J}$    \\[.15cm]
  \hline
 \end{tabular}
\end{table}

\begin{table}
  \caption{ \label{tab:current2} The constraints imposed by parity and time reversal on candidate multipole operators mediating 
  elastic $\mu \rightarrow e$ conversion, assuming nuclear ground states of good parity and CP.  Slashes and backslashes indicate
  symmetry-forbidden multipoles.}
 \begin{tabular}{|c|c|c|}
 \hline
 \rule{0cm}{0.4cm}
&  \multicolumn{2}{c|}{~ Symmetries:  Parity $\cancel{\mathrm{O}}$~Time $\bcancel{\mathrm{O}}$~~ }    \\[0.15cm]
\hline
\rule{0cm}{0.5cm}
 Current & ~~~~~~ J even~~~~~ & J odd \\[0.15cm]
 \hline
 \rule{0cm}{0.4cm}
$ \rho^{\mathrm{T}}$ & $\mathcal{M}_J$  & $\bcancel{\cancel{\mathcal{M}_J}}$    \\
 \rule{0cm}{0.4cm}
$ \rho^\slashed{\mathrm{T}}$ & $\bcancel{\mathcal{M}_J}$  & $\cancel{\mathcal{M}_J}$    \\
\rule{0cm}{0.5cm}
$ \rho_5^{\mathrm{T}}$ & $\cancel{\mathcal{M}^5_J}$  & $\bcancel{\mathcal{M}^5_J}$     \\
 \rule{0cm}{0.4cm}
$ \rho_5^\slashed{\mathrm{T}}$ & $\bcancel{\cancel{\mathcal{M}^5_J}}$  & $\mathcal{M}^5_J$     \\[.15cm]
  \hline
 \end{tabular}
\end{table}

We now add in constraints imposed by the quantum mechanical degrees of freedom available to us.
Nucleon charges generate a scalar density $\rho_{sc}(\vec{x})$.
Velocities generate a convection current density $\vec{J}_c(\vec{x})$, and spins a magnetic density $\vec{\mu}(\vec{x})$.  One can form a dot product
of spin and velocity to create a pseudoscalar density $\rho_{ps}(\vec{x})$ and a cross product to form a vector density $\vec{\omega}(\vec{x})$.
Furthermore one can form gradients of scalars and curls of currents.  Distributing these densities according to their transformation 
properties yields physical representations of the currents and charges,
\begin{eqnarray}
 \vec{J}^\mathrm{T}(\vec{x})&=& \vec{\omega}(\vec{x})+ \vec{\nabla} \rho_{sc}(\vec{x}) \nonumber \\
  \vec{J}^{\slashed{\mathrm{T}}}(\vec{x})&=& \vec{J}_c(\vec{x})+ \vec{\nabla} \times \vec{\mu}(\vec{x}) \nonumber \\
    \vec{J}_5^\mathrm{T}(\vec{x})&=& \vec{\nabla} \times \vec{\omega}(\vec{x})+ \vec{\nabla} \rho_{ps}(\vec{x}) \nonumber \\
 \vec{J}_5^{\slashed{\mathrm{T}}}(\vec{x})&=& \vec{\mu}(\vec{x})+ \vec{\nabla} \times \vec{J}_c(\vec{x}) \nonumber \\
  \rho^{\mathrm{T}}(\vec{x})&=& \rho_{sc}(\vec{x})+ \vec{\nabla} \cdot \vec{\omega}(\vec{x}) \nonumber \\
    \rho^{\slashed{\mathrm{T}}}(\vec{x})&=& \vec{\nabla} \cdot \vec{J}_c(\vec{x})   \nonumber \\
      \rho_5^\mathrm{T}(\vec{x})&=& \rho_{ps}(\vec{x})    \nonumber \\
          \rho_5^{\slashed{\mathrm{T}}}(\vec{x})&=&  \vec{\nabla} \cdot \vec{\mu}(\vec{x})  
 \end{eqnarray}
First consider the charges.  Table \ref{tab:current2} allows us to eliminate $\rho^{\slashed{\mathrm{T}}}(\vec{x})$ and 
$\rho_5^\mathrm{T}(\vec{x})$.  The divergences can be partially integrated to yield longitudinal projections of $\vec{\omega}(\vec{x})$
and $\vec{\mu}(\vec{x})$, terms that will already be generated by the currents above.  Consequently, the only contribution from the charges
comes from the even multipoles of 
\[   \rho^{\mathrm{T}}(\vec{x})= \rho_{sc}(\vec{x}). \]
This eliminates the candidate response function associated with $\mathcal{M}_j^5$: elastic $\mu \rightarrow e$ conversion cannot
probe axial-charge interactions.

For the currents, we note
\begin{enumerate}
\item Table \ref{tab:current1} establishes that $\vec{J}^\slashed{\mathrm{T}}$ and $\vec{J}_5^\mathrm{T}$ only contribute through 
the transverse magnetic projection.  Partially integrating the curls generates transverse electric projectors,
so we see both curls are redundant, as these terms are already included in the transverse electric contributions of  $\vec{J}^\slashed{\mathrm{T}}_5$
and $\vec{J}^\mathrm{T}$, respectively.
\item The gradient contribution to $\vec{J}^\mathrm{T}_5$ is longitudinal, and thus cannot contribute to a transverse magnetic multipole.
\item The gradient contribution to $\vec{J}^\mathrm{T}$ contributes to the symmetry-allowed longitudinal multipole.  But if partially 
integrated, it generates a charge multipole of $\rho_{sc}$, which we have already included.
\item The curl term in $\vec{J}^\slashed{\mathrm{T}}_5$ can only contribute to the allowed transverse electric multipole.  But on 
partially integrating, the projector becomes magnetic, and thus redundant with the convection current contribution to $\vec{J}^\slashed{\mathrm{T}}$.
\end{enumerate} 
It follows that all curls and gradients in the currents are unnecessary, so that
\begin{eqnarray}
 \vec{J}^{\slashed{\mathrm{T}}}(\vec{x})&=& \vec{J}_c(\vec{x}) \nonumber \\
 \vec{J}^\mathrm{T}(\vec{x})&=& \vec{\omega}(\vec{x})  \nonumber \\
 \vec{J}_5^{\slashed{\mathrm{T}}}(\vec{x})&=& \vec{\mu}(\vec{x})\nonumber \\
  \vec{J}_5^\mathrm{T}(\vec{x})&=& 0 
\end{eqnarray}
This eliminates the otherwise symmetry-allowed axial transverse magnetic multipole $\mathcal{T}_J^{5 \, mag}$.

We conclude that the elastic $\mu \rightarrow e$ conversion rate contains six independent response functions,
\begin{eqnarray}
\mathcal{M}_J&:&~~\mathrm{even~J} \ge 0 ~\mathrm{multipoles~of~}\rho_{sc}(\vec{x}) \rightarrow M_J,\nonumber \\
\mathcal{T}_J^{mag}&:&~~\mathrm{odd~~J} \ge 1 ~\mathrm{multipoles~of~}\vec{J}_{c}(\vec{x}) \rightarrow  \Delta_J, \nonumber \\ 
\mathcal{L}_J&:&~~\mathrm{even~J} \ge 0 ~\mathrm{multipoles~of~}\vec{\omega}(\vec{x})  \rightarrow \Phi^{\prime \prime}_J, \nonumber \\ 
\mathcal{T}_J^{el}&:&~~\mathrm{even~J} \ge 2 ~\mathrm{multipoles~of~}\vec{\omega}(\vec{x}) \rightarrow \tilde{\Phi}^\prime_J, \nonumber \\ 
\mathcal{L}_J^5&:&~~\mathrm{odd~~J} \ge 1 ~\mathrm{multipoles~of~}\vec{\mu}(\vec{x}) \rightarrow \Sigma^{\prime \prime}_J, \nonumber \\ 
\mathcal{T}_J^{5 \, el}&:&~~\mathrm{odd~~J} \ge 1 ~\mathrm{multipoles~of~}\vec{\mu}(\vec{x})  \rightarrow \Sigma^\prime_J,~~~~
\label{eq:symmetries}
\end{eqnarray}
where on the right we indicate the effective one-body operators that will generate these response functions
in the NRET formalism we introduce below.  An NRET with a full basis of operators
should generate these six response functions.  

\subsection{Motivations for NRETs}
\label{sec:sec3B}
The procedure in EFT is to construct a complete set of operators, constrained by the applicable symmetries,  up to a given order in an expansion in one or more small
parameters, with the expectation that these operators and their coefficients (the low-energy constants or LECs) will encompass
all of the physics relevant to the energy or momentum scale of the construction.   

Measurements of elastic $\mu \rightarrow e$ conversion are performed in nuclei, and, as detailed above, the constraints of $P$ and $CP$ allow six response
functions.  The NRET's charge and current operators should 
generate these  responses, providing a microscopic basis for their evaluation.
Absent a microscopic formulation, it becomes difficult to relate CLFV constraints
obtained in different nuclei. 

Another purpose of the NRET is to serve as an interface with UV formulations of CLFV.   After the nuclear
physics is done, $\mu \rightarrow e$ conversion constraints on CLFV will be encoded as limits
on specific bilinear combinations of the NRET coefficients, the LECs.  
These limits can then be used as constraints on higher-level EFTs or UV theories. 
This is generally a more efficient way to proceed than the alternative
of repeatedly selecting one UV theory and doing a ``top down" reduction to the nuclear scale, which requires one to repeat
a series of steps for each candidate UV theory being explored.  

As we noted in the Introduction, this strategy has been developed for DM direct detection over the past
decade and has proved to be very popular.   The NRET formulation of the nuclear physics has been incorporated 
into easily used scripts, which others have then used as the starting point for other scripts that treat the
matching to higher-level EFTs, such as those formulated at the light quark and gluon level.   In 
Appendix \ref{AppendixC} we describe a publicly available script for $\mu \rightarrow e$ conversion that can be utilized in a similar fashion.

\subsection{Constructing the NRET}
\label{sec:sec3C}
In $\mu \rightarrow e$ conversion the relativistic motion of the electron can be easily
removed, effectively replacing the electron's velocity by the observable $\hat{q}$, the direction of
three-momentum transfer.  This was done in writing Eq. (\ref{eq:Delectron}).  (See also Appendix \ref{AppendixA}.)  As the remaining kinematics
are nonrelativistic, the general form of the transition amplitude can be written somewhat schematically as
\begin{eqnarray}
\langle f |  \mathcal{L}_{\textrm{int}} | i \rangle    \rightarrow  \sqrt{E_e \over 2m_e}  {q_\mathrm{eff} \over q} ~~~~~~~~~~~~~~~~~~~~~~~~~~ \nn \\
\times\left\{ \begin{array}{l}   \int d\vec{r}~ \xi_{m_f}^\dagger  O_L   \xi_{m_i} 
~\phi^\mu_{1s}(\vec{r}\,)  e^{-i \vec{q}_\mathrm{eff} \cdot \vec{r}} \\[.16cm]
~~~~~~~~~~~~~~~~ \times \langle  f_N | \sum_i O_N (i) \delta(\vec{r}-\vec{r}_i) |  i_N \rangle  \\[.16cm]
  \int d\vec{r}~   \xi_{m_f}^\dagger  \vec{O}_L  \xi_{m_i} ~\phi^\mu_{1s}(\vec{r}\,)
~ e^{-i \vec{q}_\mathrm{eff} \cdot \vec{r}}~ \\[.16cm]
~~~~~~~~~~~~~~~~\cdot \langle  f_N |  \sum_i \vec{O}_N (i) \delta(\vec{r}-\vec{r}_i)   | i_N \rangle \end{array}  \right. ~~~~
\label{eq:reduced3}
\end{eqnarray}
where the upper expression is taken if the operators acting between the leptons and nucleons are scalars
and the bottom if they are vectors. 
Here $\xi_m$ is the lepton
Pauli spinor.  

The next step is to construct all of the candidate NRET operators that might contribute to the above expression.
Hermitian operators can be constructed from the lepton and nucleon
identity operators $1_L$ and $1_N$ and from the four dimensionless three-vectors
 \ba
 && i \hat{q} ={i \vec{q} \over |\vec{q}\,|}, \ \ \ \ \vec{v}, \ \ \ \ \vec{\sigma}_L, \ \ \ \ \vec{\sigma}_N.
 \ea
 Here $\hat{q}$ is the unit vector along the three-momentum transfer to the leptons and $\vec{v}$
 represents all bound-state relative velocities.
$\hat{q}$ provides a natural quantization direction, 
 relative to which one can project out the longitudinal ($\hat{q} \, \cdot$) and transverse ($\hat{q} \times$)
 components of nuclear currents.  These building blocks for the NRET transform as
 \begin{center}
\begin{tabular}{|c|ccc|}
\hline
& $\dagger$ & T & P \\
\hline
$\vec{\sigma}_L,\vec{\sigma}_N$ & +1 & $-1$ & +1 \\
$i \hat{q}$  & +1 & +1 & $-1$ \\
$ \vec{v}$  & +1 & $-1$& $-1$\\
\hline
\end{tabular}
\end{center}
and thus from these Hermitian operators we can construct various semi-leptonic CLFV interactions.
From the table one can also determine the transformation properties of the interactions
under $P$ and $T$. \\
~\\
\noindent
{\it The velocity operator.}  The electron's velocity is measured with respect to that of the
recoiling final-state nucleus.  The initial-state nucleus is assumed to be at rest.  Thus, after removal of the electron, only bound-state Jacobi velocities remain.

The standard unitary transformation to the inter-particle Jacobi velocities
\begin{eqnarray}
\dot{\vec{v}}_N(1) &=& {1 \over \sqrt{2}} \left[ \vec{v}_N(2)-\vec{v}_N(1) \right]  \nonumber \\
\dot{\vec{v}}_N(2) &=& {1 \over \sqrt{6}} \left[ 2\vec{v}_N(3)-(\vec{v}_N(1)+\vec{v}_N(2) )\right]~\dots ~~~
\end{eqnarray}
will generate $A-1$ internucleon velocities.  The last intrinsic Jacobi velocity is that of the muon, 
defined relative to the center-of-mass of nucleons in the same bound state:
\be
\dot{\vec{v}}_\mu  &=&  \left[ \vec{v}_\mu -{1 \over A} \sum_{i=1}^A \vec{v}_N(i) \right] . 
\ee
Henceforth we will drop the superscript dot, with the understanding that all velocities are
Galilean invariant and intrinsic.  

All of the velocities defined are small, with $|\vec{v}_N(i)| > |\vec{v}_\mu|$.  This hierarchy
can be utilized in constructing the NRET. The logical progression is
\begin{enumerate} 
\item Ignore all velocities, retaining only the point-nucleus operators $1_N$ and $\vec{\sigma}_N$.  
\item Retain $\vec{v}_N$ to first order but neglect the smaller $\vec{v}_\mu$, which generates the muon lower component.
\item Retain both $\vec{v}_N$ and $\vec{v}_\mu$ to first order.
\end{enumerate}
These NRET choices are intimately connected to the CLFV nuclear response functions we later generate.
The first NRET -- which we will label the allowed approximation -- generates three response functions,
corresponding to charge, longitudinal spin, and transverse spin.  
In the papers of Table \ref{tab:pastwork} prior to \cite{RulePRL}, the transverse and longitudinal spin responses were lumped together, though their form factors and dependence on nucleon-level operators are distinct.
(This is surprising as this issue in DM direct detection has been treated properly
since the early work of Engel \cite{Engel91}.)  
Thus the proper form of the lowest-level NRET for elastic $\mu \rightarrow e$ conversion has not been
used until recently \cite{RulePRL,Hoferichter:2022mna}.

The second NRET, which includes $\vec{v}_N$, is the minimal theory capable of generating all six of the nuclear response functions allowed by $P$ and $CP$,
and in our view consequently plays a special role, defining what can and cannot be learned about CLFV 
from elastic $\mu \rightarrow e$ conversion.  The inclusion of $\vec{v}_N$ makes the treatment of $\mu \rightarrow e$ conversion 
compatible with standard formulations of semi-leptonic weak interactions \cite{Walecka,Donnelly}.  It also
generates some interesting nuclear physics, including a new form of coherence that is important
for certain targets, including $^{27}$Al.  This NRET was introduced in \cite{RulePRL}; none of the earlier
papers listed in Table \ref{tab:pastwork} considered the three response functions generated by $\vec{v}_N$.
(See also \cite{Hoferichter:2022mna}.)

In the third step we extend the NRET to include $\vec{v}_\mu$.  This does not
affect selection rules or create truly distinct responses, reflecting the fact that the leptonic part of the amplitude
is not controlled by $P$ or $CP$ selection rules, as it is inelastic and as a sum over all electron distorted waves is done.
The inclusion of $\vec{v}_\mu$ does generate corrections to the nuclear response functions that we are able
to cast in a relatively simple form, thanks to our use of $q_\mathrm{eff}$.  These corrections are typically at the few percent level
for targets of current experimental interest.    In all three NRET's we will carry out 
the full multipole expansion, faithfully treating the largest of the small parameters, $y$.  
We limit our treatment of velocities to linear order as
ambiguities arise in treating velocities
in higher order in bound nuclear systems \cite{Serot}.

We begin by developing the operators for the second NRET described above.
The nucleon spin operator can be combined with $\vec{v}_N$ as
$\vec{v}_N \cdot \vec{\sigma}_N$ and $\vec{v}_N \times \vec{\sigma}_N$, but not as the rank-2 tensor
$\left[ \vec{v}_N \otimes \vec{\sigma}_N \right]_2$, 
which would not triangulate between spin-$\textstyle{1 \over 2}$ nucleon states.  As any propagator effects
in the scattering of the leptons off the nucleon would be evaluated at fixed $|\vec{q}\,|$, they can be absorbed 
into the LECs.  Consequently, the operators can be expressed as simple products of leptonic and nucleonic
operators. 

The available lepton scalars are $1_L$ and $i \hat{q} \cdot \vec{\sigma}_L$, while the vectors are $\vec{\sigma}_L$ and its transverse projection $i \hat{q} \times \vec{\sigma}_L$; the nucleon scalars are $1_N$, $\vec{v}_N \cdot \vec{\sigma}_N$, $i \hat{q} \cdot \vec{\sigma}_N$, $i \hat{q} \cdot \vec{v}_N$, 
and $i \hat{q} \cdot (\vec{v}_N \times \vec{\sigma}_N)$, while the vectors are $\vec{\sigma}_N$, $\vec{v}_N$, $\vec{v}_N \times \vec{\sigma}_N$ and their
transverse projections.  One forms all scalars, noting that products of transverse projections can be eliminated using
$\displaystyle{(i \hat{q} \times \vec{A}) \cdot (i \hat{q} \times \vec{B}) = -\vec{A} \cdot \vec{B} - i\hat{q} \cdot \vec{A} ~i \hat{q} \cdot \vec{B}}$.  We
identify a total of 16 operators:
\allowdisplaybreaks
\be
\CO_1 &=& 1_L ~1_N, \nn \\
\CO^\prime_2 &=& 1_L ~i \hat{q} \cdot \vec{v}_N, \nn \\
 \CO_3 &=& 1_L~  i \hat{q} \cdot  \left[ \vec{v}_N \times \vec{\sigma}_N \right], \nn \\
\CO_4 &=& \vec{\sigma}_L \cdot \vec{\sigma}_N, \nn \\
\CO_5 &=&  \vec{\sigma}_L \cdot \left( i \hat{q} \times \vec{v}_N \right), \nn \\
 \CO_6&=&  i \hat{q} \cdot \vec{\sigma}_L ~ i \hat {q} \cdot \vec{\sigma}_N, \nn  \\
\CO_7 &=& 1_L~ \vec{v}_N \cdot \vec{\sigma}_N, \nn \\
\CO_8 &=& \vec{\sigma}_L\cdot \vec{v}_N, \nn \\ 
 \CO_9 &=& \vec{\sigma}_L \cdot \left( i \hat{q} \times \vec{\sigma}_N \right), \nn \\
\CO_{10} &=& 1_L~ i \hat{q} \cdot \vec{\sigma}_N, \nn\\
\CO_{11} &=& i \hat{q} \cdot \vec{\sigma}_L ~ 1_N,  \nn \\
\CO_{12} &=& \vec{\sigma}_L \cdot \left[ \vec{v}_N \times \vec{\sigma}_N \right], \nn\\ 
\CO^\prime_{13} &=& \vec{\sigma}_L \cdot  \left( i \hat{q} \times \left[ \vec{v}_N \times \vec{\sigma}_N \right] \right),  \nn \\
\CO_{14} &=& i  \hat{q} \cdot \vec{\sigma}_L ~ \vec{v}_N \cdot \vec{\sigma}_N,  \nn \\
\CO_{15} &=& i \hat{q} \cdot \vec{\sigma}_L~ i \hat{q} \cdot \left[ \vec{v}_N \times \vec{\sigma}_N \right], \nn\\
\CO^\prime_{16} &=& i \hat{q} \cdot \vec{\sigma}_L ~i \hat{q} \cdot  \vec{v}_N. 
\label{eq:ops}
\ee
Alternatively, one can count the possibilities by noting that two leptonic scalars, $1_L$ and $i\hat{q} \cdot \vec{\sigma}_L$,
can be combined with two hadronic scalars, $1_N$ and $\vec{v}_N \cdot \vec{\sigma}_N$; and four leptonic
vectors, $i \hat{q}$, $\vec{\sigma}_L$, $i\hat{q} \times \vec{\sigma}_L$, and $\left[(i\hat{q} \otimes i\hat{q})_2 \otimes \vec{\sigma}_L)\right]_1$,
can be combined with three hadronic vectors, $\vec{v}_N$, $\vec{\sigma}_N$, and $\vec{v}_N \times \vec{\sigma}_N$.
The tensor in the electron's velocity that appears in $\left[(i\hat{q} \otimes i\hat{q})_2 \otimes \vec{\sigma}_L)\right]_1$ is a 
component of $i\hat{q} \cdot \vec{\sigma}_L i\hat{q}$.

The construction above is analogous to that performed earlier for DM direct detection \cite{Liam1,Liam2}, and in
fact we have adopted a similar labeling of the operators.  However there are some significant differences that
we discuss in Appendix \ref{AppendixA}, for those familiar with this earlier work.

Each operator
can have distinct couplings to protons and neutrons.  Thus the NRET interaction 
we employ in this paper takes the form
\be
 \sum_{\alpha=n,p} ~\sum_{i=1}^{16} c_i^{\alpha} \CO_i^{\alpha},
\label{eq:int}
\ee
where the unknown numerical coefficients $c_i$ 
would need to be determined by experiment or matched to some predictive UV theory.
 One can factorize 
the space/spin and proton/neutron components of Eq. (\ref{eq:int}) by introducing isospin, which
is also useful as an approximate symmetry of the nuclear wave functions.  Thus
an equivalent form for our interaction is
\be
 \sum_{i=1}^{16} ( c_i^{0} 1 + c_i^{1} \tau_3) \CO_i &=& \sum_{\tau=0,1} \sum_{i=1}^{16} c_i^\tau \CO_i t^\tau, 
\label{eq:iso}
\ee
where $c_i^{0} = {1 \over 2} (c_i^{\mathrm{p}}+c_i^{n})$ and $c_i^{1} = {1 \over 2} (c_i^{p}-c_i^{n})$. The isospin matrices are $t^0=1$ and $t^1=\tau_3$.

The NRET has a total of 32 parameters, associated with 16 space/spin operators each
of which can have distinct couplings to protons and neutrons.  If we exclude operators that are not
associated with spin-0 or spin-1 mediators, 12 space/spin operators and 24 couplings remain.
A main goal of this paper is to determine the specific constraints that a program of elastic
$\mu \rightarrow e$ conversion 
measurements can place on these LECs.

The coefficients $c_i$ in  other application of  NRET would be functions of $q^2$  \cite{Liam1}, 
\[ c_i = c_i(0) + c_i(2) q^2 + c_i(4) q^4 + \cdots \equiv F_i ({q^2 \over \Lambda^2}) \]
In DM direct detection the presence of tree-level pions in some channels would lead to $F_i \approx {1 \over q^2 + m_\pi^2}$, 
making $c_i$ a function of the observed nuclear recoil energy.  If couplings were
treated as constants in fits, a bit of information would be missing:
the deduced coupling $c_i$ would be correct only at some average momentum.  But this is not a concern in
$\mu \rightarrow e$ conversion, as $q^2 \approx m_\mu^2$ is fixed.  One
can represent $F_i$ by a constant $c_i$, and know exactly how to match to an EFT with pions.
This argument extends to interactions with massless mediators, $F_i \approx 1/q^{2}$.

As defined in Eq. (\ref{eq:iso}), the $c_i$'s carry dimensions of $1/(\mathrm{mass})^2$. 
Conventionally limits on $\mu \rightarrow e$ conversion are expressed as a branching ratio
with respect to the standard-model muon capture rate
\be
B = {\Gamma[ \mu^- + (A,Z) \rightarrow e^- + (A,Z)] \over \Gamma[ \mu^- + (A,Z) \rightarrow \nu_\mu + (A,Z-1)]} .
\ee
Consequently, it makes sense
to introduce a set of dimensionless LECs $\tilde{c}_i$ that are normalized to the weak scale,
\be
 c_i \equiv \tilde{c}_i/v^2= \sqrt{2} G_F \, \tilde{c}_i,
 \label{eq:scale}
 \ee
 where $v=246.2$ GeV is the Higgs vacuum expectation value and $G_F=1.166\times 10^{-5}$/GeV$^{2}$ is the Fermi constant.
 Alternatively, given an experimental measurement (or limit) on an LEC, one can define an energy characteristic of the CLFV,
 \[ \Lambda_i^\tau \equiv {1 \over \sqrt{|c_i^\tau}|}= {v \over \sqrt{ |\tilde{c}_i^\tau|}}  \]
 This is the energy scale that one would associate with the BSM source of the CLFV.   Often, experimental sensitivities are described
 in terms of their reach in $\Lambda_i^\tau$.
 
 \subsection{Matching}
 The NRET developed here, which follows closely a similar construction for DM phenomenology, is
 frequently used in conjunction with various ``top-down" EFT constructions.  The NRET is formulated at the nuclear level,
 with the impulse approximation employed to relate observables in different
 nuclear targets.  The operator expansion linear in $v_N $ (so including interactions of order $1/m_N$) is consistent with their use
 with nuclear wave functions derived from the many-body Schroedinger equation  \cite{Serot}.
 The coefficients of the NRET operators are constrained by fitting experiment. In
 $\mu \rightarrow e$ conversion the available information is limited in principle to the coefficients of the six nuclear response functions
 that we were able to identify by symmetry arguments. 
 It will take great effort and some good fortune to separate the response functions: if that is not possible, the total rate 
 becomes the only constraint.
 
 The NRET forms a convenient interface between the particle physics of CLFV and nuclear calculations of the 
 $\mu \rightarrow e$ rate.  The NRET operator basis fits naturally with many-body nuclear methods: formulating the physics in this
 basis makes form factor calculations more accessible to nuclear structure theorists.  The NRET operators must be 
 evaluated at $q \approx m_\mu$ and include some interactions that cannot be probed in standard-model semi-leptonic
 interactions.   The quality of the form factor calculations directly
 impacts our ability to extract meaningful constraints on CLFV from $\mu \rightarrow e$ conversion.
 
 A recent Snowmass white paper on DM effective theory \cite{Snowmass} nicely describes the particle physics interface,
 specifically, ``top-down" reductions that relate EFTs formulated at higher energy scales to our NRET.
 For example, in \cite{Bishara,Brod} an EFT formulated for operators with quarks, gluons, and photons as the external
 states, is reduced to the NRET form used here.  This reduction uses
 techniques familiar from heavy baryon effective theory, yielding in the DM case products of DM and nucleon charges
 and currents.  The end result -- see Appendix A of \cite{Bishara} -- is a dictionary of Wilson coefficients
 that relate the coefficients of the quark/gluon-level operators to the LECs $c_i$ of the NRET we develop here.   The reduction from a covariant
 EFT to a Galilean-invariant NRET leads to correlations between the NRET operators, encoded in the dictionary.  Connections
 between LO charge operators and $v_N$-dependent currents, or the reverse, necessarily accompany the nonrelativistic
 reduction.  The extensive literature on such reductions is summarized in \cite{Snowmass}. 

In this way, a theorist interested in a specific formulation of CLFV at some higher energy scale, can test its consistency
with experimental constraints encoded in the NRET LECs -- more precisely, the combinations of these LECs 
that are associated with six nuclear responses functions.
We will use the NRET to determine precisely what CLFV information might be available from future studies of $\mu \rightarrow e$ conversion.
 There are always more candidate CLFV operators in the starting EFT than nuclear observables,  some of which play no role in elastic $\mu \rightarrow e$ conversion
 due to selection rules.  Thus while $\mu \rightarrow e$ studies can establish that CLFV occurs, it will not fully determine the particle physics
 origin of the symmetry violation.
 
 The work of \cite{Bishara,Brod} was incorporated into the code DirectDM \cite{Bishara:2017nnn}
 that connects to the corresponding NRET code \cite{Liam2}.   A similar effort is described in \cite{Hof19}.
 Much of this work can be readily adapted to connect light quark/gluon CLFV EFTs to our NRET.  Work is underway \cite{Jure} to build
 a script like DirectDM  as an interface to the NRET script of Appendix C.

 \subsection{NRET variations}
 \label{sec:sec3D}
 As described earlier, there are two natural variations of this NRET construction. 
 The first is the neglect of all Jacobi velocities, the $A-1$ relative nucleon velocities and the muon velocity relative
 to the nuclear center of mass.  As the available nuclear operators are then limited to just charge and spin, this can be regarded
 as the point nucleus (or allowed) limit.  In this limit we retain only six of the operators enumerated above: $\mathcal{O}_1$, $\mathcal{O}_4$,
 $\mathcal{O}_6$, $\mathcal{O}_9$, $\mathcal{O}_{10}$, and $\mathcal{O}_{11}$.  We will later see that the coefficients of the 
 charge ($M_J$), longitudinal spin ($\Sigma_J^{\prime \prime}$), and transverse electric spin ($\Sigma_J^\prime$) response
 functions are determined by exactly the LECs of these operators.
 
 The allowed limit is a fairly simple one, and consequently it is notable that the general form of the $\mu \rightarrow e$ conversion
 rate in this limit has not appeared in past work.   While four papers listed in Table \ref{tab:pastwork} have
 considered spin-mediated interactions, all have assumed an interaction of the form of  $\mathcal{O}_4$ (Gamow-Teller).  This interaction yields a 
 specific linear combination of longitudinal and transverse electric responses. In general, however, these two responses are 
 entirely independent.  For example, as we will see in Sec. \ref{sec:sec3E}, a pseudoscalar-pseudoscalar CLFV interaction generates
 only the longitudinal response.
 
 The second variation is to include both $\vec{v}_N$ and $\vec{v}_\mu$ to first order.  The inclusion of $\vec{v}_\mu$ generates new NRET operators
 as well as corrections to nuclear response functions that, if we employ averaging, enter parametrically as $\langle f \rangle /\langle g \rangle$.   
 
 The new operators generated by this inclusion are
  \allowdisplaybreaks
\be
{\CO}^{f \, \prime}_2 &=& i \hat{q} \cdot {\vec{v}_\mu \over 2} ~1_N \nn \\
{ \CO}^f_3 &=&   i \hat{q} \cdot  \left[ {\vec{v}_\mu \over 2} \times \vec{\sigma}_L \right] ~1_N\nn \\
{\CO}^f_5 &=&   \left( i \hat{q} \times {\vec{v}_\mu \over 2} \right)  \cdot \vec{\sigma}_N \nn \\
{\CO}^f_7 &=&  {\vec{v}_\mu \over 2} \cdot \vec{\sigma}_L~1_N~ \nn \\
{\CO}^f_8 &=& {\vec{v}_\mu \over 2} \cdot \vec{\sigma}_N \nn \\ 
{\CO}^f_{12} &=& \left[ {\vec{v}_\mu \over 2} \times \vec{\sigma}_L \right] \cdot \vec{\sigma}_N \nn\\ 
{\CO}^{f \, \prime}_{13} &=&  \left( i \hat{q} \times \left[ {\vec{v}_\mu \over 2} \times \vec{\sigma}_L \right] \right)  \cdot \vec{\sigma}_N \nn \\
{\CO}^f_{14} &=&~ {\vec{v}_\mu \over 2} \cdot \vec{\sigma}_L  ~ i  \hat{q} \cdot \vec{\sigma}_N  \nn \\
{\CO}^f_{15} &=&~ i \hat{q} \cdot \left[ {\vec{v}_\mu \over 2} \times \vec{\sigma}_L \right] ~ i \hat{q} \cdot \vec{\sigma}_N \nn\\
{\CO}^{f \, \prime}_{16} &=&  i \hat{q} \cdot  {\vec{v}_\mu \over 2} ~i \hat{q} \cdot \vec{\sigma}_N
\label{eq:ops2}
\ee
where we employ $\vec{v}_\mu/2$ above, the operator that generates $f(r)$ when acting on the $1s$ muon wave function $g(r)$.
We will denote interactions associated with $\vec{v}_\mu$ (and thus the muon's lower component) by
\be
 \sum_{i} ( b_i^{0} 1 + b_i^{1} \tau_3) \CO^f_i &=& \sum_{\tau=0,1} \sum_{i} b_i^\tau \CO_i^f t^\tau, 
\label{eq:iso_lower}
\ee
The sum over $i$ extends only over the operators of Eq. (\ref{eq:ops2}), with
$b_i$ denoting the low-energy constants associated with those operators.

\begin{table*}[!]
 \caption{ \label{tab:LWL} Alternative Dirac forms of our NRET CLFV amplitudes $\CL_\mathrm{int}^j$ are related to linear combinations of the Pauli forms (the operators  $\CO_i$).
Bjorken and Drell spinor and
gamma matrix conventions are used.  Here $\chi_e=\left( \begin{array}{c} \xi_s \\ \vec{\sigma}_L \cdot \hat{q} ~\xi_s \end{array} \right)$, 
$\chi_\mu=\left( \begin{array}{c} \xi_s \\ 0  \end{array} \right)$, and $N=\left( \begin{array}{c} \xi_s \\ {\vec{\sigma}_N \cdot \vec{v}_N \over 2} \xi_s \end{array} \right)$.  The Dirac forms are expanded to first order in  $\vec{v}_N$ to maintain consistency with their use between Schr\"{o}dinger wave functions.. \\
~}
\begin{tabular}{|c|c|c|c|}
\hline
& & &  \\[-.2cm]
 $j$ & ${\CL}^j_\mathrm{int}$ & Pauli operator reduction  &~~$ \displaystyle\sum_i c_i \CO_i$ \\[0.3cm]
\hline
& & &  \\
1&$ \displaystyle{ \bar{\chi}_e \chi_\mu~\bar{N} N}$ & $ 1_L~ 1_N $&$ \CO_1$ \\[0.25cm]
2&$ \displaystyle{ \bar{\chi}_e \chi_\mu ~\bar{N}i  \gamma^5 N}$ &$  1_L~ \left( i \displaystyle{{\vec{q} \over 2 m_N} \cdot \vec{\sigma}_N} \right)$ & $\displaystyle{{q \over 2 m_N}} \CO_{10}$ \\[0.25cm]
3&$  \displaystyle{ \bar{\chi}_e i \gamma^5 \chi _\mu~\bar{N} N}$ & $ \displaystyle{ \left( -i \hat{q} \cdot \vec{\sigma}_L \right)~1_N} $ & $-\displaystyle{\CO_{11}}$ \\[0.25cm]
4&$  \displaystyle{\bar{\chi}_e  i \gamma^5 \chi_\mu \bar{N}  i \gamma^5 N} $&$ \displaystyle{ \left( - i \hat{q} \cdot \vec{\sigma}_L  \right) ~\left( i {\vec{q} \over 2 m_N} \cdot \vec{\sigma}_N \right)}$  &$\displaystyle{-{q \over 2 m_N} \CO_6}$  \\[0.25cm]
5&$  \displaystyle{\bar{\chi}_e \gamma^\mu \chi_\mu  \bar{N} \gamma_\mu N }$& $ \displaystyle{1_L 1 _N }$&$\displaystyle{ \CO_1 }$  \\[0.0cm]
~& & $\displaystyle{- \left( \hat{q} 1_L-i \hat{q} \times \vec{\sigma}_L  \right) \cdot \left( \vec{v}_N +i {\vec{q} \over 2m_N} \times \vec{\sigma}_N \right)}$ &$\displaystyle{+i \CO_2^\prime-\CO_5-{q \over 2m_N} (\CO_4+\CO_6) }$   \\[0.25cm]
6&$  \displaystyle{\bar{\chi}_e \gamma^\mu \chi _\mu \bar{N} i \sigma_{\mu \alpha} {q^\alpha \over m_N} N}$ &$  \displaystyle{- \left( \hat{q}1_L -i \hat{q} \times \vec{\sigma}_L  \right) \cdot \left( -i{\vec{q} \over m_N} \times \vec{\sigma}_N \right)}    $&$ \displaystyle{{q \over m_N} \left( \CO_4 + \CO_6 \right) } $  \\[0.25cm]
7&$  \displaystyle{\bar{\chi}_e \gamma^\mu \chi_\mu \bar{N} \gamma_\mu \gamma^5 N}$ & $ \displaystyle{ 1_L \left( \vec{v}_N \cdot \vec{\sigma}_N \right)  - \left( \hat{q}1_L -i \hat{q} \times \vec{\sigma}_L \right) \cdot \vec{\sigma}_N }$ & $ \CO_7 +i \CO_{10}- \CO_9 $   \\[0.25cm]
8& $  \displaystyle{ \bar{\chi}_e \gamma^\mu \chi_\mu  \bar{N} \sigma_{\mu \alpha} \frac{q^\alpha}{m_N}\gamma^5 N}$ & $ \displaystyle{ 1_L \left(-i {\vec{q} \over m_N} \cdot \vec{\sigma}_N \right) }$ & $ \displaystyle{ -{q \over m_N}\CO_{10}} $  \\[0.25cm]
9& $  \displaystyle{\bar{\chi}_e i \sigma^{\mu\nu} \displaystyle{q_\nu \over m_L} \chi_\mu  \bar{N} \gamma_\mu N} $ &$\displaystyle{  -{q \over m_L} 1_L~1_N }$&$ \displaystyle{ -{q \over m_L}  \CO_1} $ \\[0.0cm]
~& & $-\displaystyle{ \left( -i {\vec{q} \over m_L}\times \vec{\sigma}_L \right)  \cdot \left(\vec{v}_N+i {\vec{q} \over 2m_N} \times \vec{\sigma}_N \right)}$ &$\displaystyle{-{q \over m_L}  \left( \CO_5 +{q \over 2m_N} \left( \CO_4+\CO_6 \right) \right)}$  \\[0.25cm]
10& $ \displaystyle{\bar{\chi}_e i \sigma^{\mu\nu} \displaystyle{q_\nu \over m_L} \chi_\mu \bar{N} i \sigma_{\mu\alpha} \displaystyle{q^\alpha \over m_N} N}$ & $\displaystyle{-\left(  -i{\vec{q} \over m_L} \times \vec{\sigma}_L  \right)\cdot  \left(- i {\vec{q} \over m_N} \times \vec{\sigma}_N \right)  } $ & $\displaystyle{ {q \over m_L} {q \over m_N} \left( \CO_4+\CO_6 \right) } $  \\[0.25cm]
11& $ \displaystyle{\bar{\chi}_e i \sigma^{\mu\nu} \displaystyle{q_\nu \over m_L} \chi_\mu \bar{N} \gamma_\mu \gamma^5 N} $& $ \displaystyle{ \left(-{q \over m_L} 1_L \right) \vec{v}_N \cdot \vec{\sigma}_N  -\left(-i {\vec{q} \over m_L} \times \vec{\sigma}
_L \right) \cdot \vec{\sigma}_N}$ & $ \displaystyle{-{q \over m_L} \left( \CO_7+\CO_9 \right)  }$ \\[0.25cm]
12 & $  \displaystyle{\bar{\chi}_e i \sigma^{\mu\nu} \displaystyle{q_\nu \over m_L} \chi_\mu  \bar{N} \sigma_{\mu \alpha} \frac{q^\alpha}{m_N} \gamma^5 N}$ & $\displaystyle{\left( -{q \over m_L} 1_L \right) \left( -i {\vec{q} \over m_N} \cdot \vec{\sigma}_N \right)}$  &$ \displaystyle{ {q \over m_L} {q \over m_N} \CO_{10} } $ \\[0.25cm]
13 & $ \displaystyle{\bar{\chi}_e \gamma^\mu \gamma^5 \chi _\mu \bar{N}\gamma_\mu N }$&$\displaystyle{ \left(\hat{q} \cdot \vec{\sigma}_L \right) 1_N -\vec{\sigma}_L \cdot \left(\vec{v}_N + i {\vec{q} \over 2m_N} \times \vec{\sigma}_N \right)}$ &$ \displaystyle{-i \CO_{11} -\CO_8 -{q \over 2m_N} \CO_9 }$ \\[0.3cm]
14 &$  \displaystyle{\bar{\chi}_e \gamma^\mu \gamma^5 \chi_\mu \bar{N} i \sigma_{\mu\alpha} \displaystyle{q^\alpha \over m_N} N }$&$ \displaystyle{ -\vec{\sigma}_L  \cdot \left( -i {\vec{q} \over m_N} \times \vec{\sigma}_N \right)} $ & $ \displaystyle{q \over m_N}\CO_9$  \\[0.25cm]
15 & $ \displaystyle{\bar{\chi}_e \gamma^\mu \gamma^5 \chi_\mu \bar{N} \gamma_\mu \gamma^5 N}$ &$\displaystyle{\left( \hat{q} \cdot \vec{\sigma}_L \right) \left(\vec{v}_N  \cdot \vec{\sigma}_N \right) }-\vec{\sigma}_L \cdot \vec{\sigma}_N $ & $-i \CO_{14}- \CO_4$   \\[0.25cm]
16 & $ \displaystyle{\bar{\chi}_e \gamma^\mu \gamma^5 \chi_\mu  \bar{N} \sigma_{\mu \alpha} \frac{q^\alpha}{m_N}\gamma^5 N}$ & $ \displaystyle{\left(  \hat{q} \cdot \vec{\sigma}_L \right)  \left( -i {\vec{q} \over m_N} \cdot \vec{\sigma}_N \right)}$  &$ \displaystyle{i {q \over m_N} \CO_6}$  \\[0.25cm]
17 & $ \displaystyle{  \bar{\chi}_e \sigma^{\mu \nu} \frac{q_\nu}{m_L} \gamma^5 \chi_\mu \bar{N}\gamma_\mu  N} $& $ \displaystyle{\left( -i {\vec{q} \over m_L} \cdot \vec{\sigma}_L \right) 1_N }$ & $ \displaystyle{-{q \over m_L}  \CO_{11}}$   \\[0.0cm]
~& & $ \displaystyle{-  i{q \over m_L} \left(\vec{\sigma}_L -\hat{q} \hat{q} \cdot \vec{\sigma}_L  \right) \cdot \left( \vec{v}_N +i {\vec{q} \over 2m_N} \times \vec{\sigma}_N \right) }$ &  $ \displaystyle{-{q \over m_L} \left( i \CO_8+i {q \over 2m_N}\CO_9 +i\CO_{16}^\prime \right)}$   \\[0.25cm]
18 & $ \displaystyle{ \bar{\chi}_e  \sigma^{\mu \nu} \frac{q_\nu}{m_L} \gamma^5 \chi_\mu \bar{N} i \sigma_{\mu \alpha} \displaystyle{{q^\alpha \over m_N}} N} $ & $ \displaystyle{  - i{q \over m_L}  \left(\vec{\sigma}_L - \hat{q} \hat{q} \cdot \vec{\sigma}_L  \right) \cdot \left(-i {\vec{q} \over m_N} \times \vec{\sigma}_N \right) }$& $ \displaystyle{i {q \over m_L} {q \over m_N} \CO_9} $   \\[0.25cm]
19& $  \displaystyle{ \bar{\chi}_e  \sigma^{\mu \nu} \frac{q_\nu}{m_L} \gamma^5 \chi_\mu \bar{N} \gamma_\mu \gamma^5 N} $& $ \displaystyle{\left( -i {\vec{q} \over m_L} \cdot \vec{\sigma}_L \right) \left(\vec{v}_N \cdot \vec{\sigma}_N \right)}$ & $ \displaystyle{-{q \over m_L}  \CO_{14} }$  \\[0.0cm]
&  & $ \displaystyle{-  i{q \over m_L} \left(\vec{\sigma}_L - \hat{q} \hat{q} \cdot \vec{\sigma}_L  \right) \cdot \vec{\sigma}_N }$ & $ \displaystyle{-{q \over m_L} \left( i \CO_4 +i \CO_6 \right)}$  \\[0.25cm]
20 & $  \displaystyle{ \bar{\chi}_e  \sigma^{\mu \nu} \frac{q_\nu}{m_L} \gamma^5 \chi _\mu \bar{N} \sigma_{\mu \alpha} \frac{q^\alpha}{m_N} \gamma^5 N}$   & $ \displaystyle{\left( -i {\vec{q} \over m_L} \cdot \vec{\sigma}_L \right) \left( -i {\vec{q} \over m_N} \cdot \vec{\sigma}_N \right)}$ &$ \displaystyle{{q \over m_L} {q \over m_N} \CO_6}$  \\
 & &  & \\
 \hline
\end{tabular}
\end{table*}

\subsection{Dirac spinor NRET equivalent}
\label{sec:sec3E}
 In some DM and $\mu \rightarrow e$ conversion work operators are expressed
 as matrix elements between Dirac spinors.   For example, DM experimentalists conducting
 generalized analyses of spin-dependent scattering often write magnetic, electric dipole, and anapole operators in the 
 Dirac forms familiar from textbooks  \cite{PandaXII}.  In Table \ref{tab:LWL} we write our NRET amplitudes in the Dirac form, connecting 
 this representation to the Pauli spinor equivalent.  
 The interactions included in the table are limited to those generated through scalar or vector mediators.  We stress that this Dirac form
 is merely a rewriting of our NRET: the expansion remains one linear in $v_N$.
 
The amplitudes are constructed from
the available leptonic scalars 
\[ \bar{\chi}_e \chi_\mu,~~\bar{\chi}_e i \gamma^5 \chi_\mu \]
and four-vectors
\[ \bar{\chi}_e \gamma^\mu \chi_\mu,~~\displaystyle{\bar{\chi}_e i \sigma^{\mu \nu}  {q_\nu \over m_L}  \chi_\mu},~~\bar{\chi}_e \gamma^\mu \gamma^5 \chi_\mu,~~\displaystyle{\bar{\chi}_e  \sigma^{\mu \nu} {q_\nu \over m_L} \gamma^5 \chi_\mu} \] 
which can be combined with their nucleon counterparts to form all possible scalar interactions.  We can assume a four-fermion interaction: any
propagator effects can be absorbed into the LECs because $q$ is fixed, as noted earlier.  Thus there are
$2^2 + 4^2=20$ combinations.   

Including $\vec{v}_N$ but neglecting $\vec{v}_\mu$ in the spinors,
\[ \chi_e=\left( \begin{array}{c} \xi_s \\ \vec{\sigma}_L \cdot \hat{q} ~\xi_s \end{array} \right) ~~N=\left( \begin{array}{c} \xi_s \\ {\vec{\sigma}_N \cdot \vec{v}_N \over 2} \xi_s \end{array} \right)~~\chi_\mu=\left( \begin{array}{c} \xi_s \\ 0  \end{array} \right),\]
a Pauli reduction of
the 20 Dirac forms yields various linear combinations of 12 of the 16 NRET operators, shown in the
last column of Table \ref{tab:LWL}.   Thus the Dirac representation of our NRET is rather inefficient.
The four spin-velocity operators are not needed.  The most
general Dirac form of the interaction is
\[ {\CL}_\mathrm{int} \equiv \sum_{j=1}^{20} ~d_j {\CL}^j_\mathrm{int}  \]
Isospin labels
can be added to this expression.  Note that expressions for the $c_i$ in terms of the $d_i$ can be read off the table, e.g.,
\[ c_1 = d_1 -{q \over m_L} d_9~~~~~~c_2=i d_5~~~~~~ \mathrm{etc.} \]
so that a rate given in terms of the $c _i$ can be easily converted.

Despite the redundancy in the $d_i$'s, one might wonder whether the Dirac operator representation offers an advantage by connecting certain
$o(1)$ charge operators with their associated $o(1/m_N)$ currents, or the reverse.  This is not the case.  Of the eight cases in Table \ref{tab:LWL} where
this situation arises, four involve axial couplings, but nuclear selection rules prevent the axial charge from playing any role in elastic
$\mu \rightarrow e$ conversion.  The remaining cases are vector, where coherence enhances the charge coupling, which typically relegates the 
convection current rate contribution to the level $\lesssim 10^{-4}$. Thus, although the vector charge and convection current contributions are isolated in
distinct response functions, it is impractical to extract the convection current piece.  The situation can be summarized by stating
that the embeddings of these interactions in the nucleus destroys any operator connections through covariance.

\begin{table*}[!]
\caption{ \label{tab:LWL2} As in Table \ref{tab:LWL}, but listing the additional terms generated when the linear expansion in velocities includes 
$\vec{v}_\mu$, so that
$\chi_\mu=\left( \begin{array}{c} \xi_s \\ {\vec{\sigma}_L \cdot \vec{v}_\mu \over 2} \xi_s  \end{array} \right)$.  \\
~}
\begin{tabular}{|c|c|c|c|}
\hline
& & &  \\[-.2cm]
 $j$ & ${\CL}^j_\mathrm{int}$ & Pauli operator reduction  &~~$ \displaystyle\sum_i b_i \CO^f_i$   \\[0.3cm]
\hline
& & &  \\
1&$ \displaystyle{ \bar{\chi}_e \chi_\mu~\bar{N} N}$ & $ -\displaystyle{1 \over 2} \hat{q} \cdot \vec{v}_\mu~ 1_N -\displaystyle{i \over 2} \hat{q} \cdot [\vec{v}_\mu \times \vec{\sigma}_L] ~ 1_N$&$ \displaystyle{i } \CO_2^{f \, \prime} - \CO_3^f $  \\[0.25cm]
3&$  \displaystyle{ \bar{\chi}_e i \gamma^5 \chi _\mu~\bar{N} N}$ & $ \displaystyle{ i \over 2} \vec{v}_\mu  \cdot \vec{\sigma}_L  1_N$ & $ \displaystyle{i} \CO_7^f$ \\[0.25cm]
5&$  \displaystyle{\bar{\chi}_e \gamma^\mu \chi_\mu  \bar{N} \gamma_\mu N }$& $\displaystyle{1 \over 2} \hat{q} \cdot \vec{v}_\mu ~1_N+\displaystyle{i \over 2} \hat{q} \cdot [\vec{v}_\mu \times \vec{\sigma}_L] ~1_N$&  $ -\displaystyle{i} \CO_2^{f \, \prime} +\CO_3^f $    \\[0.25cm]
7&$  \displaystyle{\bar{\chi}_e \gamma^\mu \chi_\mu \bar{N} \gamma_\mu \gamma^5 N}$ & $-\displaystyle{1 \over 2} \vec{v}_\mu \cdot \vec{\sigma}_N -\displaystyle{i \over 2} [ \vec{v}_\mu \times \vec{\sigma}_L] \cdot \vec{\sigma}_N$ & $-  \CO_8^f - \displaystyle{i }  \CO^f_{12} $   \\[0.3cm]
9& $  \displaystyle{\bar{\chi}_e i \sigma^{\mu\nu} \displaystyle{q_\nu \over m_L} \chi_\mu  \bar{N} \gamma_\mu N} $ & $\displaystyle{q \over 2 m_L} \left(  \hat{q} \cdot \vec{v}_\mu ~1_N +i \hat{q} \cdot [\vec{v}_\mu \times \vec{\sigma}_L] ~1_N \right)$&$ \displaystyle{ q \over m_L}  \left( -i  \CO_2^{f \, \prime} + \CO_3^f \right)  $\\[0.25cm]
11& $ \displaystyle{\bar{\chi}_e i \sigma^{\mu\nu} \displaystyle{q_\nu \over m_L} \chi_\mu \bar{N} \gamma_\mu \gamma^5 N} $& $ \displaystyle{q \over 2 m_L} \left( \vec{v}_\mu \cdot \vec{\sigma}_N +i [\vec{v}_\mu \times \vec{\sigma}_L] \cdot \vec{\sigma}_N \right. $ & $ \displaystyle{{q \over  m_L} \left( \CO_8^f+i \CO^f_{12} \right.~~~~  }$ \\[0.0cm]
 & & $~\left.  -i \hat{q} \cdot [\vec{v}_\mu \times \vec{\sigma}_L ] \hat{q} \cdot \vec{\sigma}_N-\hat{q} \cdot \vec{v}_\mu \hat{q} \cdot \vec{\sigma}_N  \right)$  & $~~~~~~\left.  +i  \CO_{15}^f + \CO_{16}^{f \, \prime}\right) $  \\[0.25cm]
13 & $ \displaystyle{\bar{\chi}_e \gamma^\mu \gamma^5 \chi _\mu \bar{N}\gamma_\mu N }$&$ \displaystyle{1 \over 2} \vec{v}_\mu \cdot \vec{\sigma}_L ~ 1_N$ & $ \CO_7^f$  \\[0.25cm]
15 & $ \displaystyle{\bar{\chi}_e \gamma^\mu \gamma^5 \chi_\mu \bar{N} \gamma_\mu \gamma^5 N}$ & $\displaystyle{i \over 2} [ \hat{q} \times \vec{v}_\mu] \cdot \vec{\sigma}_N-\displaystyle{1 \over 2} ( \hat{q} \times [\vec{v}_\mu \times \vec{\sigma}_L]) \cdot \vec{\sigma}_N  $ & $  \CO^f_{5}+\displaystyle{i }  \CO_{13}^{f \, \prime}$   \\[0.0cm]
 & & $-\displaystyle{1 \over 2} \vec{v}_\mu \cdot \vec{\sigma}_L ~\hat{q} \cdot \vec{\sigma}_N $& $+ \displaystyle{i } \CO_{14}^f $   \\[0.25cm]
17 & $ \displaystyle{  \bar{\chi}_e \sigma^{\mu \nu} \frac{q_\nu}{m_L} \gamma^5 \chi_\mu \bar{N}\gamma_\mu  N} $ & $ \displaystyle{ i q \over 2 m_L} \vec{v}_\mu \cdot \vec{\sigma}_L ~1_N $ & $ \displaystyle{i q \over  m_L}  \CO^f_7 $   \\[0.25cm]
19& $  \displaystyle{ \bar{\chi}_e  \sigma^{\mu \nu} \frac{q_\nu}{m_L} \gamma^5 \chi_\mu \bar{N} \gamma_\mu \gamma^5 N} $& $ \displaystyle{ q \over 2 m_L} ( [\hat{q} \times \vec{v}_\mu] \cdot \vec{\sigma}_N+(i \hat{q} \times [\vec{v}_\mu \times \vec{\sigma}_L]) \cdot \vec{\sigma}_N) $ & $ \displaystyle{q \over  m_L}  (-i \CO_5^f +\CO_{13}^{f \, \prime}) $  \\
 & &  &  \\
 \hline
\end{tabular}
\end{table*}

If the muon's lower component is included in the reduction,
\[ \chi_\mu  = \left( \begin{array}{c} \xi_s \\ {\vec{\sigma}_L \cdot \vec{v}_\mu \over 2} \xi_s  \end{array} \right), \]
and the expansion repeated to first order in velocities, the additional terms shown in Table \ref{tab:LWL2} are
obtained.   Note that all ten NRET operators of Eq. (\ref{eq:ops2}) are generated.  This occurs because
the leptonic spin-velocity operator $\vec{v}_\mu \times \vec{\sigma}_L$ arises from the coupling of leptonic
lower components, as the electron is relativistic.  In contrast, such coupling on the nucleon side
would be second order in the nucleon velocity.

As already noted, the spin-velocity current $\vec{v}_N \times \vec{\sigma}_N$  does not appear because of the restriction to scalar and vector mediators, eliminating the
NRET operators
$\CO_3$, $\CO_{12}$, $\CO_{13}^\prime$, and $\CO_{15}$.
If we include tensor-mediated interactions, e.g.,
\be
\bar{\chi}_e i  \sigma^{\mu \nu} \gamma^5 \chi_\mu \bar{N} i \sigma_{\mu \nu} \gamma^5 N,
\ee
they do.  The equivalent Pauli form is
\begin{eqnarray}
&&{q \over m_N} 1_L 1_N+2i \hat{q} \cdot[\vec{v}_N \times \vec{\sigma}_N] -2 \vec{\sigma}_L \cdot \vec{\sigma}_N  \nonumber \\
&&~~~-2 \vec{\sigma}_L \cdot (\hat{q} \times [\vec{v}_N \times \vec{\sigma}_N]) -i (\hat{q} \times \vec{v}_\mu) \cdot \vec{\sigma}_N  \nonumber \\
&&~~~+\vec{v}_\mu \cdot \vec{\sigma}_L \hat{q} \cdot \vec{\sigma}_N + (\hat{q} \times [\vec{v}_\mu \times \vec{\sigma}_L]) \cdot \vec{\sigma}_N \nonumber \\
&&= {q \over m_N} \CO_1 +2 \CO_3-2 \CO_4+ 2i \,  \CO_{13}^\prime  \nonumber \\
&&~~~-2 \CO_5^f-2 i \CO_{13}^{f \prime } -2 i \CO_{14}^f
\label{eq:tensorexch}
\end{eqnarray}
generating $\CO_3$ and  $\CO_{13}^\prime$.  In contrast to Table \ref{tab:LWL}, we will later find
that the nuclear embedding of the spin-velocity current leads to a novel enhancement of this $v_N$-dependent operator.  Furthermore, that enhancement can be turned
on and off by picking appropriate targets.  Consequently, if $\mu \rightarrow e$ conversion were dominated by tensor-mediated interactions,
covariance could be one of the sources of measurable operator correlations.

\subsection{Nuclear-level LECs as constraints on CLFV}   
Later in this paper we will describe shell-model treatments of the nuclear many-body physics of $\mu \rightarrow e$ conversion.
We know the underlying UV theory of the strong interaction (QCD) and have available rather precise potentials, tuned
to experimental phase shifts, describing the interactions between composite nucleons.  Yet despite this knowledge, when we
treat the nuclear problem in the truncated spaces that current technology allows in the shell model, we rely on 
phenomenological effective interactions.   While some approaches may begin with the $NN$ interaction and the $g$-matrix it generates,
the use of highly restricted Hilbert spaces excludes a great deal of high-momentum physics and induces many-body
forces that are difficult to estimate theoretically.  Consequently, parameters in interactions are treated phenomenologically, adjusted to improve
agreement between shell-model predictions and measured nuclear properties, such as energy levels.  The end results
are often quite impressive; a good example of the
state of the art is \cite{GCN2850}.  Yet the relationship between the effective interaction and QCD is not simple.

The same issues will arise for the NRET interactions we have developed at the nucleon level, but employ in the complex nuclei where
$\mu \rightarrow e$ conversion experiments are done.  As described below, experience gained from the strong interaction suggests
that the NRET can still be a very successful phenomenology for analyzing CLFV, provided we consider the LECs in the operator
expansion to be effective.  Because the operator basis is complete, it has the flexibility to account for the majority of the renormalization and operator mixing
that arises from Hilbert space truncation, provided the operator coefficients are treated as parameters.  This is in part a consequence
of the averaging of many-body charges and currents when they are evaluated in nuclei with inert cores and a few active valence nucleons near the Fermi
surface. 

That is, while one could envision extending the nucleon-level NRET just constructed to include a complete set of two- or three-nucleon
Galilean-invariant interactions, the effects of these more complicated charge and current operators can be largely absorbed by
adjusting the one-body LECs.  In fact if one were to simplify the shell-model description of $^{27}$Al,
treating the ground state as a hole in a $^{28}$Si core, 
two-body operators would exactly reduce to density-dependent
effective one-body operators.  In more realistic descriptions, some true two-body corrections would persist, but 
qualitatively the argument would remain valid, reflecting the combinatorics of having more core than valence nucleons.
As nuclear saturation leads to little variation in nuclear density, the density-dependent renormalization will be only weakly
dependent on $A$: nucleons in $^{27}$Al and in Cu locally see a similar nuclear medium.

This argument is not specific to CLFV, but instead familiar in many nuclear contexts.
One standard-model example is the axial-vector coupling $g_A$, which for a free nucleon has the
value $\approx$ 1.27, while the empirical value deduced in the $2s1d$ shell from $\beta$ decay is $\approx 1.0$ \cite{Brown}. 
Limited Hilbert spaces not only lead to operator renormalization of this sort, but also to mixing among operators
of the same symmetry.  For example, the axial-vector spin can mix with the orbital angular momentum $\vec{\ell}$
and spin-tensor $\left[ Y_2(\Omega_r) \otimes \vec{\sigma} \right]_1$ operators.  One nice aspect of an NRET is that it
includes all operators up to a designated order, so that the construction should include all of the candidate mixing partners.

Truncated shell model spaces are not the only source of renormalization and mixing.
Even if a complete basis of Slater determinants is used in a nuclear calculation, new charge and current operators
would arise from explicit pions, vector mesons, and other degrees of freedom not present in such a basis.
For example, in standard-model electroweak interactions, important pion exchange currents arise for the vector three-current
and the axial charge, as the one- and two-body interactions arise at the same order in $v/c$.  Again, because of the averaging 
properties of nuclei, almost all of the effects of these multi-nucleon currents
can be absorbed into one-body operators, if the operator LECs are regarded as effective.  For example, the axial charge operator
and its two-nucleon pion-range correction are both of order $\vec{v}_N$.  Phenomenologically, the contributions of the latter
can be absorbed into an effective coupling of the axial charge operator, significantly enhancing this coupling \cite{Park,HaxtonPNC}. 

Consequently, there is a great deal of empirical evidence that the current NRET formulation will be a reliable phenomenology for analyzing
and correlating $\mu \rightarrow e$ conversion results obtained over a range of nuclei.  The fitted LECs will 
encode the available CLFV information.  The difficulty though, will come in extracting from those phenomenological LECs
constraints on one's favorite UV theory of CLFV.

As described in the
introduction, CLFV requires physics beyond the minimal standard model,
and we know such physics exists: the discovery of flavor violation in the oscillations of light neutrinos allows $\mu \rightarrow e$
conversion to proceed through a $W$-boson neutrino loop, though at a rate that is not currently measurable.  However there are
many other sources of new ultraviolet (UV) physics -- supersymmetry, heavy neutrinos, a more complicated Higgs 
sector, leptoquarks, compositeness, a new heavy Z -- that can produce observable $\mu \rightarrow e$ conversion.  What
can we learn about those sources, once we have a set of NRET LECs in hand?  This is the point where the complicated
renormalization and operator mixing effects discussed above must be confronted head on.
All of the sources of ``effectiveness" that renormalize and mix nuclear-level operators would have to be unwound.

Some of the nucleon-level issues described above, such as the absence of the pion, other mesons, or baryonic resonances in the nuclear
Hilbert space, can be addressed, and in fact have been extensively studied in the
kinematically analogous process of WIMP-nucleus scattering.
At the three-momentum transfer relevant to elastic $\mu\rightarrow e$ conversion $|\vec{q}\,|\approx m_{\mu}\approx m_{\pi}$, pionic contributions to single-nucleon form factors could be significant, depending on details of couplings and their isospins.  As these form factors are evaluated at fixed $\vec{q}^{\,2}$, and as the magnitude of the three-momentum transfer varies by only a few percent across the light- and medium-mass nuclei of interest, their effects can be absorbed into the LECs, yielding another source of renormalization. As demonstrated for the analogous DM effective theory, the impact of single-nucleon form factors on the LECs can be explicitly calculated by matching
the NRET to, for example, results from
heavy baryon chiral perturbation theory (HBChPT) \cite{Fan,Bishara,Brod,Fan,Cirelli,Barello,Hill}. When consistency of the chiral theory requires the introduction of two-nucleon (or higher-body) interactions, these corrections may be averaged to effective single-nucleon operators,
generating couplings that can be fairly matched to those determined from the NRET nuclear analysis.

However, a number of issues can arise in such matching.  Current
experience with QCD provides some context.  Renormalization via ChPT does successfully produce softer strong interactions more suitable
for nuclear physics, but the improvement ceases at a momentum scale $\approx$ 450 MeV, far above the shell-model scale.  The
softer interactions obtained are also in some tension with the saturation density derived from experiment.  
The reasons for this behavior are still under study (and are certainly
beyond the scope of present discussions) but could involve issues like the incompatibility of the plane-wave bases used in ChPT 
and the harmonic oscillator bases required in nuclear physics to preserve translational invariance.  Some relevant discussion can
be found in \cite{Drischler}.  Significant progress has been made in developing numerical methods of renormalization
that can begin at 450 MeV and continue to lower momenta \cite{Similarity}.

But the essential point for us is the expectation that the NRET operator expansion should provide a very successful formalism
for $\mu \rightarrow e$ conversion analysis, correctly capturing in its LECs the information needed to constrain UV CLFV.   Once a pattern of
couplings emerges from experiment, the challenge of making the detailed connection to UV theories would raise all of the issues discussed above.
As we have noted that the NRET can be developed in various levels of sophistication, we now turn to the 
question of what level should be employed to yield a rate formula providing the right balance between completeness and simplicity.

~\\
\section{The $\mu \rightarrow e$ Conversion Rate \label{sec:sec4}}
The nuclear amplitude $\CM$ for $\mu \rightarrow e$ conversion is obtained in the impulse approximation from
the NRET interactions described above.  We will derive the full result for the rate using the $q_\mathrm{eff}$ 
representation of distorted electron waves, where the muon's upper and lower wave functions $g(r)$ and $f(r)$
appear in transition densities. We then take advantage of the much slower variation of the muon's wave function
to simplify this result by averaging, a step that also allows us to evaluate analytically all matrix elements of
multipole operators between harmonic-oscillator Slater determinants.  For each
NRET operator, one can efficiently evaluate the full nuclear response function, while employing large-basis state-of-the-art
shell-model wave functions. With this treatment of the muon wave function, and after averaging over initial
lepton spins and nuclear magnetic quantum numbers and summing over final 
magnetic quantum numbers, we find
 \begin{eqnarray}
&&\frac{1}{2}\frac{1}{2j_N+1}\sum_{\textrm{spins}} |\mathcal{M}|^2  \nonumber \\
&&~ \approx\sum_{k}  \sum_{\tau=0,1} \sum_{\tau^\prime=0,1} R_k \left( \left\{c_i^\tau c_j^{\tau^\prime} \right\} \right)
~W_k^{\tau \tau^\prime}(y)~~~~
\label{eq:RS}
\end{eqnarray}
The nuclear response functions $W_k$ are quantities that vary depending on the character of the nuclear ground
state, including its angular momentum and isospin, as well as aspects of its internal structure, such as the spin
and angular momentum content of valence nucleons.  Thus an experimentalist can ``dial" this knob by selecting 
an optimal nucleus.  The leptonic response functions contain the particle physics: they determine which
combinations of the LECs $c_i$ can and cannot be probed in $\mu \rightarrow e$ conversion. \\

\subsection{NRET form of the transition amplitude}
\label{sec:sec3D}
The effective interaction derived in the previous section can be written
\begin{eqnarray}
[ l_0^\tau 1_N +l_0^{A \, \tau} \vec{v}_N \cdot \vec{\sigma}_N+\vec{l}_5^\tau \cdot \vec{\sigma}_N +\vec{l}_M^\tau \cdot \vec{v}_N ~~~\nonumber \\
+~i \vec{l}_E^\tau \cdot  ( \vec{\sigma}_N \times \vec{v}_N)]~ t^\tau~~~
\end{eqnarray}
where the leptonic factors obtained in our standard NRET (where $\vec{v}_N$ is retained but the smaller $\vec{v}_\mu$ neglected) are
\begin{eqnarray}
&&l_0^\tau = c_1^\tau 1_L +c_{11}^\tau  i \hat{q} \cdot \vec{\sigma}_L, \nonumber \\
&&l_0^{A \, \tau} = c_7^\tau 1_L +c_{14}^\tau i \hat{q} \cdot \vec{\sigma}_L, \nonumber \\
&&\vec{l}_5^\tau = c_4^\tau \vec{\sigma}_L + c_6^\tau i \hat{q} \cdot \vec{\sigma}_L i\hat{q} -c_9^\tau i \hat{q} \times \vec{\sigma}_L +c_{10}^\tau i \hat{q} 1_L, \nonumber \\
&&\vec{l}_M^\tau = c_2^\tau i \hat{q} 1_L -c_5^\tau i \hat{q} \times \vec{\sigma}_L +c_8^\tau \vec{\sigma}_L +c_{16}^\tau i \hat{q} \cdot \vec{\sigma}_L i \hat{q},  \nonumber \\
&&\vec{l}_E^\tau = -c_3^\tau \hat{q} 1_L +c_{12}^\tau i \vec{\sigma}_L +c_{13}^\tau \hat{q} \times \vec{\sigma}_L -i c_{15}^\tau \hat{q} \cdot \vec{\sigma}_L \hat{q}.  \nonumber \\
&&~~
\end{eqnarray}

If in addition one sets $\vec{v}_N \equiv 0$, the point-nucleus NRET limit, $l_0^\tau$ and $\vec{l}_5^\tau$ are unchanged, while
$l_0^{A \, \tau}$, $\vec{l}_M^\tau$, and $\vec{l}_E^\tau$ all $\rightarrow 0$.   Alternatively, if one includes $\vec{v}_\mu$, the muon's lower
component generates corrections $\approx f$ to the point-nucleus leptonic factors, 
\begin{eqnarray} 
&&l_0^\tau \rightarrow l_0^\tau +b_2^\tau i \hat{q} \cdot {\vec{v}_\mu \over 2}+b_3^\tau i\hat{q}\cdot [{\vec{v}_\mu \over 2} \times \vec{\sigma}_L] +b_7 {\vec{v}_\mu \over 2} \cdot \vec{\sigma}_L \nonumber \\
&&~~~ \equiv l_0^\tau + {l}_{0  f}^\tau, \nonumber \\
&&\vec{l}_5^\tau \rightarrow \vec{l}_5^\tau+b_5^\tau i\hat{q} \times {\vec{v}_\mu \over 2} + b_8^\tau {\vec{v}_\mu \over 2}  +b_{12}^\tau {\vec{v}_\mu \over 2} \times \vec{\sigma}_L  \nonumber \\
&&~~~~+b^\tau_{13} i \hat{q} \times [{\vec{v}_\mu \over 2} \times \vec{\sigma}_L] +b^\tau_{14}  {\vec{v}_\mu \over 2}  \cdot  \vec{\sigma}_L i \hat{q} \nonumber \\
&&~~~~+b^\tau_{15} i \hat{q} \cdot [{\vec{v}_\mu \over 2} \times \vec{\sigma}_L] i \hat{q} + b^\tau_{16} i \hat{q} \cdot {\vec{v}_\mu \over 2} i \hat{q} \nonumber \\
&&~~~\equiv \vec{l}^\tau_5 + \vec{l}^\tau_{5  f}.
\end{eqnarray}
Henceforth, $l_0^\tau$ and $\vec{l}^\tau_{5}$ will denote the leptonic factors with $\vec{v}_\mu \equiv 0$, while
 $l_{0 f}^\tau$ and $\vec{l}^\tau_{5 f}$ will denote the corrections due to $\vec{v}_\mu$.

The prescription for embedding this interaction in the nucleus is the replacement of the nucleon
operators above by their one-body equivalents, summed over all nucleons in a nucleus.  This yields the coordinate-space 
operator
\begin{widetext}
\begin{eqnarray}
   \sum_{\tau=0,1} \left[
 ( l_0^\tau+l_{0 f}^\tau)~ \sum_{i=1}^A  ~  \delta(\vec{x}-\vec{x}_i)
~+~l_0^{A\tau}~ \sum_{i=1}^A ~ {1 \over 2m_N} \left(-{1 \over i} \overleftarrow{\nabla}_i \cdot  \vec{\sigma}_N(i) ~ \delta(\vec{x}-\vec{x}_i) +   \delta(\vec{x}-\vec{x}_i) ~ \vec{\sigma}_N(i)  \cdot  {1 \over i} \overrightarrow{\nabla}_i \right) \right.  \nonumber \\
 ~+~ (\vec{l}_5^\tau+\vec{l}_{5f}^\tau) \cdot  \sum_{i=1}^A  ~\vec{\sigma}_N(i) ~  \delta(\vec{x}-\vec{x}_i)
 ~+~ \vec{l}_M^\tau \cdot  \sum_{i=1}^A  ~ {1 \over 2m_N} \left(-{1 \over i} \overleftarrow{\nabla}_i  ~ \delta(\vec{x}-\vec{x}_i)  +   \delta(\vec{x}-\vec{x}_i)~ {1 \over i} \overrightarrow{\nabla}_i \right)  ~~~~~~~\nonumber \\
~+~ \vec{l}_E^\tau \left. \cdot  \sum_{i=1}^A ~ {1 \over 2m_N} \left( \overleftarrow{\nabla}_i \times \vec{\sigma}_N(i) ~ \delta(\vec{x}-\vec{x}_i)  +   \delta(\vec{x}-\vec{x}_i)~  \vec{\sigma}_N(i) \times \overrightarrow{\nabla}_i \right)   \right]_{int} t^\tau(i) ~~~~~~~~~~~~~~~~
\label{eq:fulldensities}
\end{eqnarray}
Here the subscript $int$ denotes that the $A$ single-nucleon velocities appearing in the above expression should be replaced by $A-1$
relative (or Jacobi) velocities: the same issue has been discussed in connection with DM direct detection \cite{Liam2}.
Provided one takes care to remove the over-completeness of the shell-model basis by projecting the center-of-mass into the
harmonic oscillator $1s$ state, the difference between the correct intrinsic operator and the usual impulse-approximation one-body
operator is a recoil correction, suppressed by ${1/A}$ (that is, by the target mass).  The nuclear recoil velocity is much smaller
than the internucleon velocity $\vec{v}_N$ and beyond the order of our NRET expansion.  Consequently, we can use the simple
one-body form of operators.  This is equivalent to declaring the nucleus as infinitely heavy.

The matrix element of the decay amplitude can then be obtained by substituting this expression into Eq. (\ref{eq:reduced3}). Including
contributions from $\vec{v}_\mu$ we find
\begin{eqnarray}\CM &=& \sqrt{{E_e \over 2 m_e}} ~{q_\mathrm{eff} \over q} \sum_{\tau=0,1}
  \langle \tfrac{1}{2}  {s_f}; j_N m_f |  \left[
  \sum_{i=1}^A  ~e^{-i \vec{q}_\mathrm{eff} \cdot \vec{x}_i} \left[  l_0^\tau {i g(x_i) \over \sqrt{4 \pi}} +  l_{0 f}^\tau(\hat{x}_i) {f(x_i) \over \sqrt{4 \pi}}\right]  ~  \right. \nonumber \\
&+& l_0^{A\tau}~ \sum_{i=1}^A ~ {1 \over 2m_N} \left(-{1 \over i} \overleftarrow{\nabla}_i \cdot  \vec{\sigma}_N(i)~ e^{-i \vec{q}_\mathrm{eff} \cdot \vec{x}_i} {i g(x_i) \over \sqrt{4 \pi}} + {i g(x_i) \over \sqrt{4 \pi}}e^{-i \vec{q}_\mathrm{eff} \cdot \vec{x}_i}  \vec{\sigma}_N(i)  \cdot  {1 \over i} \overrightarrow{\nabla}_i \right)  \nonumber \\
&+&  \sum_{i=1}^A  ~\vec{\sigma}_N(i)~e^{-i \vec{q}_\mathrm{eff} \cdot \vec{x}_i} \cdot \left[  \vec{l}_5^\tau {i g(x_i) \over \sqrt{4 \pi}} + \vec{l}_{5 f}^\tau(\hat{x}_i) {f(x_i) \over \sqrt{4 \pi}} \right]   \nonumber \\
&+&  \vec{l}_M^\tau \cdot  \sum_{i=1}^A  ~ {1 \over 2m_N} \left(-{1 \over i} \overleftarrow{\nabla}_i e^{-i \vec{q}_\mathrm{eff} \cdot \vec{x}_i}  {i g(x_i) \over \sqrt{4 \pi}} + {i g(x_i) \over \sqrt{4 \pi}}e^{-i \vec{q}_\mathrm{eff} \cdot \vec{x}_i} {1 \over i} \overrightarrow{\nabla}_i \right)  \nonumber \\
&+&  \vec{l}_E^\tau \left. \cdot  \sum_{i=1}^A ~ {1 \over 2m_N} \left( \overleftarrow{\nabla}_i \times \vec{\sigma}_N(i) e^{-i \vec{q}_\mathrm{eff} \cdot \vec{x}_i}  {i g(x_i) \over \sqrt{4 \pi}} + {i g(x_i) \over \sqrt{4 \pi}}e^{-i \vec{q}_\mathrm{eff} \cdot \vec{x}_i}  \vec{\sigma}_N(i) \times \overrightarrow{\nabla}_i \right)   \right]_{int} t^\tau(i) ~ |  \tfrac{1}{2} {s_i}; j_N m_{i} \rangle~~~
\label{eq:IANuc}
\end{eqnarray}
where the muon's $p$-wave lower component generates an additional angular dependence through $\hat{x}_i$,
\begin{eqnarray} 
\label{eq:newlep}
&&l_{0 f}^\tau(\hat{x}_i) \equiv b_2^\tau i \hat{q} \cdot \hat{x}_i +b_3^\tau i \hat{q} \cdot [\hat{x}_i \times \vec{\sigma}_L] +b_7^\tau \hat{x}_i \cdot \vec{\sigma}_L \\
&&\vec{l}_{5f}^\tau(\hat{x}_i) \equiv b_5^\tau i\hat{q} \times {\hat{x}_i} + b_8^\tau {\hat{x}_i}  +b_{12}^\tau {\hat{x}_i } \times \vec{\sigma}_L 
+b^\tau_{13} i \hat{q} \times [{\hat{x}_i } \times \vec{\sigma}_L] +b^\tau_{14}  {\hat{x}_i }  \cdot  \vec{\sigma}_L i \hat{q} 
+b^\tau_{15} i \hat{q} \cdot [{\hat{x}_i } \times \vec{\sigma}_L] i \hat{q} + b^\tau_{16} i \hat{q} \cdot {\hat{x}_i } i \hat{q}  \nonumber 
\end{eqnarray}
Note that the terms involving gradients can be partially integrated without concern about differentiating $g(x_i)$, as this would produce terms
second order in velocities.   The NRET with $\vec{v}_\mu \equiv 0$ is obtained by setting $l_{0 f}^\tau(\hat{x}_i) = \vec{l}_{5f}^\tau(\hat{x}_i)=0$,
while the point nucleus NRET corresponds to retaining only $l_0^\tau$ and $\vec{l}_5^\tau$.

As we have previously discussed, the full multipole expansion of this operator should be carried out to avoid errors that can be $o(1)$:
As $q_\mathrm{eff} \approx m_\mu \approx 1/R_\mathrm{nuc}$ where $R_\mathrm{nuc}$ is the nuclear radius, the momentum transfer in $\mu \rightarrow e$
conversion is sufficient to excite various angular and radial modes of the nucleus.  The relevant operators carrying definite parity and angular momentum are obtained by 
completing a standard multipole decomposition, expanding
plane waves in Eq. (\ref{eq:IANuc}) in spherical components, combining these with the single-nucleon operators to form the needed multipoles.  We quantize along $\hat{q}$.  For elastic processes,
as we described earlier in this section, 
the nearly exact parity and $CP$ of the ground state impose selection rules that eliminate certain operators entirely and restrict the allowed 
angular momenta of others to either even or odd $J$.  Some of these results are standard in semi-leptonic weak interactions, though terms associated with
$\vec{v}_\mu$ require some additional work.  We obtain
\begin{eqnarray}
\label{eq:Helastic}
&&\CM = \sqrt{{E_e \over 2 m_e}} ~{q_\mathrm{eff} \over q} \sum_{\tau=0,1}
  \langle \tfrac{1}{2} {s_f}; j_N m_f | \nonumber \\
&&~~\Bigg[ \sum_{J=0,2,...}^\infty  \sqrt{4 \pi (2J+1)} (-i)^J  \left[i  l_0^\tau M^g_{J0;\tau}(q_\mathrm{eff})+{l}_{0}^{ \tau   (2)} {M}^{ (2)f}_{J0;\tau}(q_\mathrm{eff}) + i {q_\mathrm{eff} \over m_N} l_{E0}^\tau \Phi_{J0;\tau}^{ \prime \prime \, g}(q_\mathrm{eff})     \right]   \nonumber \\
&&~~+\sum_{J=1,3,...}^\infty \sqrt{2 \pi (2J+1)} (-i)^J \sum_{\lambda=\pm 1}  \Big[   l_{5 \lambda}^\tau \Sigma_{J-\lambda;\tau}^{ \prime \, g}(q_\mathrm{eff})+i  l_{5 \lambda}^{\tau (0)} \Sigma_{J-\lambda;\tau}^{ \prime \,(0) f}(q_\mathrm{eff})+i  l_{5 \lambda}^{\tau (2)} \Sigma_{J-\lambda;\tau}^{ \prime \, (2) f}(q_\mathrm{eff}) \nonumber \\
&&~~~~~~~~~~~~~~~~~~~~~~~~~~~~~~~~~~~~~~~~~~~~~~~~~~~~~~ -{q_\mathrm{eff} \over m_N} l_{M\lambda}^\tau \lambda \Delta^g_{J-\lambda;\tau}(q_\mathrm{eff}) \Big] \nonumber \\
&&~~+\sum_{J=2,4,...}^\infty \sqrt{2 \pi (2J+1)} (-i)^J \sum_{\lambda=\pm 1}  \Big[  {l}_{\lambda }^{\tau (1)}  {M}^{ (1) f}_{J-\lambda;\tau}(q_\mathrm{eff})-i{q_\mathrm{eff} \over m_N} l_{E\lambda}^\tau \tilde{\Phi}_{J-\lambda;\tau}^{\prime \, g}(q_\mathrm{eff}) \Big] \nonumber \\
&&~~+\sum_{J=1,3,...}^\infty \sqrt{4 \pi (2J+1)} (-i)^J   \Big[ - l_{5 0}^\tau  \Sigma_{J0;\tau}^{\prime \prime \, g }(q_\mathrm{eff})+ i l_{5 0}^{\tau (0)}  \Sigma_{J0;\tau}^{\prime \prime \,(0)  f }(q_\mathrm{eff}) + i l_{5 0}^{\tau (2)}  \Sigma_{J0;\tau}^{\prime \prime \, (2) f }(q_\mathrm{eff})   \Big] \Bigg]~| \tfrac{1}{2} {s_i}; j_N m_i \rangle~~~~
\label{eq:HamAA}
\end{eqnarray}
\end{widetext}
Note that $l_0^A$ does not appear above: it is associated with a pseudoscalar nuclear operator that cannot contribute to elastic scattering, as we demonstrated earlier from the $P$ and $CP$ selection rules.
Equation (\ref{eq:HamAA}) divides the amplitude into four terms, corresponding to longitudinal/transverse nuclear
operators and even/odd angular momenta.  When the spin sums are completed in the rate evaluation, the Wigner-Eckart theorem eliminates interferences among these terms.

We calculate the transition probability, squaring the amplitude in Eq. (\ref{eq:HamAA}), averaging over initial nuclear spins, summing over final nuclear spins, and evaluating the phase space.
Details are given in Appendix \ref{AppendixB}.  The resulting ``master rate formula" is presented in two forms, 
by  Eqs. (\ref{eq:amp2}) and (\ref{eq:amp3}).

As we established quantitatively earlier in this paper, the nuclear matrix elements can be simplified with very little loss of accuracy, put into
a form where they can be evaluated analytically as functions of $y$, once one computes the needed shell-model one-body density matrices.
Recalling our definition
\be
 \hat{O}_{JM}^g(q)  &\equiv& \sum_{i=1}^A  {1 \over \sqrt{4 \pi}}g(x_i ) \hat{O}_{JM}(q x_i)  
\ee
and similarly for $f(r)$, 
we replace $g(r)$ and $f(r)$ by the average values obtained by the procedure described in Sec. \ref{sec:sec2},
\begin{eqnarray}
\label{eq:averaged}
\hat{O}_{JM}^g(q)   &\rightarrow&   |\phi_{1s}^{Z_\mathrm{eff}}(\vec{0})| \sum_{i=1}^A  \hat{O}_{JM}(q x_i), \nonumber \\ 
 \hat{O}_{JM}^f(q)   &\rightarrow&   |\phi_{1s}^{Z_\mathrm{eff}}(\vec{0})| ~ {\langle f \rangle \over \langle g \rangle }~ \sum_{i=1}^A  \hat{O}_{JM}(q x_i).  ~~~
 \end{eqnarray}
The effects of the muon's lower component are encoded in the ratio $\langle f \rangle / \langle g \rangle$  The effects of higher multipoles, $\vec{v}_N$, and $\vec{v}_\mu$ are then reflected in the
dependence of our rate formula on the parameters 
$y$, ${q_\mathrm{eff}/ m_N}$, and $\langle f \rangle / \langle g \rangle$.  Our master rate formula, rewritten in this averaged form as Eq. (]\ref{eq:amp3A}), then shows the competition among
these parameters.    In the next section, we emphasize the importance of such rate representations
 in the ``discovery phase" of CLFV, when the goals are to measure nonzero rates and use them to deduce the underlying operator structure
 of the CLFV. 


In Eq. (\ref{eq:HamAA}), the six nuclear operators associated with the muon's upper component are familiar from standard-model electroweak interaction theory.  They are constructed
from the Bessel spherical harmonics
and vector spherical harmonics,
$M_{JM}(q \vec{x}) \equiv j_J(q x) Y_{JM}(\Omega_x)$ and $\vec{M}_{JL}^M(q\vec{x}) \equiv j_L(q x) \vec{\mathcal{Y}}_{JLM}(\Omega_x)$:
\begin{widetext}
\allowdisplaybreaks
\begin{eqnarray}
\label{eq:operators}
M_{JM;\tau}(q) &\equiv& \sum_{i=1}^A M_{JM}(q \vec{x}_i)~ t^\tau(i), \nonumber \\
\Delta_{JM;\tau}(q ) &\equiv&\sum_{i=1}^A \vec{M}_{JJ}^M(q \vec{x}_i) \cdot {1 \over q} \vec{\nabla}_i ~t^\tau(i), \nonumber \\
\Sigma^\prime_{JM;\tau}(q) &\equiv& -i \sum_{i=1}^A \bigg[ {1 \over q} \vec{\nabla}_i \times \vec{M}_{JJ}^M (q \vec{x}_i) \bigg] \cdot \vec{\sigma}_N(i)~t^\tau(i) \nonumber \\ 
&=& \sum_{i=1}^A  \bigg[ -{\textstyle \sqrt{J \over 2J+1}}~\vec{M}_{JJ+1}^M(q \vec{x}_i) + {\textstyle \sqrt{J+1 \over 2J+1}}~\vec{M}_{JJ-1}^M(q \vec{x}_i) \bigg] \cdot \vec{\sigma}_N(i) ~t^\tau(i),\nonumber \\
 \Sigma^{\prime \prime}_{JM;\tau}(q) &\equiv& \sum_{i=1}^A  \bigg[ {1 \over q} \vec{\nabla}_i ~ M_{JM} (q \vec{x}_i) \bigg] \cdot \vec{\sigma}_N(i)~t^\tau(i) \nonumber \\
  &=& \sum_{i=1}^A  \bigg[ {\textstyle  \sqrt{J+1 \over 2J+1}}~\vec{M}_{JJ+1}^M(q \vec{x}_i) + {\textstyle\sqrt{J \over 2J+1}}~\vec{M}_{JJ-1}^M(q \vec{x}_i) \bigg] \cdot \vec{\sigma}_N(i) ~t^\tau(i),\nonumber  \\
\tilde{\Phi}^{\prime}_{JM;\tau}(q) &\equiv& \sum_{i=1}^A \bigg[ \left( {1 \over q} \vec{\nabla}_i \times \vec{M}_{JJ}^M(q \vec{x}_i) \right) \cdot \left(\vec{\sigma}_N(i) \times {1 \over q} \vec{\nabla}_i \right) + {1 \over 2} \vec{M}_{JJ}^M(q \vec{x}_i) \cdot \vec{\sigma}_N(i) \bigg]~t^\tau(i), \nonumber \\
\Phi^{\prime \prime}_{JM;\tau}(q ) &\equiv& i  \sum_{i=1}^A\left( {1 \over q} \vec{\nabla}_i  M_{JM}(q \vec{x}_i) \right) \cdot \left(\vec{\sigma}_N(i) \times {1 \over q} \vec{\nabla}_i \right)~t^\tau(i).
\end{eqnarray}
Six new nuclear operators for elastic $\mu \rightarrow e$ conversion arise from the muon's lower component:
\begin{eqnarray}
\label{eq:operators2}
&&M^{(1)}_{JM;\tau}(q) \equiv  \sum_{i=1}^A {\textstyle\sqrt{J(J+1)}} {1 \over q x_i} j_J(q x_i)Y_{JM}(\Omega_{x_i}) ~t^\tau(i) ~~~~~~~M^{(2)}_{JM;\tau}(q) \equiv  \sum_{i=1}^A {d \over d q x_i} j_J(q x_i)  Y_{JM}(\Omega_{x_i} )~ t^\tau(i), \nonumber \\
&&\Sigma^{\prime(0)}_{JM;\tau} \equiv \sum_{i=1}^A j_J(qx_i) \left[ {\textstyle \sqrt{J \over 2J+1}} \vec{\mathcal{Y}}_{JJ+1M}(\Omega_i) \cdot \vec{\sigma}_N(i) +{\textstyle  \sqrt{J+1 \over 2J+1} }\vec{\mathcal{Y}}_{JJ-1M}(\Omega_i) \cdot \vec{\sigma}_N(i) \right], \nonumber \\
&&\Sigma^{\prime(2)}_{JM;\tau} \equiv \sum_{i=1}^A  \left[-{d \over dqx_i} j_{J+1}(q x_i)~{\textstyle \sqrt{J \over 2J+1} } \vec{\mathcal{Y}}_{JJ+1M}(\Omega_i) \cdot \vec{\sigma}_N(i) + {d \over  d qx_i} j_{J-1}(qx_i)~{\textstyle \sqrt{J+1 \over 2J+1}} \vec{\mathcal{Y}}_{JJ-1M}(\Omega_i) \cdot \vec{\sigma}_N(i) \right], \nonumber \\
&&\Sigma^{\prime \prime (0)}_{JM;\tau} \equiv \sum_{i=1}^A j_J(qx_i) \left[ -{\textstyle \sqrt{J+1 \over 2J+1} }\vec{\mathcal{Y}}_{JJ+1M}(\Omega_i) \cdot \vec{\sigma}_N(i) + {\textstyle \sqrt{J \over 2J+1}} \vec{\mathcal{Y}}_{JJ-1M}(\Omega_i) \cdot \vec{\sigma}_N(i) \right], \nonumber \\
&&\Sigma^{\prime \prime(2)}_{JM;\tau} \equiv \sum_{i=1}^A  \left[{d \over dqx_i} j_{J+1}(q x_i)~ {\textstyle \sqrt{J+1 \over 2J+1}} \vec{\mathcal{Y}}_{JJ+1M}(\Omega_i) \cdot \vec{\sigma}_N(i) + {d \over  d qx_i} j_{J-1}(qx_i) ~{\textstyle \sqrt{J \over 2J+1}} \vec{\mathcal{Y}}_{JJ-1M}(\Omega_i) \cdot \vec{\sigma}_N(i) \right], ~~~~~
\end{eqnarray}
\end{widetext}
with the superscripts $(0),~(1),~(2)$ denoting the radial dependence.   In all case above we have given the operators with $\langle g \rangle$ and $\langle f \rangle$ 
removed.  This rather elegant result for the lower-component Coulomb physics is
another benefit of introducing $q_\mathrm{eff}$.  In treating $\vec{v}_\mu$, the muon's lower component yields $\hat{x}_i$, which is a ladder
operator in angular momentum, while in treating $\vec{v}_N$, the nucleon's lower component yields 
$\vec{\nabla}$, which is a ladder operator for both angular momentum and the radial functions $j_J(r)$.   This difference leads to the contrasting
structures of Eqs. (\ref{eq:operators}) and (\ref{eq:operators2}).
The parity and time-reversal properties of the new operators $M^{(i)}_J$, $\Sigma'^{(i)}_J$, and $\Sigma''^{(i)}_J$ reflect those of their namesakes in Eq. (\ref{eq:operators}).  If the
nuclear physics is done with Slater determinants in a harmonic oscillator basis, matrix elements of all
operators in Eq. (\ref{eq:operators2}) can be evaluated analytically -- though not in terms of finite polynomials in $y$, as is possible for the operators of Eq. (\ref{eq:operators}).

\subsection{NRET decay rates}
\label{sec:NRETrates}
In 1936, Gamow and Teller \cite{GT} concluded from the systematics of $\beta$ decay that Fermi's charge (vector) weak amplitude was
accompanied by a spin-dependent (axial) amplitude of comparable strength.  Yet more than two decades were to pass before it could be concluded
that the low-energy weak interaction was V-A, with scalar, pseudoscalar, and tensor interactions eliminated as leading operators.  One 
motivation for developing an NRET of $\mu \rightarrow e$ conversion is to provide a general framework that will allow experimentalists --
once CLFV is detected -- to analyze their data in a systematic way, to deduce which NRET operators are
responsible for the CLFV.  The NRET provides a complete set of operators, organized in a hierarchy.
One can test for the presence of leading operators, exploiting nuclear physics properties, very much as Gamow and Teller used nuclear
selection rules to demonstrate the presence of an allowed axial interaction in $\beta$ decay.  We will see below that secondary interactions can also
be explored, given the number of nuclear ``knobs" available through judicious choices of nuclear targets.\\

\noindent
{\it The ``allowed" NRET response.}   We define the allowed approximation as the rate one obtains by omitting velocity-dependent operators from the NRET,
$\vec{v}_N \equiv 0$ and $\vec{v}_\mu \equiv 0$.  From Eq. (\ref{eq:amp2}) we have
\begin{widetext}
\allowdisplaybreaks
\be
 \omega = {G_F^2 \over \pi} ~ {q_\mathrm{eff}^2 \over 1+{q \over M_T}} ~ |\phi_{1s}^{Z_\mathrm{eff}}(\vec{0})|^2 ~
\sum_{ \tau=0,1} \sum_{\tau^\prime = 0,1}  ~ \Big[  ~\tilde{R}_{M M}^{\tau \tau^\prime}~ W_{M M}^{\tau \tau^\prime}(q_\mathrm{eff}) +\tilde{R}_{\Sigma^{\prime } \Sigma^{\prime }}^{\tau \tau^\prime}~ W_{\Sigma^{\prime } \Sigma^{\prime }}^{\tau \tau^\prime}(q_\mathrm{eff})  + \tilde{R}_{\Sigma^{\prime \prime } \Sigma^{\prime \prime }}^{\tau \tau^\prime} W_{\Sigma^{\prime \prime } \Sigma^{\prime \prime }}^{\tau \tau^\prime}(q_\mathrm{eff}) \Big] ~~~~~ 
\label{eq:ampallowed}
\ee
where
\be
 \tilde{R}_{M M}^{\tau \tau^\prime} = \tilde{c}_1^\tau \tilde{c}_1^{\tau^{\prime} * } + \tilde{c}_{11}^\tau \tilde{c}_{11}^{\tau^\prime * },~~~~~~~~~~~\tilde{R}_{\Sigma^{\prime } \Sigma^{\prime }}^{\tau \tau^\prime} =\tilde{c}_4^\tau \tilde{c}_4^{\tau^{\prime} * } + \tilde{c}_{9}^\tau \tilde{c}_{9}^{\tau^\prime * },~~~~~~~~~~~~\tilde{R}_{\Sigma^{\prime \prime } \Sigma^{\prime \prime }}^{\tau \tau^\prime} &=&  ( \tilde{c}
 _4^\tau- \tilde{c}_6^{\tau })  ( \tilde{c}_4^{\tau^\prime *}- \tilde{c}_6^{\tau^\prime *}) +\tilde{c}_{10}^\tau \tilde{c}_{10}^{\tau^\prime *}.    \nonumber
\ee
Alternatively, the leptonic responses can be expressed in terms of the covariant couplings,
\begin{eqnarray}
\tilde{R}_{M M}^{\tau \tau^\prime} &=&  (\tilde{d}_1^\tau +\tilde{d}_5^\tau-{q \over m_L} \tilde{d}_9^\tau)   (\tau \rightarrow \tau^\prime) +(\tilde{d}^\tau_3+{q \over m_L} \tilde{d}^\tau_{17}) (\tau \rightarrow \tau^\prime)+\tilde{d}^\tau_{13} \tilde{d}^{\tau^\prime}_{13}, \nonumber \\
 \tilde{R}_{\Sigma^{\prime } \Sigma^{\prime }}^{\tau \tau^\prime} 
 &=& (\tilde{d}_{15}^\tau +{q \over 2 m_N} (\tilde{d}_5^\tau-2 \tilde{d}_6^\tau+{q \over m_L} (\tilde{d}_9^\tau-2 \tilde{d}_{10}^\tau))) (\tau \rightarrow \tau^\prime) +{q^2 \over m_L^2} \tilde{d}_{19}^\tau \tilde{d}_{19}^{\tau^\prime}  \nonumber \\
&& ~~~~~~~~~~~~~~+(\tilde{d}_7^\tau+{q \over m_L} \tilde{d}_{11}^\tau+{q \over 2 m_N} (\tilde{d}_{13}^\tau -2 \tilde{d}_{14}^\tau))(\tau \rightarrow \tau^\prime) +{q^2 \over 4 m_N^2} \frac{q^2}{m_L^2} (\tilde{d}_{17}^\tau-2 \tilde{d}_{18}^\tau)(\tau \rightarrow \tau^\prime), \nonumber \\
 \tilde{R}_{\Sigma^{\prime \prime } \Sigma^{\prime \prime }}^{\tau \tau^\prime} &=&  (\tilde{d}_{15}^\tau -{q \over 2 m_N} (\tilde{d}_4^\tau-{q \over m_L} 2 \tilde{d}_{20})) (\tau \rightarrow \tau^\prime) +{q^2 \over m_N^2} \tilde{d}_{16}^\tau \tilde{d}_{16}^{\tau^\prime}+\tilde{d}_7^\tau \tilde{d}_7^{\tau^\prime}+{q^2 \over 4 m_N^2} (\tilde{d}_2^\tau -2 \tilde{d}_8^\tau+{q \over m_L} 2 \tilde{d}_{12}^\tau)(\tau \rightarrow \tau^\prime), \nonumber
\end{eqnarray}
\end{widetext}
where $(\tau \rightarrow \tau^\prime)$ indicates that the second term is obtained from the first by the indicated isospin 
index change.
The nuclear response functions $W$, given by Eq. (\ref{eq:nucresponse}), involve sums over even multipoles for the charge operator and odd multipoles for the spin operators,
a consequence of the $P$ and $CP$ selection rules.  We note
\begin{enumerate} [leftmargin=1em]
\item All of the work reported in Table \ref{tab:pastwork} prior to \cite{RulePRL} employed the simplest point operators, the interactions $\CO_1$ and $\CO_4$.
These operators generate responses of the allowed form.
They couple to the nucleus's charge and spin, with sensitivity
to the nucleus's internal structure arising only through the response functions
$W^{\tau \tau^\prime}(q_\mathrm{eff})$, sampled at a momentum transfer $q_\mathrm{eff} \approx m_\mu$.
\item The two spin-dependent responses -- transverse electric $\Sigma^\prime$ and  longitudinal $\Sigma^{\prime \prime}$ --
have distinct form factors, allowing one in principle to distinguish longitudinal operators, e.g., $\CO_6$ and $\CO_{10}$, from transverse ones, e.g., 
$\CO_{9}$.  Measurements in multiple targets with specific nuclear properties would be needed.
\item In standard-model weak interactions like beta decay, the allowed limit generally is synonymous with retention
of just the Fermi and Gamow-Teller operators.
But because the momentum transfer in $\mu \rightarrow e$ conversion is fixed, other operators can mimic these responses.
This is apparent from the expressions for $\tilde{R}^{\tau \tau^\prime}_{\Sigma^\prime \Sigma^\prime}$ and $\tilde{R}^{\tau \tau^\prime}_{\Sigma^{\prime \prime} \Sigma^{\prime \prime}}$
in terms of the $\tilde{d}_i$.  These  responses are also generated by
convection currents, magnetic moments, etc. 
\end{enumerate}

We noted previously that in past work on Coulomb effects in $\mu \rightarrow e$ conversion (see Table \ref{tab:pastwork}),
Coulomb corrections to the allowed response were made by including the muon's lower component while simultaneously 
excluding electron partial waves with $|\kappa| >1$ (and also the effects of $\vec{v}_N$).  We now illustrate why this
choice does not properly respect the hierarchy of small parameters.

We take the example of the scalar operator $\CO_1$, which because of its simplicity 
has been frequently studied (see Table \ref{tab:pastwork}).   We also assume the operator
isospin is isovector.  Eq. (\ref{eq:amp3A}) yields a rate
proportional to
\be 
(\tilde{d}_1^1)^2 \sum_{J=0,2,...}  \left[ \langle M_J- {\langle f \rangle \over \langle g \rangle} M_J^{(2)} \rangle^2 +{\langle f \rangle^2 \over \langle g \rangle^2}  \langle M_J^{(1)} \rangle^2 \right]. \nonumber
\ee
In $^{27}$Al the parameters are $\langle f \rangle/\langle g \rangle \approx$ - 0.027 and $y \approx 0.27$.  Evaluating
the structure functions, one finds
\begin{eqnarray}
\braket{M_0}^2 &\approx& 0.477+o(y) \approx  0.125, \nonumber \\
\braket{M_2}^2 &\approx& 0.160y^2+o(y^3) \approx  0.006, \nonumber \\
-2\frac{\braket{f}}{\braket{g}}\braket{M_0}\braket{M_0^{(2)}} &\approx& -2  \frac{\braket{f}}{\braket{g}}  y^{1 \over 2} \left(-0.568 +o(y) \right) \nonumber \\
&\approx& -0.006. \nonumber
\label{eq:al27_scalar_mediated}
\end{eqnarray}
One sees that the last line -- the included correction -- is equal and opposite to the contribution
from the quadrupole response function, which is excluded by the restriction $|\kappa|$=1.  The magnitude
of each correction is about 5\%.
One would have been better off making no correction for the muon's relativity, rather than including this correction at the
cost of limiting the electron's partial wave expansion.  This is true even for a nucleus
with a relatively weak quadrupole response that contributes 
only as a probability, not through interference.

In fact, the consequences of selectively including corrections, in the absence of a systematic expansion,
can be considerably more severe.  Several of the potential pitfalls can be illustrated with the tensor-mediated interaction 
of Eq. (\ref{eq:tensorexch}).  For an isoscalar coupling of strength $\tilde{d}_T^{\, 0}$ one has
\begin{eqnarray} 
&& (2 \tilde{d}_T^{\, 0})^2 \bigg[ \sum_{J=0,2,...}  \langle {q \over 2 m_N} M_J- { q_\mathrm{eff} \over m_N}  \Phi^{\prime \prime}_J \rangle^2   \nonumber \\
&& ~+  \sum_{J=1,3,...}  \bigg[  \langle \Sigma_J^\prime  - {\langle f \rangle \over \langle g \rangle} \Sigma_J^{\prime (0)} \rangle^2 + \langle \Sigma_J^{\prime \prime}   + {\langle f \rangle \over \langle g \rangle} \Sigma_J^{\prime \prime (0)} \rangle^2  \bigg] \nonumber \\
&& ~~~~~~+  \sum_{J=2,4,...}  \langle {q_\mathrm{eff} \over m_N} \tilde{\Phi}^\prime _J \rangle^2  \bigg] 
\label{eq:Texample}
\end{eqnarray}
On expanding in the small parameters and evaluating response functions, one finds for $^{27}$Al
\begin{eqnarray}
\frac{q^2}{4m_N^2}\braket{M_0}^2&\approx& \frac{q^2}{4m_N^2}\left(348.1+o(y)\right)\approx 0.392 \nonumber \\
\frac{q_\mathrm{eff}^2}{m_N^2}\braket{\Phi''_0}^2 &\approx& \frac{q_\mathrm{eff}^2}{m_N^2}\left(11.9+o(y)\right)\approx 0.077 \nonumber \\
-\frac{q_\mathrm{eff}}{m_N}\frac{q}{m_N}\braket{M_0}\braket{\Phi''_0}&\approx& -\frac{q_\mathrm{eff}}{m_N}\frac{q}{m_N}\left(-64.3+o(y) \right) \nonumber \\
&\approx& 0.348 \nonumber \\
\braket{\Sigma'_1}^2&\approx& 0.237 + o(y) \approx 0.046 \nonumber \\
\braket{\Sigma''_1}^2&\approx& 0.118 + o(y) \approx 0.051 \nonumber
\end{eqnarray}
Here we see that (1) the allowed spin operators that have $\vec{v}_\mu$ corrections are themselves relatively weak, so that
the corrections are well below the 1\% level and thus are ignored above; (2) the allowed response is dominated by the coherent
isoscalar charge operator, though for the tensor-mediated operator under study, there is a $q/m_N$ suppression; (3) consequently, operators
generated by $\vec{v}_N$ are of the same order in $1/m_N$; (4) the leading such operator numerically is $\Phi^{\prime \prime}_0$,
which both interferes with the charge operator and exhibits its own coherence (see Sec. \ref{sec:sec5}); and (5) the error one makes from ignoring $\vec{v}_N$-dependent
operators beyond the allowed approximation is $o(1)$.  

Repeating this calculation for Cu, a heavier target with properties similar to $^{27}$Al,  yields
\begin{eqnarray}
\frac{q^2}{4m_N^2}\braket{M_0}^2&\approx& \frac{q^2}{4m_N^2}\left(1290+o(y)\right) \approx 0.629, \nonumber \\
\frac{q_\mathrm{eff}^2}{m_N^2}\braket{\Phi''_0}^2 &\approx& \frac{q_\mathrm{eff}^2}{m_N^2}\left(65.2+o(y)\right) \approx  0.248, \nonumber \\
-\frac{q_\mathrm{eff}}{m_N}\frac{q}{m_N}\braket{M_0}\braket{\Phi''_0}&\approx& -\frac{q_\mathrm{eff}}{m_N}\frac{q}{m_N}\left(-289.4+o(y)\right) \nonumber \\
&\approx&  0.790, \nonumber \\
\braket{\Sigma'_1}^2&\approx& 0.326 + o(y) \approx 0.028, \nonumber \\
\braket{\Sigma''_1}^2&\approx& 0.163 + o(y) \approx 0.026. \nonumber
\end{eqnarray}
Corrections beyond the allowed limit associated with $\vec{v}_N$ enhance the rate by nearly a factor of 3.\\
~\\
\noindent
{\it The NRET response to $o(\vec{v}_N)$.}  A conclusion one can draw from the above examples is that, because
the CLFV operator structure is unknown, the best guide for experimental investigations is symmetry; the $P$ and
$CP$ constraints that apply to the elastic process do limit the possible responses to the six we have defined.
This also makes practical sense, as the principal tool available to diagnose the source of CLFV is variation of the
nuclear target, with the various response functions serving as knobs: the experimentalist can control target spin, isospin, its spin-orbit structure, its valence structure
(e.g., whether the angular momentum is dominated by intrinsic spin or is orbital), etc.
But we cannot insist, without a theory of CLFV, that we have a hierarchy of allowed and 
first-forbidden responses that make certain operators more important than others.   One cannot tell whether an
LEC is $o(1)$ or $o(q/m_N)$; one shortcoming of $\mu \rightarrow e$ conversion 
is that it allows us to sample physics at only one momentum
transfer $\approx m_\mu$, 

As we previously pointed out \cite{RulePRL} and also discuss below, this makes an NRET that includes the effects of $\vec{v}_N$ but
not $\vec{v}_\mu$ of particular interest:
\begin{widetext}
\allowdisplaybreaks
\begin{eqnarray}
\label{eq:ratemn}
&&\omega = {G_F^2 \over \pi} ~ {q_\mathrm{eff}^2 \over 1+{q \over M_T}}  ~ |\phi_{1s}^{Z_\mathrm{eff}}(\vec{0})|^2  ~ 
\sum_{ \tau=0,1} \sum_{\tau^\prime = 0,1} 
 \Bigg\{  ~\left[ \tilde{R}_{MM}^{\tau \tau^\prime}~W_{MM}^{\tau \tau^\prime}(q_\mathrm{eff}) +\tilde{R}_{\Sigma^{\prime \prime} \Sigma^{\prime \prime}}^{\tau \tau^\prime} ~W_{\Sigma^{\prime \prime} \Sigma^{\prime \prime}}^{\tau \tau^\prime}(q_\mathrm{eff})  +   \tilde{R}_{\Sigma^\prime \Sigma^\prime}^{\tau \tau^\prime} ~ W_{\Sigma^\prime \Sigma^\prime}^{\tau \tau^\prime}(q_\mathrm{eff}) \right]  \nonumber \\  
&&~~~~~~~~~~~~~~~~~~~~~~~~~~~~~~~~~~~~~~~~~~~~~~+ {q_\mathrm{eff}^{~2} \over m_N^2} ~\left[ \tilde{R}_{\Phi^{\prime \prime} \Phi^{\prime \prime}}^{\tau \tau^\prime} ~ W_{\Phi^{\prime \prime} \Phi^{\prime \prime}}^{\tau \tau^\prime}(q_\mathrm{eff})  
+   \tilde{R}_{\tilde{\Phi}^\prime \tilde{\Phi}^\prime}^{\tau \tau^\prime}~W_{\tilde{\Phi}^\prime \tilde{\Phi}^\prime}^{\tau \tau^\prime}(q_\mathrm{eff})  + \tilde{R}_{\Delta \Delta}^{\tau \tau^\prime}~ W_{\Delta \Delta}^{\tau \tau^\prime}(q_\mathrm{eff})   \right]\nonumber \\
&&~~~~~~~~~~~~~~~~~~~~~~~~~~~~~~~~~~~~~~~~~~~~~-  {2 q_\mathrm{eff} \over m_N}~\left[ \tilde{R}_{ \Phi^{\prime \prime}M}^{\tau \tau^\prime} ~W_{ \Phi^{\prime \prime}M}^{\tau \tau^\prime}(q_\mathrm{eff}) 
 +  \tilde{R}_{\Delta \Sigma^\prime}^{\tau \tau^\prime} ~W_{\Delta \Sigma^\prime}^{\tau \tau^\prime}(q_\mathrm{eff})   \right]  \Bigg\}~~~~~~
\end{eqnarray}
\end{widetext}
Here we have chosen to remove the muon's upper component wave function $g$ from the multipole operators, using Eq. (\ref{eq:amp3A}) rather than Eq. (\ref{eq:amp3}).
The leptonic tensors not already defined in Eq. (\ref{eq:ampallowed}) are
\begin{eqnarray}
\label{eq:Rs}
 \tilde{R}_{\Phi^{\prime \prime }  \Phi^{\prime \prime }  }^{\tau \tau^\prime}&=& \tilde{c}_3^\tau \tilde{c}_3^{\tau^\prime * } + (\tilde{c}_{12}^\tau- \tilde{c}_{15}^\tau ) ( \tilde{c}_{12}^{\tau^\prime *}-\tilde{c}_{15}^{\tau^\prime *} ),  \nonumber \\
 \tilde{R}_{\Phi^{\prime \prime } M}^{\tau \tau^\prime} &=& \mathrm{Re} [ \tilde{c}_3^\tau \tilde{c}_1^{\tau^\prime *} -  \left( \tilde{c}_{12}^\tau - \tilde{c}_{15}^\tau \right) \tilde{c}_{11}^{\tau^\prime *} ], \nonumber \\ 
      \tilde{R}_{\Delta \Delta}^{\tau \tau^\prime} &=&\tilde{c}^\tau_{5}  \tilde{c}_{5}^{\tau^\prime *} + \tilde{c}_{8}^\tau \tilde{c}_{8}^{\tau^\prime * }, \nonumber \\
      \tilde{R}_{\Delta \Sigma^{ \prime }}^{\tau \tau^\prime} &=&\mathrm{Re}[\tilde{c}^\tau_{5}  \tilde{c}_{4}^{\tau^\prime *} + \tilde{c}_{8}^\tau\tilde{ c}_{9}^{\tau^\prime * }], \nonumber \\                 
  \tilde{R}_{\tilde{\Phi}^{\prime } \tilde{\Phi}^{\prime }}^{\tau \tau^\prime}&=& \tilde{c}_{12}^\tau \tilde{c}_{12}^{\tau^\prime * }+\tilde{c}_{13}^\tau \tilde{c}_{13}^{\tau^\prime *}.   \nonumber
\end{eqnarray}
The terms accompanied by factors of ${q_\mathrm{eff} \over m_N}$ depend on the intrinsic nucleon velocity operator
and thus are a reflection of the finite size of the nucleus.

Arguably, this result provides the optimal formalism for analyzing $\mu \rightarrow e$ conversion
as experimentalist seek to discover CLFV:
\begin{enumerate} [leftmargin=1em]
\item The result is complete.  It provides leading-order expressions for each of the six response functions and two interference
terms that, as we argued in the discussion around Eq. (\ref{eq:symmetries}), are allowed by symmetry arguments.
\item These eight terms can in principle be separately measured: their associated nuclear structure functions $W$ 
are ``knobs" the experimentalist can turn by selecting nuclear targets with appropriate properties.   The leptonic tensors accompanying these eight terms
thus represent the available information that can be extracted from a comprehensive program of $\mu \rightarrow e$ conversion experiments.  In the
current impulse approximation NRET formalism, this
information consists of constraints on the $R$'s, bilinears
in the LECs of the nucleon-level NRET.
\item This result is effectively equivalent to the full result of Eq. (\ref{eq:amp3}), if we interpret the deduced LECs as effective, incorporating
small form-factor corrections associated with $\vec{v}_\mu$ and thus proportional to $\langle f \rangle/\langle g \rangle$.  Until CLFV is measured with such precision, 
the effects of such corrections will not be apparent.  In contrast to $\vec{v}_\mu$,
$\vec{v}_N$ generates new operators that can, as we have seen, generate leading contributions to CLFV.
\end{enumerate}
In Appendix \ref{AppendixC} we describe a publicly available script that evaluates $\mu \rightarrow e$ conversion rates for the targets
investigated in this paper, using Eq. (\ref{eq:ratemn}).\\
~\\
\noindent
{\it The NRET response linear in velocities.}  For completeness the rate linear in both velocities is given in two forms in Eqs. (\ref{eq:amp2}) and (\ref{eq:amp3}), with
the muon's upper and lower components included in the transition densities, and in Eq. (\ref{eq:amp3A}) in a factorized form, where
$f$ and $g$ are replaced by their multipole averages.  The impact of including $\vec{v}_\mu$ is to add a Coulomb correction to nuclear form factors
proportional to the ratio of the muon's upper and lower components, $\langle f \rangle/\langle g \rangle$.  See Appendix B for discussion.\\

\section{$\mu \rightarrow e$ Conversion Nuclear Physics}
\label{sec:sec5}
Here we discuss the nuclear physics of the response functions that govern elastic $\mu \rightarrow e$ conversion.  We
discuss differences between the nuclear-level theory for elastic $\mu \rightarrow e$ conversion and  the nucleon-level NRET:
the nucleus imposes selection rules that suppress certain interactions and generates two types of coherence that enhances sensitivities to others.
We also describe the shell model
calculations performed for $^{27}$Al and the other nuclear targets considered here.\\

\subsection{Response function properties}
The leading multipole operators for the three response functions included in the allowed rate of Eq. (\ref{eq:ampallowed}) 
take on the following familiar forms in the long-wavelength ($q \rightarrow 0$) limit, 
\begin{eqnarray}
\sqrt{4 \pi} ~M_{00}(0) =  \sum_{i=1}^A~ 1(i) ~~\sqrt{6 \pi} ~ \Sigma^\prime_{1M}(0 ) &=&  \sum_{i=1}^A~ \sigma_{1M}(i) \nonumber \\
\sqrt{12 \pi} ~ \Sigma^{\prime \prime}_{1M}(0) = \sum_{i=1}^A~ \sigma_{1M}(i)
 \label{eq:LWL}
\end{eqnarray}
These  operators would be present even if all effects of nuclear size are ignored.
The coherent operator $M_{00}$ can be isolated by using a $j_N=0$ target like $^{12}$C, while the transverse electric and longitudinal operators $\Sigma^\prime$
and $\Sigma^{\prime \prime}$, respectively, 
require $j_N \ge {1 \over 2}$.  As these spin-dependent operators appear
as an incoherent sum in our rate equations, they would appear to be difficult to separate, implying that the $c_i$'s associated with
these operators should be treated as a single set.  However, when the full form factors are considered, there can be significant variations in the relative strengths of $\Sigma^\prime$ and $\Sigma^{\prime \prime}$
from target to target, opening up possibilities for separating them.  The incoherent sum arises from completing the traces over leptonic
spins; we mention in Appendix \ref{AppendixB} that muon hyperfine interactions can affect this trace.

\begin{figure*}[ht]   
\centering
\includegraphics[scale=0.45 ]{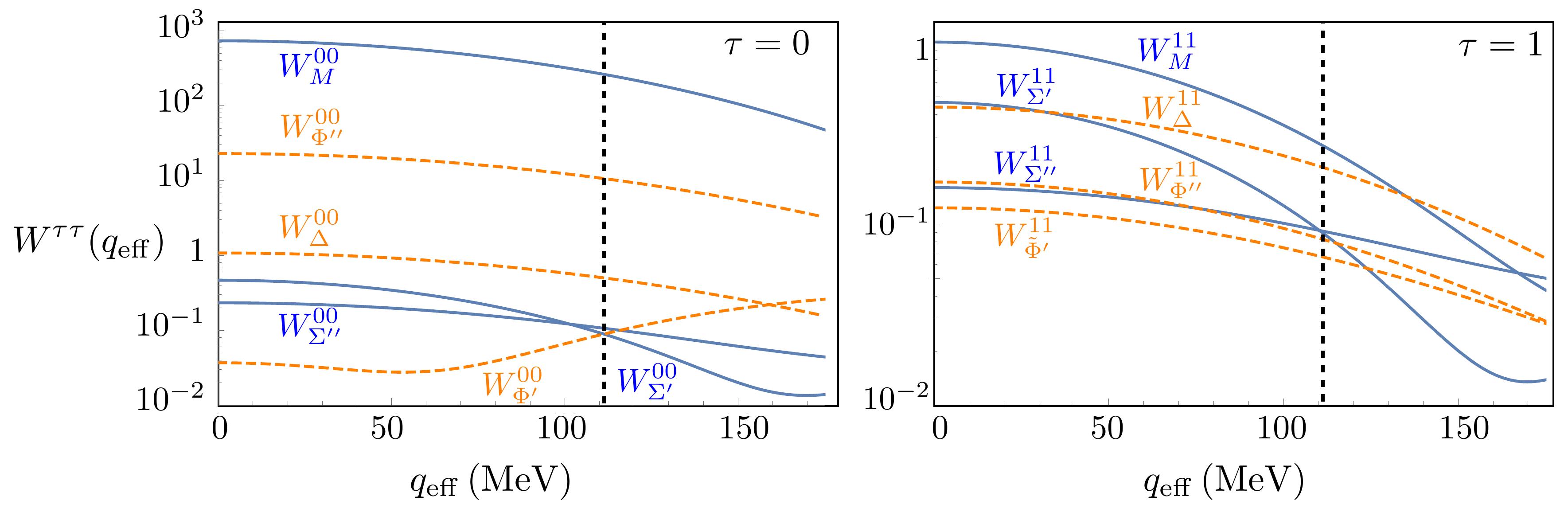}
\caption{The nuclear response functions for the six operators contributing to elastic $\mu \rightarrow e$ conversion on $^{27}$Al.
Here $W_{O}^{\tau \tau}(q_\mathrm{eff})$ denotes $W_{O O}^{\tau \tau}(q_\mathrm{eff})$.
The left (right) panel gives the results for isoscalar (isovector) coupling.  The response functions are needed at the  $q_\mathrm{eff}$ 
indicated by the dashed line.  The results in blue correspond to charge and spin couplings, while those in orange correspond to the velocity-dependent
operators where the response functions are accompanied by the additional factor $q_\mathrm{eff}^2/m_N^2 \approx 0.014$. }
\label{fig:AlFF}
\end{figure*}

\begin{figure*}[ht]   
\centering
\includegraphics[scale=0.45 ]{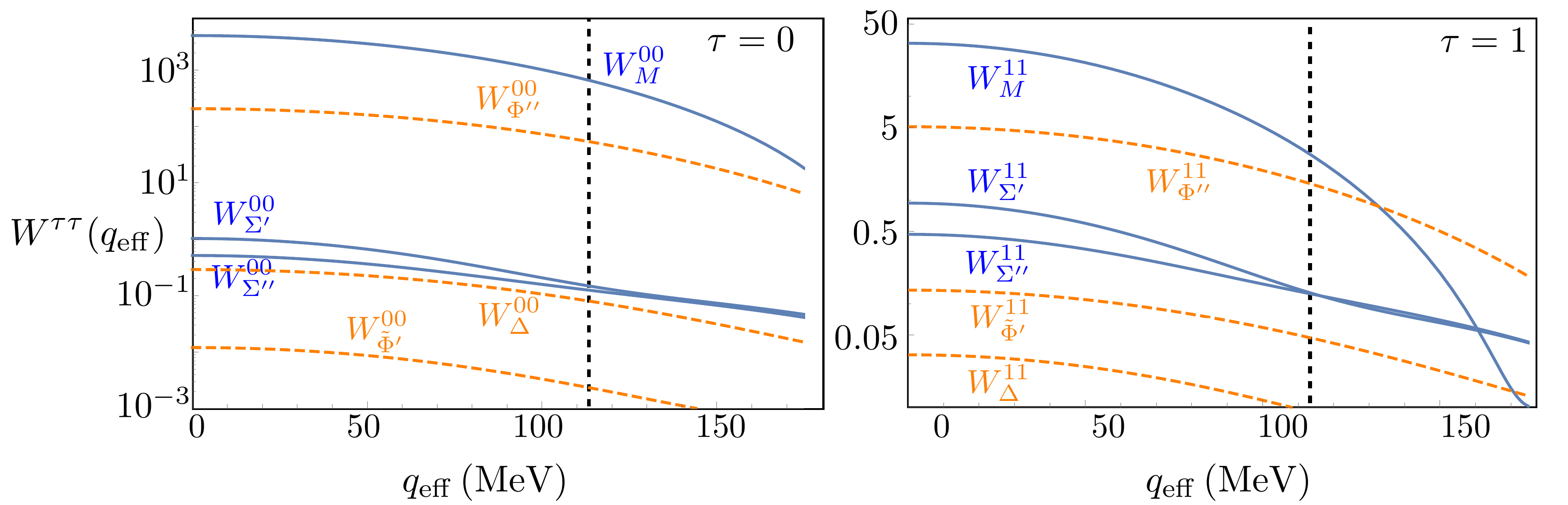}
\caption{As in Fig. \ref{fig:AlFF} but for Cu.}
\label{fig:CuFF}
\end{figure*}

The remaining three operators are associated with couplings to nucleon relative velocities, and thus arise from
nuclear compositeness. In the expression for the decay rate Eq. (\ref{eq:ratemn}), each such operator is accompanied by a factor $q_\mathrm{eff}/m_N$ that vanishes in the long-wavelength limit.  For small but non-zero $q_\mathrm{eff}/m_N$, these operators reduce to 
\begin{widetext}
\begin{eqnarray}
\Delta_{1M}(0) =-\textstyle{1 \over \sqrt{24 \pi}}  \sum_{i=1}^A~\ell_{1M}(i), ~~~~~~~~
 \tilde{\Phi}^{\prime}_{2M}(0) = -\textstyle{1 \over \sqrt{20 \pi}} \sum_{i=1}^A~\left[ x(i) \otimes \left(\vec{\sigma}(i) \times {1 \over i} \vec{\nabla}(i) \right)_1 \right]_{2M}, \nonumber \\
\Phi^{\prime \prime}_{JM}(0 ) =  \left\{ \begin{array}{lr}- \textstyle{1 \over 6 \sqrt{\pi}} \sum\limits_{i=1}^A ~ \vec{\sigma}(i) \cdot \vec{\ell}(i), &~~J=0,  \\[0.4cm]
 -{1 \over \sqrt{30 \pi}} \sum\limits_{i=1}^A ~\left[ x(i) \otimes \left(\vec{\sigma}(i) \times {1 \over i} \vec{\nabla}(i) \right)_1 \right]_{2M},  &~~ J=2, \end{array} \right.~~~~~~~~~~~~~~~~~~~~~ \nonumber \\
 ~~
 \label{eq:NLO}
\end{eqnarray}
\end{widetext}
where $\vec{\ell}$ is the orbital angular momentum operator. 

The response functions $W_{OO^\prime}^{\tau \tau}(q_\mathrm{eff})$ for $^{27}$Al are given in Fig. \ref{fig:AlFF}, taken from $2s1d$ shell-model
calculations using the USDA interaction (to be described later).  
For comparison, we show results for Cu in Fig. \ref{fig:CuFF} -- a case analogous to Al as the stable isotopes of Cu have
an unpaired proton -- evaluated from wave functions obtained by diagonalizing the GCN2850 interaction in the shell-model space $2p_{3/2}1f_{5/2} 2p_{1/2}1g_{9/2}$.   One might 
na\"ively expect that the contribution to the rate -- and thus sensitivity to the associated $c^2_i$'s -- of the velocity-dependent response functions
would be typically reduced by a factor of 100, due to the accompanying $q_\mathrm{{eff}}^2/m_N^2$.   But the figures show a more interesting pattern.

For isoscalar coupling, the contribution of $M_{00}$ to $W_{MM}$ is enhanced by a factor $\lesssim A$: with increasing $A$, the coherent sum over the core
involves more nucleons, but the associated point $q \approx m_\mu$ on the elastic form factor moves closer to the first diffraction minimum.  The isoscalar response per target  nucleon
is maximized for medium-mass nuclei in the neighborhood of Cu.   No such coherence operates for the spin and convection current operators: the responses 
$W_{\Sigma^\prime\Sigma'}^{00}$, $W_{\Sigma^{\prime \prime}\Sigma''}^{00}$, and $W_{\Delta\Delta}^{00}$  have roughly the expected single-particle value.   This pattern
is seen in both  Al and Cu, with differences reflecting nuclear structure details, e.g., whether the angular momentum carried by
the unpaired nucleon is dominated by orbital angular momentum or spin.  

In contrast, $W_{\Phi^{\prime \prime} \Phi^{\prime \prime}}^{00}$, which is
associated with the scalar spin-orbit operator $\vec{\sigma} \cdot \vec{\ell}$, exhibits a coherence in the two cases illustrated.
The na\"ive shell-model picture of $^{27}$Al corresponds to a nearly full $1d_{5/2}$ shell, but an empty $1d_{3/2}$; similarly, the
$1f_{7/2}$ is closed for Cu but the $1f_{5/2}$ mostly open.   This is a consequence of the strong spin-orbit contribution to the nuclear mean field in these mid-shell nuclei.
If the operator $\vec{\sigma} \cdot \vec{\ell}$ is summed over all $2(\ell+1)$ states of the $j=\ell+\frac{1}{2}$ subshell and all $2\ell$ states of the $j=\ell-\frac{1}{2}$ subshell, it vanishes.
But the sum is coherent if done over half shells.  Al and Cu are nearly optimal cases for exploiting this coherence.  If CLFV is seen in a $J=0$ target, it will be very easy to 
determine which of the two candidates scalar operators is responsible,
due to this distinctive nuclear physics.

Consequently, in cases like $^{27}$Al and Cu, where this spin-orbit coherence operates, we find a hierarchy of nuclear operators that differs from that of the nucleon-level NRET:
$W_{MM}^{00} \gg \left\{ W^{00}_{\Sigma^{\prime} \Sigma^{\prime}} , W^{00}_{\Sigma^{\prime \prime} \Sigma^{\prime \prime}}, {q_\mathrm{eff}^2 \over m_N^2} W_{\Phi^{\prime \prime} \Phi^{\prime \prime}}^{00} \right\} \gg \left\{ {q_\mathrm{eff}^2 \over m_N^2} W_{\Delta\Delta}^{00}, {q_\mathrm{eff}^2 \over  m_N^2} W_{{\tilde{\Phi}}^{ \prime}{\tilde{\Phi}}'}^{00} \right\}$.  $W_{\Phi^{\prime \prime} \Phi^{\prime \prime}}^{00} $ is ``elevated" by the nucleus, with the coherent enhancement  sufficient
to overcome the ${q_\mathrm{eff}^2 \over m_N^2}$ suppression, making $\Phi^{\prime \prime}$ numerically as relevant as the allowed spin operators.   As this operator is associated with the spin-velocity current $\vec{v}_N \times \vec{\sigma}_N$
and thus a set of LECs distinct from those accompanying $M$, $\Sigma^\prime$, or $\Sigma^{\prime \prime}$,
it provides experimentalists with a nice probe of tensor-mediated and other more exotic interactions.

\begin{figure*}[ht]   
\centering
\includegraphics[scale=0.32]{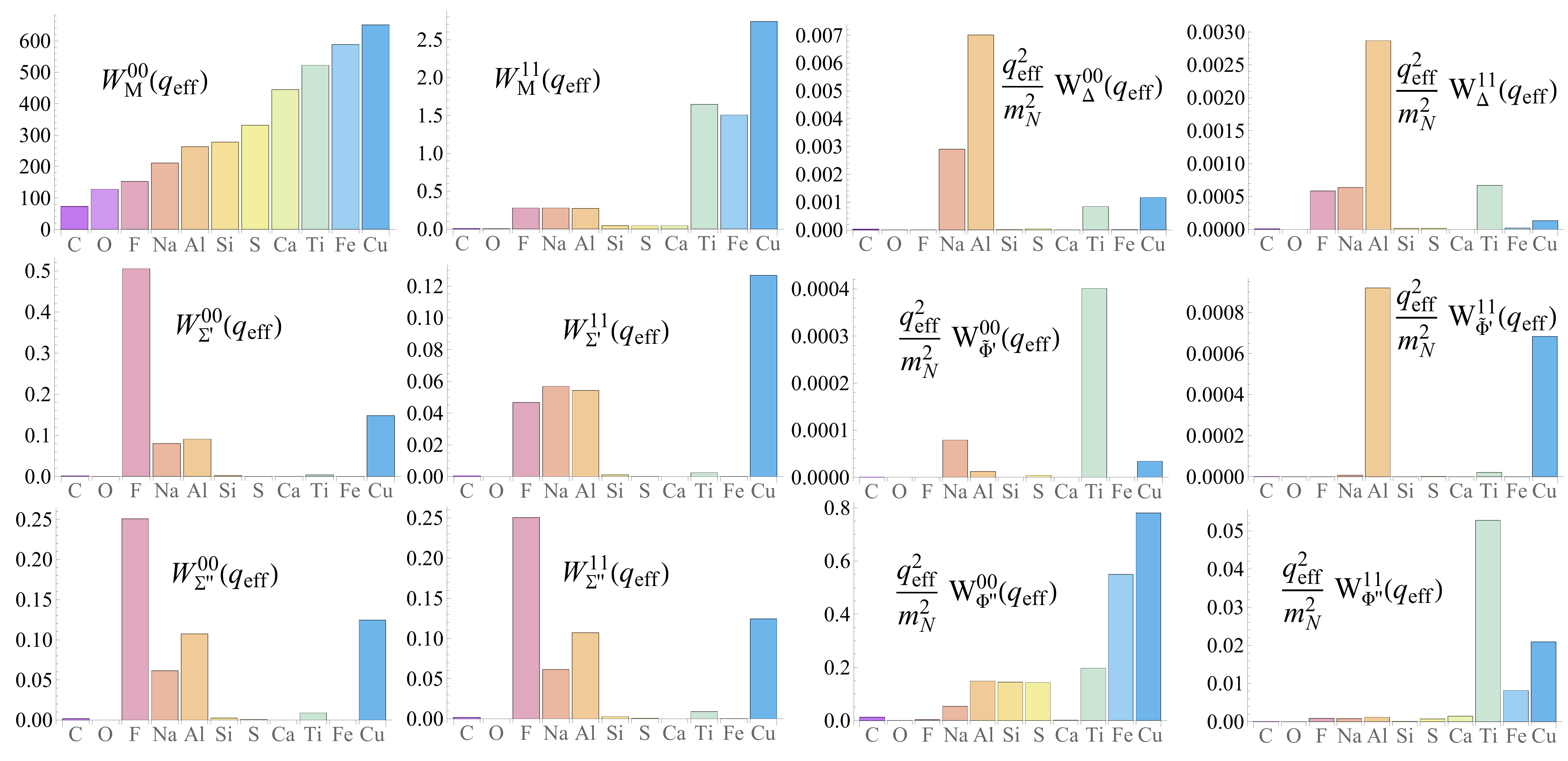}
\caption{The nuclear responses to both allowed (left two panels) and velocity-induced (right) operators, 
for the 11 targets considered in this paper.  Target properties can significantly alter operator responses,
providing an important diagnostic tool for determining the source of CLFV, should it be observed.}
\label{fig:MEs}
\end{figure*}

In the case of isovector coupling, an isospin symmetric core makes no contribution, and consequently for light targets like Al
the operator hierarchy is determined simply by $ {q_\mathrm{eff }\over m_N}$, with $ \left\{ W_{MM}^{11}, W^{11}_{\Sigma^{\prime}\Sigma'}, W^{11}_{\Sigma^{\prime \prime}\Sigma''} \right\} \gg \left\{ {q_\mathrm{eff}^2 \over m_N^2} W_{\Delta\Delta}^{11}, {q_\mathrm{eff}^2 \over m_N^2} W_{{\tilde{\Phi}}^{ \prime}{\tilde{\Phi}}'}^{11},{q_\mathrm{eff}^2 \over m_N^2} W_{\Phi^{\prime \prime}\Phi''}^{11} \right\}$.   This pattern does change a bit as the neutron excess grows
in heavier nuclei, as the coherence grows over that excess (other conditions being satisfied).  This is apparent in the comparison between
the panels on the right in Figs. \ref{fig:AlFF} and \ref{fig:CuFF}. 

Of course, all of these general expectations may vary considerably from nucleus to nucleus, reflecting specific aspects of the structure: properties that potentially could be exploited to determine the operators responsible for CLFV.  In Fig. \ref{fig:MEs} we compare the operator sensitivities
of the eleven targets discussed here.  The spin-orbit coherent amplification of $\Phi^{\prime \prime}$ is present in mid-shell nuclei like Al, Si, and Cu but absent in closed-shell nuclei like O and Ca.
The valence nucleon structure of F produces a strong response to spin, but a weak one to orbital motion; Al is the reverse.  O and Ca are
effectively sensitive to only $M$.  Were CLFV to be discovered, a lot could be learned from the relative responses of different
targets about its operator origin.

\begin{figure}[ht]   
\centering
\includegraphics[scale=0.265]{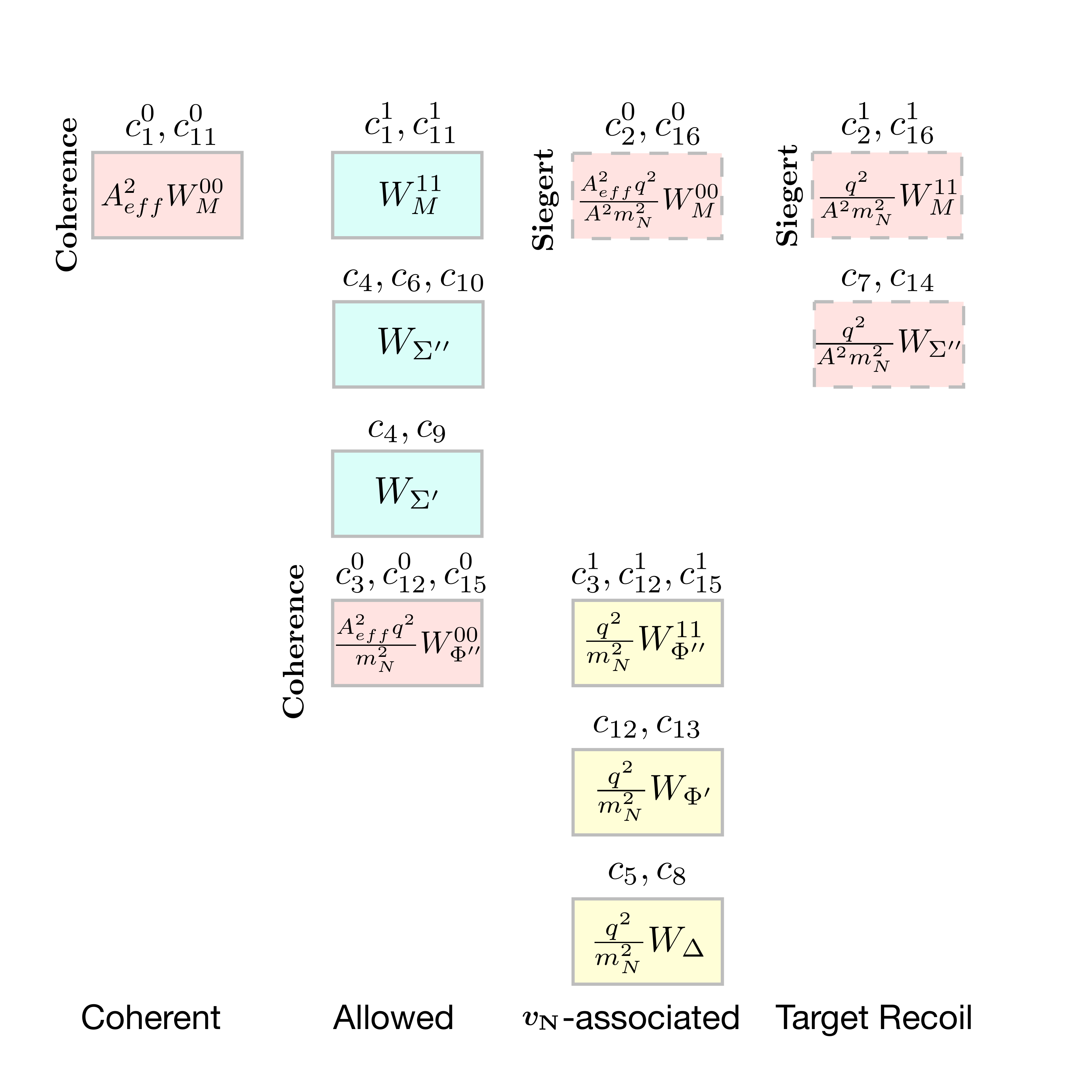}
\caption{Nuclear- vs nucleon-level NRET, with strong operators to the left and weak ones to the right.  Na\"ively, NRET operators are 
either allowed (blue-green) or velocity suppressed (yellow).  The nucleus alters
this pattern, as indicated in pink.  Distinct forms of nuclear coherence 
enhance the responses $W_M^{00}$ and $W_{\Phi^{\prime \prime}}^{00}$ by factors typically $\approx$ 100 in $^{27}$Al..  Other nucleon-level NRET operators,
$\CO_2$, $\CO_7$, $\CO_{14}$, and $\CO_{16}$, are forbidden by selection rules or current conservation, apart from small
nuclear recoil effects, suppressing these operators by $\approx {1 \over 100}$.}
\label{fig:nuclearEFT}
\end{figure}

In addition to sources of coherence, the other important nuclear aspect of elastic $\mu \rightarrow e$ conversion is its blindness to certain LECs of the NRET,
as a consequence of $P$ and $CP$ selection rules.
The LECs $\{c_2,c_{16} \}$ are associated with the longitudinal projection of $\vec{v}_N$,
but the only symmetry-allowed multipoles of $\vec{v}_N$ are odd-$J$ transverse magnetic ones.
The LECs $\{c_7,c_{14} \}$ are associated with the axial charge operator $\vec{v}_N \cdot \vec{\sigma}_N$, but Coulomb multipoles 
of this operator are parity odd when $J$ is even and time-reversal odd when $J$ is odd.  Thus there are no surviving multipoles.
The $P$ and $T$ symmetry constraints utilized above will be broken very slightly by weak corrections that generate small parity or $CP$
admixtures in the nuclear ground state, and also in some cases by nuclear recoil corrections, which we have neglected as they are
of order $m_\mu/Am_N \lesssim 0.01$ in amplitude.  The effects of nuclear embedding on the NRET are summarized
in Fig. \ref{fig:nuclearEFT}.

Such symmetry constraints do not limit $\mu \rightarrow e$ conversion accompanied by nuclear excitation.  Despite the increased background
from free muon decay, forthcoming experiments should be able to extract significant limits on inelastic $\mu \rightarrow e$ conversion involving
low-lying nuclear states.  The formalism presented here can be generalized to include inelastic excitation:  the NRET then generates additional
operators as well as additional multipoles of the operators defined here.  We discuss this generalization elsewhere \cite{RuleInelastic}.

\subsection{Nuclear response function evaluation}
\label{sec:secE}
Here we describe the evaluations we have done of the nuclear response functions for 11 selected targets ranging up to Cu.  We assume
natural targets and sum responses over all significant isotopes.  In each case we use empirically determined effective interactions, usually
exploring several when multiple good choices exist.   We limit ourselves to cases where full-basis calculations can be done, retaining all
Slater determinants that can be formed in the spaces appropriate to the effective interaction.   This is the proper use of the effective interaction.

Our calculations were performed with BIGSTICK \cite{Johnson:2013bna,Calvin}, a Lanczos-algorithm code capable of handling Hilbert spaces of dimension up to 10$^{11}$.
While effective interactions can be determined without referencing the underlying single-particle basis, we use a harmonic oscillator single-particle basis
with a size parameter $b$ chosen to reproduce the nuclear charge density, as deduced from electron scattering.  This basis choice is unique
in allowing exact projection of center-of-mass motion, preserving translational invariance, provided the requisite complete sets of Slater
determinants are employed, which is the case in the present work.  A second advantage of the harmonic oscillator basis is that, if one chooses to replace $g(r)$ 
and $f(r)$ by average values,  all nuclear matrix elements can be evaluated analytically.   The momentum
dependence of all form factors is encoded into a single dimensionless parameter
$y=(q_\mathrm{eff} b/2)^2$, a result we have utilized repeatedly in our NRET discussions,
$ W_i^{\tau \tau^\prime} \rightarrow W_i^{\tau \tau^\prime}(y)$.

We have employed the impulse approximation, though, as discussed earlier, this approach is more general in work that includes
a complete basis of single-particle operators, as we do here.   The
nuclear core effectively averages multi-body operators to density-dependent one-body forms -- forms that would be in our operator basis --
effectively renormalizing the impulse approximation LECs.   It is these renormalized
couplings that would be determined from fitting to experiment, were CLFV observed.

The many-body matrix element of any one-body operator $\hat{O}_{J;T}$ of good $J$ and $T$ can be expressed in terms of the one-body density matrix
\begin{eqnarray}
\langle f~ \vdots \vdots~ \sum_{i=1}^A \hat{O}_{JT}(q_\mathrm{eff} \vec{x}_i) ~ \vdots \vdots ~i \rangle~~~~~~~  \nonumber \\
~~~~~ =\sum_{|\alpha|,|\beta|} \Psi_{|\alpha|,|\beta|} ^{fi; JT} ~\langle |\alpha|~ \vdots \vdots ~O_{JT}(q_\mathrm{eff} \vec{x}) ~\vdots \vdots~ |\beta| \rangle 
\label{eq:densitymatrix}
\end{eqnarray} 
where $i$ and $f$ denote the initial and final nuclear states, $\vdots \vdots$ indicates a matrix element reduced in angular momentum and isospin, 
and the sum extends over a complete set of non-magnetic single-particle matrix elements $|\alpha|$ and $|\beta|$. 
This result is exact: the density matrix extracts
from fully correlated many-body nuclear wave functions just the information necessary to exactly evaluate any one-body spherical
tensor operator of the indicated rank in $J$ and $T$.  
The shell model approximates this result by truncating the summations in
$|\alpha|$ and $|\beta|$ to the shell-model valence space and core, hopefully capturing the physics most important to the
evaluation of low-momentum operators. 
As the operators considered here have isospin 1 or $\tau_3$, one obtains
\begin{equation}
\begin{split}
&\langle |\alpha|~ \vdots \vdots~ \hat{O}_{JT}(q_\mathrm{eff} \vec{x}) ~\vdots \vdots~ |\beta| \rangle \\
 &= [\textstyle{1 \over 2}] [T]~ \langle n_\alpha (\ell_\alpha 1/2)j_\alpha ~|| \hat{O}_J (q_\mathrm{eff} \vec{x})  ||~ n_\beta (\ell_\beta 1/2) j_\beta \rangle, ~~~
\label{eq:densitymatrixfinal}
\end{split}
\end{equation}
where $\hat{O}_J$ denotes the space-spin part of the operator and $[T]\equiv \sqrt{2T+1}$.


For the harmonic oscillator single-particle basis we employ, the reduced matrix elements
of $O_J = \{M_J$, $\Sigma^\prime_J$, $\Sigma^{\prime \prime}_J$, $\Delta_J$, $\tilde{\Phi}^\prime_J$, $\Phi^{\prime \prime}_J \}$ can be obtained analytically as
\be
\langle n_\alpha (\ell_\alpha 1/2)j_\alpha || O_J(q \vec{x}) || n_\beta (\ell_\beta 1/2)j_\beta \rangle  \nonumber \\
={1 \over \sqrt{4 \pi}} y^{(J-K)/2} e^{-y} p(y)~~~~~~~~~
\ee
where $K=2$ for the normal parity operators $M_J$, $\tilde{\Phi}_J^\prime$, and $\Phi_J^{\prime \prime}$ and $K=1$ for the abnormal parity operators
$\Delta_J$, $\Sigma^\prime_J$, and $\Sigma^{\prime \prime}_J$.  Here $p(y)$ is a finite polynomial in $y$.  Thus the
nuclear response functions $W$ displayed in Figs. \ref{fig:AlFF} and \ref{fig:CuFF}  have this polynomial
form \cite{Donnelly}.  The additional operators generated from the muon's lower component -- designated by superscripts (0), (1), or (2) --
can also be evaluated analytically but yield hypergeometric functions that cannot be represented as finite polynomials.

As detailed in Appendix \ref{AppendixC}, for most targets we were able to generate wave functions and one-body
density matrices using several available, well-tested effective interactions.    
While the nuclear physics is model based, thus ruling out quantitative error estimation, we can get some feel for uncertainties
by exploring the differences associated with the choice of interaction.   
Thus, while the results we present in Sec. \ref{sec:sec6} use specific effective interactions --
the Cohen and Kurath $1p_{3/2}-1p_{1/2}$ interaction \cite{CK} for C; USDB $1d_{5/2}-2s_{1/2}-1d_{3/2}$ interaction \cite{USDB}for Al, Si, and S;
the KB3G $1f_{7/2}-2p_{3/2}-2p_{1/2}-1f_{5/2}$ 
interaction \cite{KB3G} for Ti; and the GCN2850 $2p_{3/2}-2p_{1/2}-1f_{5/2}-1g_{9/2}$ interaction \cite{GCN2850} for Cu --
the code described in Appendix \ref{AppendixC} allows users to compute rates with density matrices derived from other effective interactions.
Appendix \ref{AppendixC} also details
other aspects of the nuclear physics, such as the procedure for choosing oscillator parameters consistent
with electron scattering determinations of nuclear charge radii. 

\begin{table}[!]
\caption{Existing limits on branching ratios for $\mu \rightarrow e$ conversion, taken from the tabulation of \cite{COMET}.}
\label{tab:symmetry3}
{\renewcommand{\arraystretch}{1.4}
\begin{tabular}{lll}
\hline
Process &~~ Limit &~~~Lab/Reference  \\
\hline
$\mu^-$+$^{32}$S $\rightarrow e^-$+$^{32}$S~ &~~ $7 \times 10^{-11}$~ &~~~SIN \cite{SIN1}~ ~ \\
$\mu^-$+Ti $\rightarrow e^-$+Ti~ &~~ $1.6 \times 10^{-11}$~ &~~~TRIUMF \cite{TRIUMF1}~ ~ \\
$\mu^-$+Ti $\rightarrow e^-$+Ti~ &~~ $4.6 \times 10^{-12}$~ &~~~TRIUMF \cite{TRIUMF2}~ ~ \\
$\mu^-$+Ti $\rightarrow e^-$+Ti~ &~~ $4.3 \times 10^{-12}$~ &~~~PSI \cite{PSI1}~ ~ \\
$\mu^-$+Ti $\rightarrow e^-$+Ti~ &~~ $6.1 \times 10^{-13}$~ &~~~PSI \cite{TiLim}~ ~ \\
$\mu^-$+Cu $\rightarrow e^-$+Cu~ &~~ $1.6 \times 10^{-8}$~ &~~~SREL \cite{SREL1}~ ~ \\
$\mu^-$+Au $\rightarrow e^-$+Au~ &~~ $7 \times 10^{-13}$~ &~~~PSI \cite{AuLim}~ ~ \\
$\mu^-$+Pb $\rightarrow e^-$+Pb~ &~~ $4.9 \times 10^{-10}$~ &~~~TRIUMF \cite{TRIUMF2}~ ~ \\
$\mu^-$+Pb $\rightarrow e^-$+Pb~ &~~ $4.6 \times 10^{-11}$~ &~~~PSI \cite{PSI4}~ ~ \\
\hline
\end{tabular}
}
\end{table} 

\section{LEC Limits: Mu2e, COMET, and DeeMe Impacts}
\label{sec:sec6}
Table \ref{tab:symmetry3} summarizes existing limits on $\mu \rightarrow e$ conversion.  We focus here on measurements employing
lighter nuclei in the table, those less massive than Au, especially
Mu2e and COMET measurements using Al, as well as future DeeMe measurements using graphite and silicon carbide (SiC) targets.   
For such targets full-space, translationally invariant wave functions can be computed, as just described.
As noted previously, the coherent response tends to peak
near Ti due to competition between increasing $A^2$ and the decreasing nuclear form factor.  Thus
light- and medium-mass nuclei are attractive experimental choices, in addition to being simpler in terms of their nuclear physics.

The branching ratios for $\mu \rightarrow e$ conversion are given with respect to the corresponding muon capture rates.
We adopt the following values
\be
\omega_{\mu \rightarrow \nu_\mu} = \left\{ \begin{array}{lr}  0.0378 & \mathrm{C}   \\0.703 & \mathrm{Al}  \\ 0.865  & \mathrm{Si}  \\ 1.351  & \mathrm{S}    \\2.592  & \mathrm{Ti} \\ 5.673 & \mathrm{Cu}  \end{array} \right\} \times 10^6/\mathrm{s}~~~~
\ee
which were obtained by computing the weighted averages of the measurements compiled in Ref. \cite{Suzuki}.  These values can be used to convert $\mu \rightarrow e$ rates to branching ratios.

Over the next five years, new experiments employing high-intensity pulsed muon beams should lead to
substantial improvements on $\mu \rightarrow e$ conversion limits.
The COMET experiment at J-PARC is expected to reach a branching ratio sensitivity of $ 7 \times 10^{-15}$ 
(90\% C.L.) in Phase I \cite{COMETI} and ultimately a Phase II sensitivity of $\approx 10^{-17}$ \cite{COMETII}.
The Mu2e experiment at Fermilab is expected to reach a branching ratio sensitivity of $7 \times 10^{-17}$ (90\% C.L.), 
and a proposed follow-up experiment Mu2e-II, which will take advantage of future beam upgrades at Fermilab, could 
improve this limit by another order of magnitude, so to $7 \times 10^{-18}$ (90\% C.L.) \cite{Abusalma}.
Both COMET and Mu2e will employ Al targets.   A second J-PARC experiment has been proposed
by the DeeMe Collaboration, to be mounted at the Materials and Life Sciences Facility.  Its
goal is a branching ratio of $1 \times 10^{-13}$ for a graphite target.  A follow-up experiment using a 
silicon carbide target has also been discussed.  In this phase, the branching ratio goal would be 
$2 \times 10^{-14}$ \cite{Teshima}.

Our rate calculations were performed with the code described in Appendix \ref{AppendixC}, which is available
in both \textit{Mathematica} and Python versions.   The formalism follows standard
multipole treatments of semileptonic weak interactions \cite{Donnelly,Walecka}.   
The code's nuclear physics consists of a library of density matrices, computed as described previously for
the effective interactions listed in Appendix \ref{AppendixC}.  

\begin{table*}[!]
\caption{Limits on the  dimensionless CLFV LECs $|\tilde{c}_i^\tau|$ (defined relative to the weak scale) and the associated energy scale $\tilde{\Lambda}_i^\tau \equiv \Lambda_i^\tau/\mathrm{TeV} =0.246/\sqrt{|\tilde{c}_i^\tau|}$ (see text), 
imposed by the indicated $\mu \rightarrow e$ conversion branching ratios. 
Branching ratios anticipated in upcoming experiments are indicated by $^\dagger$. }
\label{tab:LEClimits}
\begin{tabular}{|c|l|l|l|c|l|c|}
\hline
 & & & & & &  \\[-7.5pt]
Target  &~~~~~~~~~~ Al&~~~~~~~~~~C&~~~~~~~~SiC&$^{32}$S~~~~~&~~~~~~~~~Ti&Cu\\[1.6pt]
\hline 
 & & & & & &  \\[-7.5pt]
 \diaghead{\theadfont DColumnnnHead II}%
 {Limits \\$|\tilde{c}^\tau_i|$ / $\tilde{\Lambda}^\tau_i$   }{Branch~\\ratio~\\} &  \thead{$10^{-17 \, \dagger}$} & \thead{$10^{-13 \, \dagger}$} & \thead{$2 \times 10^{-14 \, \dagger}$} & \thead{$7 \times 10^{-11}$\cite{SIN1}} & \thead{$6.1 \times 10^{-13}$\cite{TiLim}} & \thead{$1.6 \times 10^{-8}$\cite{SREL1}} \\
 & & & & & &  \\[-7.5pt]
\hline
 & & & & & &  \\[-5.5pt]
$i=1,11~\tau=0$ &\,4.0$\cdot 10^{-10}$/1.2$\cdot 10^4$\,& \,5.1$\cdot 10^{-8}$/1.1$\cdot 10^3$\,&\,1.8$\cdot 10^{-8}$/1.9$\cdot 10^3$\,&\,1.0$\cdot 10^{-6}$/240\,&\,7.4$\cdot 10^{-8}$/910\,&\,1.2$\cdot 10^{-5}$/\,71\,\\
$i=1,11~\tau=1$ &\,1.2$\cdot 10^{-8~}$/2.2$\cdot 10^3$\,& \,6.3$\cdot 10^{-6}$/~~~98\,&\,1.4$\cdot 10^{-6}$/$~~210$\,& - &\,1.3$\cdot 10^{-6}$/$210$\,&\,1.9$\cdot10^{-4}$/\,18\,\\
$i=3,15 ~\tau=0$ &\,1.6$\cdot 10^{-8~}$/1.9$\cdot 10^3$\,& \,4.0$\cdot 10^{-6}$/$~~120$\,&\,7.3$\cdot 10^{-7}$/$~~290$\,&\,4.5$\cdot 10^{-5}$/\,37\,\,&\,3.8$\cdot 10^{-6}$/130\,&\,3.5$\cdot10^{-4}$/\,13\,\\
$i=3,15 ~\tau=1$ &\,1.9$\cdot 10^{-7~}$/$~~570$\,& \,1.3$\cdot10^{-4}$/~~~21\,&\,4.0$\cdot 10^{-5}$/~~~39 \, & - &\,7.3$\cdot 10^{-6}$/~91\,&\,2.1$\cdot 10^{-3}$/5.3\,\\
$i=4~~~~~ \tau=0$ &\,1.4$\cdot 10^{-8~}$/2.1$\cdot 10^3$\,& \,9.4$\cdot 10^{-6}$/~~~80 \,&\,4.1$\cdot 10^{-6}$/$~~120$\,& - &\,1.5$\cdot 10^{-5}$/~63\,&\,5.9$\cdot10^{-4}$/\,10\,\\
$i=4 ~~~~~ \tau=1$ &\,1.7$\cdot 10^{-8~}$/1.9$\cdot 10^3$\,& \,1.1$\cdot 10^{-5}$/~~~76 \,&\,4.9$\cdot 10^{-6}$/$~~110$\,& - &\,1.7$\cdot 10^{-5}$/~59\,&\,6.1$\cdot10^{-4}$/9.9\,\\
$i =5,8 ~~ \tau=0$  &\,7.8$\cdot 10^{-8~}$/$~~880$\,& \,9.6$\cdot 10^{-5}$/~~~25\,&\,7.1$\cdot 10^{-5}$/~~~29 \,& - &\,5.8$\cdot 10^{-5}$/~32\,&\,9.0$\cdot 10^{-3}$/2.6\,\\
$i=5,8 ~~ \tau=1$  &\,1.2$\cdot 10^{-7~}$/$~~720$\,& \,1.6$\cdot10^{-4}$/~~~20\,&\,7.3$\cdot 10^{-5}$/~~~29 \,& - &\,6.5$\cdot 10^{-5}$/~30\,&\,2.7$\cdot 10^{-2}$/1.5\,\\
$i=6,10~ \tau=0$  &\,2.0$\cdot 10^{-8~}$/1.8$\cdot 10^3$\,& \,1.1$\cdot 10^{-5}$/~~~75\,&\,5.5$\cdot 10^{-6}$/$~~110$\,& - &\,1.8$\cdot 10^{-5}$/~59\,&\,8.7$\cdot10^{-4}$/8.3\,\\
$i=6,10~\tau=1$  &\,2.2$\cdot 10^{-8~}$/1.7$\cdot 10^3$\,& \,1.2$\cdot 10^{-5}$/~~~71\,&\,6.1$\cdot 10^{-6}$/~~~99 \,& - &\,2.0$\cdot 10^{-5}$/~55\,&\,8.7$\cdot10^{-4}$/8.3\,\\
$i=9~~~~~\tau=0$   &\,2.1$\cdot 10^{-8~}$/1.7$\cdot 10^3$\,& \,1.9$\cdot 10^{-5}$/~~~57\,&\,6.2$\cdot 10^{-6}$/~~~99 \,& - &\,2.8$\cdot 10^{-5}$/~47\,&\,8.0$\cdot10^{-4}$/8.7\,\\
$i=9~~~~~\tau=1$  &\,2.8$\cdot 10^{-8~}$/1.5$\cdot 10^3$\,& \,2.2$\cdot 10^{-5}$/~~~52\,&\,8.1$\cdot 10^{-6}$/~~~87 \,& - &\,3.4$\cdot 10^{-5}$/~42\,&\,8.7$\cdot10^{-4}$/8.4\,\\
$i=12~~~~\tau=0$ &\,1.6$\cdot 10^{-8~}$/1.9$\cdot 10^3$\,& \,4.0$\cdot 10^{-6}$/$~~120$\,&\,7.3$\cdot 10^{-7}$/$~~290$\,&\,4.5$\cdot 10^{-5}$/\,37\,\,&\,3.8$\cdot 10^{-6}$/130\,&\,3.5$\cdot10^{-4}$/\,13\,\\
$i=12~~~~\tau=1$  &\,1.4$\cdot 10^{-7~}$/$~~660$\,& \,1.3$\cdot10^{-4}$/~~~21 \,&\,4.0$\cdot 10^{-5}$/~~~39 \,& - &\,7.3$\cdot 10^{-6}$/~91\,&\,2.1$\cdot 10^{-3}$/5.4\,\\
$i=13~~~~\tau=0$ &\,1.8$\cdot 10^{-6~}$/$~~180$\,&~~~~~~~~~~~ - &~~~~~~~~~~~ - & - &\,8.4$\cdot 10^{-5}$/~27\,&\,5.3$\cdot 10^{-2}$/1.1\,\\
$i=13~~~~\tau=1$ &\,2.1$\cdot 10^{-7~}$/$~~540$\,&~~~~~~~~~~~ - &~~~~~~~~~~~ - & - &\,3.7$\cdot10^{-4}$/~13\,&\,1.2$\cdot 10^{-2}$/2.3\,\\[2.6pt]
\hline
\end{tabular}
\end{table*}

\subsection{LEC Analysis}
The resulting limits on CLFV are given in Table \ref{tab:LEClimits}, expressed 
in terms of the magnitude of the dimensionless couplings $|\tilde{c}_i^\tau|$, where $\tilde{c}_i^\tau=1$ corresponds to the weak scale,
and alternatively in terms of the CLFV energy scale $\Lambda_i^\tau$.  Recall
\[ c_i^\tau = \sqrt{2} G_F \tilde{c}_i^\tau = {1 \over (\Lambda_i^{\tau } )^2} \]
As 12 LECs contribute to elastic
$\mu \rightarrow e$ conversion and each operator has two isospin components, there are 24 degrees of freedom.
We have explored each LEC separately, as an exercise to assess the comparative
sensitivity of past and anticipated experiments.

Some observations follow about the tables:
\begin{enumerate}
\item In addition to its favorable properties as a muon stopping target, $^{27}$Al is a versatile choice 
theoretically, with a ground-state angular momentum of ${5 \over 2}^+$ that allows all NRET operators to contribute, and with
an isospin of ${1 \over 2}$ that tests both isoscalar and isovector operators.  
\item We have chosen to explore isoscalar and isovector, rather than proton and
neutron, isospin directions.  While the former is more natural theoretically, one would see greater contrast
in operator sensitivity, in cases where the valence nucleon is an odd proton or neutron, by choosing the
latter, incases where operators act primarily on valence nucleons.  The script described in Appendix \ref{AppendixC}
can treat arbitrary isospin choices.
\item The Ti branching ratio we have used is $6.1 \times 10^{-13}$ from Ref. \cite{TiLim}, a conference proceeding.  If one were to instead use the SINDRUM II bound of $4.3 \times 10^{-12}$,
the Ti entries for $\Lambda_i^\tau$ in Table \ref{tab:LEClimits} would need to be decreased by a factor of 1.63, weakening the bounds.
\item The calculations for silicon carbide (SiC) were performed under the assumption that 30\% of the
muons capture on C and 70\% on Si, as the capture probability in a composite material is expected to be
proportional to Z \cite{Fermi:1947uv}.  From the free lifetime of the muon and the muon capture rates given
earlier, one finds that only 7.6\% of the muons bound to C are captured (the rest decay in orbit), while 66\%
of the Si muons are captured.  Similarly, $\mu \rightarrow e$ conversion rates for C are typically one to two orders of magnitude
lower than for Si, depending on the operator choice.  So while the table treats the target as composite, the results are nearly equivalent to 
those one would obtain for a Si target.
\item The nuclear density matrices were calculated from standard effective interactions that conserve
isospin.  This is reflected in the table, e.g., $^{32}$Si as a $T=0$ target is blind to all isovector operators.
Were the Coulomb or other isospin-violating interactions included in the nuclear effective interaction, these operators would be quite
suppressed, but not forbidden.
\end{enumerate}

\subsection{Electromagnetic couplings}
Our $\mu \rightarrow e$ conversion NRET formalism includes cases where the CLFV resides in an electromagnetic coupling to the leptons,
induced by a loop, which then allows a coherent interaction with the nucleus through the exchange of a virtual photon. 
But this source of CLFV could also be explored in purely leptonic
processes such as $\mu \rightarrow e \gamma$ and $\mu \rightarrow 3e$.  Here we consider this scenario, 
evaluating the comparative sensitivities of $\mu \rightarrow e$ and these leptonic processes.

 The most general CLFV electromagnetic vertex is
\begin{eqnarray}
 \Gamma^{\mu }_{\mu \rightarrow e} &=& {1 \over m_\mu^2}  (q^2 \gamma^\mu - q^\mu \slashed{q} ) \left[ f_R^{\mu \rightarrow e}(q^2)  + i f_A ^{\mu \rightarrow e}(q^2)  \gamma_5 \right] \nonumber \\
  &&+i \sigma^{\mu \nu} {q_\nu \over m_\mu} \left[ f_M^{\mu \rightarrow e}(q^2) + i f_E^{\mu \rightarrow e}(q^2) \gamma_5 \right]
  \label{eq:EM}
 \end{eqnarray}
 Here the subscripts $R$, $A$, $M$, and $E$ denote the induced (dimensionless) CLFV charge radius, anapole, magnetic dipole, and electric dipole
 form factors, respectively.   This leads to the nucleon-level relativistic amplitude for electromagnetic  $\mu \rightarrow e$ conversion,
 \be
 {4 \pi \alpha \over q^2} ~ \bar{\chi}_e \Gamma_{\mu \rightarrow e}^\mu(q^2) \chi_\mu ~\bar{N} \gamma_\mu \left({1 + \tau_3 \over 2} \right) N
 \ee
 where in this process the square of the four-momentum $q^2 \approx - m_\mu^2$.   
 
 The nonrelativistic reduction produces
 two coherent EFT operators $\CO_1$ and $\CO_{11}$ that couple to protons with strengths:
 \be
 c_1^0 &=&c_1^1 ={2 \pi \alpha \over m_\mu^2} \left[  f_R^{\mu \rightarrow e}(-m_\mu^2) + f_M^{\mu \rightarrow e}(-m_\mu^2)  \right], \nonumber \\
 c_{11}^0 &=&c_{11}^1 ={2 \pi \alpha \over m_\mu^2} \left[  f_A^{\mu \rightarrow e}(-m_\mu^2) - f_E^{\mu \rightarrow e}(-m_\mu^2)  \right]. ~~~~
 \ee
As the CLFV couplings $f_i$ are suppressed, induced by leptonic loops controlled by some heavy scale $\Lambda$, it is helpful to make this
explicit by defining a new set of dimensionless couplings $\tilde{f}_i \approx$ 1 through the relationship
\[ f_i \equiv {m_\mu^2 \over \Lambda^2} ~ \tilde{f}_i, ~~i=R,A,M,E, \]
in terms of which the dimensionless $\tilde{c}_i$ used in our rate formula are
 \be
 \tilde{c}_1^0 &=&\tilde{c}_1^1 =2 \pi \alpha \frac{v^2}{\Lambda^2}\left[  \tilde{f}_R^{\mu \rightarrow e}(-m_\mu^2) + \tilde{f}_M^{\mu \rightarrow e}(-m_\mu^2)  \right], \nonumber \\
 \tilde{c}_{11}^0 &=&\tilde{c}_{11}^1 =2 \pi \alpha   \frac{v^2}{\Lambda^2}\left[ \tilde{f}_A^{\mu \rightarrow e}(-m_\mu^2) - \tilde{f}_E^{\mu \rightarrow e}(-m_\mu^2)  \right]. ~~~~~
 \ee
The coherent rate is generated through the leptonic tensor $\tilde{R}_{MM}^{\tau \tau^\prime}$ which yields, after summing over isospin,
the coupling to target protons
\[ \left(  |\tilde{c}_1^p|^2+|\tilde{c}_{11}^p|^2 \right), ~~\mathrm{where}~~~\tilde{c}_i^p \equiv \tilde{c}_i^0+\tilde{c}_i^1 = 2c_i^0 . \]
 Thus the contributions of the LECs associated with the two coherent operators cannot be disentangled.  
 
 New information could come from $\mu \rightarrow e \gamma$, as the charge-radius and anapole terms in Eq. (\ref{eq:EM}) vanish
 for on-shell photons.  A calculation of the decay rate for the electromagnetic vertex of Eq. (\ref{eq:EM}) yields
 \begin{eqnarray}
 \omega^{\mu \rightarrow e \gamma} &=& \alpha~ {m_\mu \over 2} \left( |f_M^{\mu \rightarrow e}(0)|^2 + |f_E^{\mu \rightarrow e}(0)|^2 \right) \nonumber \\
 &=& \alpha~ {{m_\mu^5  \over 2 \Lambda^4}} \left( |\tilde{f}_M^{\mu \rightarrow e}(0)|^2 + |\tilde{f}_E^{\mu \rightarrow e}(0)|^2 \right) 
 \end{eqnarray}

\begin{figure*}[ht]   
\centering
\includegraphics[scale=0.95 ]{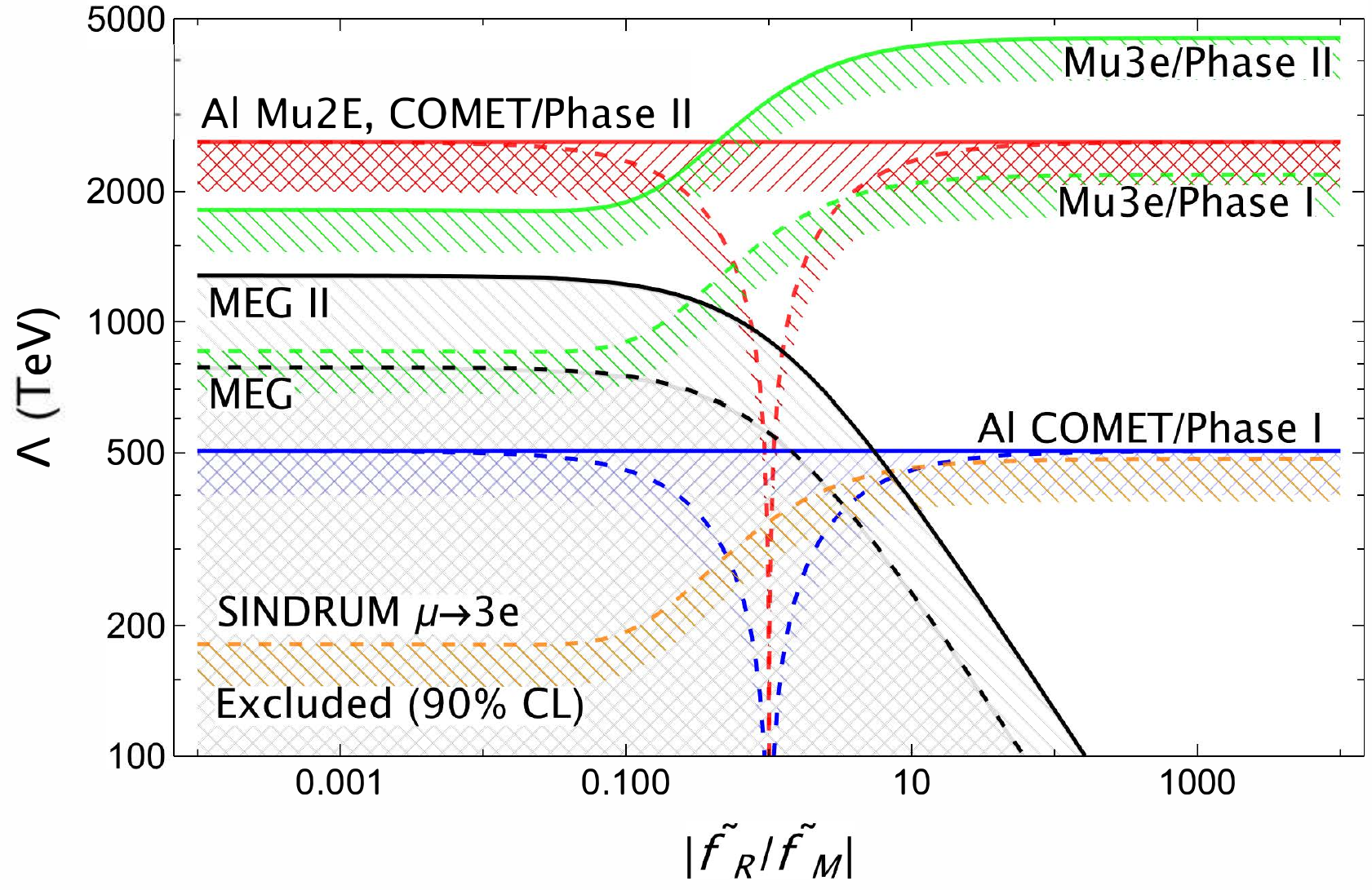}
\caption{Exclusion curves for the CLFV electromagnetic coupling considered in Eq. (\ref{eq:EM}), for the case $\tilde{f}_A=\tilde{f}_E=0$ and $|\tilde{f}_R|+|\tilde{f}_M|=1$. The dashed (solid) black curve shows the (expected) limit for on-shell $\mu\rightarrow e\gamma$ conversion obtained from MEG (MEG II). The branching ratio limits are $B(\mu\rightarrow e\gamma)<4.2\times 10^{-13}$ \cite{MEG} and $6\times 10^{-14}$  \cite{MEGII}, respectively. The dashed (Phase I) and solid (Phase II) green curves show the 
$\mu\rightarrow 3e$  results expected from Mu3e.  The respective branching ratios limits are $< 2.0 \times 10^{-15}$ and $10^{-16}$.  The orange curve is the SINDRUM result, $B(\mu\rightarrow 3e)<1.0\times 10^{-12}$ \cite{SinMu3e}. The solid red (blue) curve corresponds to a $\mu\rightarrow e$  branching ratio limit $B(\mu\rightarrow e)< 10^{-17}$ ($7\times 10^{-15}$) where $\tilde{f}_R$ and $\tilde{f}_M$ contribute with the same sign. The dashed red (blue) curve is the case where $\tilde{f}_R$ and $\tilde{f}_M$ are opposite in sign. In this case, the charge radius and magnetic dipole contributions to coherent conversion on nuclei cancel when $|\tilde{f}_R/\tilde{f}_M|=1$. }
\label{fig:muegamma}
\end{figure*}
 
 Under the assumption that the CLFV arises {\it only} from the coupling in Eq. (\ref{eq:EM}), Fig. \ref{fig:muegamma} compares the 
 constraints obtained from $\mu \rightarrow e$ conversion and $\mu \rightarrow e \gamma$ when $\tilde{f}_M$ and $\tilde{f}_R$ {are} nonzero,
 but their ratio is varied (or equivalently, when $\tilde{f}_E$ and $\tilde{f}_A$ are nonzero, but their ratio is varied). Furthermore it is assumed that $|\tilde{f}_R|+|\tilde{f}_M|=1$. Then the MEG and MEG II experiments, which look for $\mu\rightarrow e\gamma$ with a final state photon, provide a more stringent constraint than Phase I COMET measurements of $\mu\rightarrow e$ conversion mediated by a virtual photon, provided that the magnetic dipole form factor $\tilde{f}_M$ is not small compared to the charge radius form factor $\tilde{f}_R$. On the other hand, a branching ratio limit $B(\mu\rightarrow e)\leq 10^{-17}$ for an Al target will provide a more stringent constraint on CLFV electromagnetic couplings than MEG II, unless significant cancellation occurs between the two couplings $\tilde{f}_R$ and $\tilde{f}_M$.
 
\subsection{Connection to $\mu\rightarrow 3e$}
The case of electromagnetic CLFV can also be related to the process $\mu^{\pm}\rightarrow e^{\pm}e^+e^-$. Like $\mu\rightarrow e$ conversion, improvements in $\mu\rightarrow 3e$ branching ratio sensitivities of potentially four orders of magnitude are expected within the next decade, due to the Mu3e experiment at PSI \cite{Mu3e}. The CLFV $\mu\rightarrow 3e$ process can be mediated by the effective Lagrangian
\begin{equation}
\mathcal{L}=\frac{1}{\Lambda^2}\left(C^L_{\mu eee}\bar{e}^c_L\mu_L \bar{e}_Le^c_L+C^R_{\mu eee}\bar{e}^c_R\mu_R\bar{e}_Re^c_R\right),
\label{eq:mu3e}
\end{equation}
with a corresponding branching ratio
\begin{equation}
B(\mu\rightarrow 3e)=\frac{1}{2}\left(\frac{v}{\Lambda}\right)^4\left(|C^L_{\mu eee}|^2+|C^R_{\mu eee}|^2\right).
\end{equation}
In general, there are dimension-6 operators that mediate $\mu\rightarrow 3e$ beyond those included in Eq. (\ref{eq:mu3e}), but we have restricted to these operators unique in that at one-loop order they generate an effective $\mu\rightarrow e\gamma$ vertex that is enhanced by a large logarithm;
the resulting electromagnetic couplings are \cite{Raidal, Cirigliano2, Cirigliano3}
\begin{equation}
\begin{split}
\tilde{f}_R^{\mu\rightarrow e}(q^2)&=-\frac{1}{(4\pi)^2}\frac{2}{3}\left(C^L_{\mu eee}+C^R_{\mu eee}\right)\ln\frac{-q^2}{\Lambda^2},\\
\tilde{f}_A^{\mu\rightarrow e}(q^2)&=-i\frac{1}{(4\pi)^2}\frac{2}{3}\left(C^L_{\mu eee}-C^R_{\mu eee}\right)\ln\frac{-q^2}{\Lambda^2},
\end{split}
\end{equation}
where we have retained only the large logarithm contributions. Assuming that these are the only contributions to the charge radius and anapole form factors, these relations can be inverted to write the $\mu\rightarrow 3e$ branching ratio in terms of the induced couplings,
\begin{equation}
\begin{split}
&B(\mu\rightarrow 3e)=\left(\frac{v}{\Lambda}\right)^4\left(\frac{12\pi^2}{\ln\left(m_{\mu}^2/\Lambda^2\right)}\right)^2\\
&\times\left(|\tilde{f}_R^{\mu\rightarrow e}(-m_{\mu}^2)|^2+|\tilde{f}^{\mu\rightarrow e}_A(-m_{\mu}^2)|^2\right).
\end{split}
\end{equation}
As before, let us take $\tilde{f}_A=\tilde{f}_E=0$ and vary the couplings $\tilde{f}_R$ and $\tilde{f}_M$ subject to the constraint $|\tilde{f}_R|+|\tilde{f}_M|=1$. Although the four fermion $\mu\rightarrow 3e$ vertex in Eq. (\ref{eq:mu3e}) contributes only to the charge radius form factor, in the limit that $|\tilde{f}_M|>>|\tilde{f}_R|$ the $\mu\rightarrow e\gamma$ magnetic dipole dominates, and $\mu\rightarrow 3e$ then proceeds through a virtual photon decaying to an electron/positron pair. In this limit, the branching ratios satisfy \cite{Hisano}
\begin{equation}
B(\mu\rightarrow 3e)\approx \frac{\alpha}{3\pi}\left(\ln\frac{m_{\mu}^2}{m_e^2}-2\right)B(\mu\rightarrow e\gamma).
\end{equation}

For this scenario,  Fig. \ref{fig:muegamma} compares the constraints obtained from $\mu\rightarrow 3e$, $\mu\rightarrow e$, and $\mu\rightarrow e\gamma$.  If the charge radius is dominant, it is very unlikely for any signal to be detected at MEG II, but signals are possible in both $\mu\rightarrow e$ at Mu2e/COMET and $\mu \rightarrow 3e$ at Mu3e.  There is a modest region just above $\Lambda \approx 3000$ TeV where a signal could arise only in Mu3e.  If the charge-radius and magnetic dipole couplings are roughly equal, $|\tilde{f}_R/\tilde{f}_M| \approx 1$, then multiple scenarios are possible. For relatively low $\Lambda \lesssim 1000$ TeV, signals could be seen in both MEG II and Mu3e. Detection at Mu2e/COMET then depends on the relative sign of $\tilde{f}_R$ and $\tilde{f}_M$.   For $\tilde{f}_R \approx -\tilde{f}_M$ signals could be seen in MEG II and Mu3e, but not Mu2e/COMET.  (Note that Mu3e limits are based on single-event sensitivities: small changes in Fig. \ref{fig:muegamma} would occur if these are converted to 90\% C.L. limits \cite{Rule}.)

For a dominant magnetic dipole coupling, signals could be seen in all three processes, only $\mu\rightarrow e$ and $\mu\rightarrow 3e$, or only $\mu\rightarrow e$.  While in these examples we have assumed that the only contribution to the electromagnetic charge radius form factor is that induced by $\mu\rightarrow 3e$ at one loop, this scenario is sufficient to establish the complementarity of these different probes of CLFV.

\section{Summary and Discussion}
\label{sec:sec7}
The main goal of this paper has been the construction of a nuclear-level EFT for 
$\mu \rightarrow e$ conversion that exploits the largely nonrelativistic behavior of the bound nucleons and muon.
Because experiments are performed at low energy with composite nuclear targets, an NRET formulation provides the most
efficient phenomenology for their analysis: the operator basis is complete but not over-complete.  The
nucleon-level NRET is compatible with standard nonrelativistic treatments of the nuclear physics,
providing through the impulse approximation a convenient way to relate measurements made
in different targets.  We have discussed how this
approach is in fact more general, due to the core averaging -- the mean field physics -- that occurs
naturally in heavier nuclei, when multi-nucleon charge and current operators are introduced.  
The LECs being fit can thus be regarded as the effective LECs of a nuclear-level NRET.

We have emphasized the hierarchy of scales that arises in elastic $\mu \rightarrow e$ conversion,
$y > |\vec{v}_N| > |\vec{v}_\mu|$ for all targets of current interest.  This allows 
one to develop consistent nucleon-level NRETs of varying levels of sophistication.  We have formulated
such NRETs, embedding them in the nucleus using complete operator bases, retaining all electron
partial waves, thus properly generating both the full nuclear response functions $W(y)$ and all
effects linear in the Jacobi velocities.  Our formulation exploits an important
simplification: the Dirac Coulomb electron partial waves can be very accurately represented by 
a shifted and enhanced plane wave governed by a $q_\mathrm{eff}$ we determine
from the intra-nucleus Coulomb potential.

Calculations employing a full basis of operators, a full summation over electron partial
waves, a proper treatment of the muon's lower component, and the nucleon velocity operator have not
been presented previously.   We stress that even in the lowest level of the NRET -- the allowed approximation,
in which only the point-nucleus charge and spin operators are retained --
there are three independent response functions, with the transverse electric and longitudinal spin
operators probing largely distinct underlying CLFV operators.  The addition of the next most important
scale, $|\vec{v}_N|$, is new and proved exceedingly interesting.  This is the minimal theory that generates
all six nuclear response functions allowed by symmetry: the coefficients of these response functions 
thus provide the answer to an important question, what can and cannot be learned about CLFV
from elastic $\mu \rightarrow e$ conversion?   As noted below, this NRET also produces some surprises,
including the promotion of one velocity-dependent operator to a strength comparable to or even exceeding
that of the allowed spin operators.  Finally, we develop an NRET linear in both $\vec{v}_N$ and $\vec{v}_\mu$, thereby
including the muon's lower component.

Some previous efforts have treated  $\vec{v}_\mu$ but in a way that does not respect the
hierarchy of scales $y>|\vec{v}_N|>|\vec{v}_\mu|$.  Specifically, the inclusion of $\vec{v}_\mu$ has been
accompanied by a severe restriction on the electron partial waves, $|\kappa|=1$.   From our full
result it can be seen that operationally, the inclusion of $\vec{v}_\mu$ generates no new operator
physics but instead adds new components to response functions, which alter the net response by
an amount $\approx$ 5\%, for targets like $^{27}$Al.   But the truncation $|\kappa|=1$, limiting the
angular momentum transfer to the nucleus, eliminates higher multipoles, leading to errors that are typically $o(y^2)$.
In $^{27}$Al, $y \approx 0.27  \gg |\vec{v}_\mu| \approx 0.027$.

We showed that the embedding of a nucleon-level NRET in a nucleus has a significant
impact on operator physics, enhancing some interactions while suppressing others.  
Two sources of coherence, one of which was previously
unrecognized, boost sensitivity to certain LECs.  The new source of coherence arises from the strong spin-orbit interaction,
which in some nuclei leads to the occupation of the magnetic substates of just one
of the two spin-orbit partners.  Both
$^{27}$Al and Cu are cases where this occurs.  The resulting enhancement elevates a velocity-dependent
operator that nominally is suppressed by ${q_\mathrm{eff}/ m_N}$ to allowed strength, competitive with or even
stronger than the allowed spin operators.
We also find that four NRET interactions do not contribute to elastic $\mu \rightarrow e$ 
conversion due to $P$ and $CP$ selection rules, apart from tiny nuclear recoil corrections.  As these
selection rules would not operate for inelastic nuclear transitions, additional CLFV constraints
could be obtained by looking for electrons slightly below the kinematic maximum, where backgrounds
from free muon decay are still suppressed.  $^{27}$Al has low-lying ${1 \over 2}^+$ and ${3 \over 2}^+$ excited states
at 0.842 and 1.013 MeV, respectively.

We intend to generalize the present NRET analysis for inelastic $\mu \rightarrow e$ conversion \cite{RuleInelastic}.
The relaxation of the $P$ and $CP$ selection rules leads to five new operators as well as additional multipoles of the
operators we have introduced for elastic $\mu \rightarrow e$ conversion.

The formulation we developed allows one to factor the nuclear response function 
physics from the CLFV physics, which is isolated in leptonic response functions $R$ that are bilinear in the LECs.
This identifies precisely what CLFV physics can be extracted from elastic $\mu \rightarrow e$ conversion
and what cannot.  Once these constraints are obtained, they can be matched to any
higher-level theory, where more degrees of freedom will exist.   We provided an example,
the matching to a relativistic nucleon-level interaction, though one limited to relativistic
interactions mediated by scalar or vector exchanges. 
In fact, because the NRET formulation of elastic $\mu \rightarrow e$ conversion is so similar to that
developed a decade ago for WIMP direct detection, much of the work done to match UV DM
theories to NRET  \cite{Bishara,Brod,Cirelli,Barello,Hill} carries over to $\mu \rightarrow e$ conversion.

Were $\mu \rightarrow e$ conversion observed, a program of
measurements in properly selected targets could, in principle, isolate the various 
leptonic response functions $R$.  These coefficients would then serve as 
constraints on UV formulations of CLFV.  The accompanying nuclear response functions can be 
viewed as experimental ``knobs" that can be turned by selecting targets with specific properties.
These include the ground-state angular momentum,
the character of the valence nucleon in odd-$A$ targets (e.g., whether dominated by
spin or orbital angular momentum), and target isospin.  It would also include the spin-orbit structure
of the ground state.
We discussed the physics of these
knobs, describing the threshold behavior of the operators that govern the various nuclear response functions.   The level of success
possible in such a program will
depend on details: a favorable case would be a pattern of LECs generating comparable
responses in several channels,  which could be disentangled by selecting targets that isolate those channels.

While the full rate formula is given by Eq. (\ref{eq:amp2}), or alternatively Eq. (\ref{eq:amp3}), the forms presented in
the text replace the muon wave functions $g(r)$ and $f(r)$ by average values.   This is done to expose the
small parameters ${q_\mathrm{eff} \over m_N}$ and $\langle f \rangle/\langle g \rangle$ generated by the
velocity operators $\vec{v}_N$ and $\vec{v}_\mu$, respectively.   The averaging done by others (see Table \ref{tab:pastwork})
has followed that for muon capture, a more complex, inclusive process where averaging is typically done after performing an approximate
summation over excited states by closure.  This procedure also generates a $Z_\mathrm{eff}^4$ scaling of rates,
in contrast to the $Z_\mathrm{eff}^3$ Coulomb scaling of $\mu \rightarrow e$ conversion.  
In contrast, the averaging we do is with respect
to the dominant $\mu \rightarrow e$ conversion multipole, and thus exactly reproduces the result that would have been obtained by 
including $g(r)$ [or $f(r)$] in that multipole.  Errors only result from the use of the same average in other
multipoles.  We have quantified this error for both $\langle g \rangle$ and $\langle f  \rangle$, finding typical 
deviations in rates on the few percent level, and thus small compared to the nuclear structure uncertainties that arise in multipole matrix element evaluation.

We recommend using Eq. (\ref{eq:ratemn}), which neglects the muon's lower component and employs averaging, during the search and early discovery phase of CLFV.
That result is consistent with the general form of the rate we deduced from symmetry considerations, but has the
phase-space effects of the Coulomb interaction (encoded in $Z_\mathrm{eff}$ and $q_\mathrm{eff}$), the particle physics
(encoded in the LEC bilinears $R$), and the nuclear physics (encoded in the nuclear response functions $W$) all nicely
factored.  When the field has progressed to the point that the CLFV operator structure is known and the LECs 
are determined to better than 10\%, then small corrections due to $\vec{v}_\mu$ can be included.   Even then, as these corrections
effectively alter the nuclear form factors --  see Eqs. (\ref{eq:amp3}) and (\ref{eq:amprel}) -- their inclusion would have impact
only if the nuclear structure uncertainties affecting these same form factors have been reduced to below 10\%.

We examined the impact of future experiments that may achieve branching ratio sensitivities of $\lesssim 10^{-17}$, such as 
the $^{27}$Al experiments Mu2e and COMET.  While the choice of $^{27}$Al as a target is driven
largely by experimental considerations, it is also quite a nice choice for theoretical reasons.  As a single isotope
with a ground-state spin/parity of ${5 \over 2}^+$ and an unpaired valence proton, it responds to all of the NRET operators that can contribute to elastic $\mu \rightarrow e$ conversion,
including both isoscalar and isovector interactions.  We have noted its favorable spin-orbit structure that leads to
a new form of coherence.  Our results are presented as limits on the couplings $\tilde{c}_i$ and the associated energy scales
$\Lambda_i$.  The energy scales probed are operator dependent
and range from $\approx 10^2$ to  $\approx 10^4$ TeV.

These calculations were performed with the \textit{Mathematica} script (and its Python analog) described in Appendix \ref{AppendixC} and currently available.

Finally, for $\mu \rightarrow e$ conversion generated by a lepton-photon vertex,  we have examined  the relation of
this process to $\mu \rightarrow e \gamma$ and $\mu \rightarrow 3e$.   There is sufficient freedom in the four
available electromagnetic couplings to allow CLFV discovery in any of these channels, motivating efforts to
improve all of their sensitivities.

\section*{Acknowledgments} 
This work was supported in part by the U.S. Department of Energy under Grants No. DE-SC0004658, No. DE-SC0015376, and No. DE-AC02-05CH11231,
by the National Science Foundation under cooperative agreements 2020275 and 1630782, and by the Heising-Simons Foundation under Award No. 00F1C7.
\clearpage

\appendix

\section{Relationship to dark matter NRETs}
\label{AppendixA}
  The NRET for DM scattering off nuclei 
has a similar operator structure \cite{Liam1,Liam2}, and we have tried to retain here that earlier notation.  The thirteen 
unprimed operators of Eq. (\ref{eq:ops}) are identical to, or proportional to, those employed in dark 
matter studies.   Differences arise from the use of Pauli spin matrices for the leptons
(in the DM work the WIMP spin is arbitrary) and from
the fully relativistic electron velocity $\hat{q}$ (the scattered WIMPs are nonrelativistic).
We have reordered several of the operators to more clearly display the role of $i \hat{q}$ as a longitudinal or
transverse projector.

The three primed operators are new, connected with the treatment of the nuclear convection current
$\vec{v}_N$.  In the DM direct detection EFT work a modified current was introduced 
\begin{eqnarray}
 \vec{v}^\perp &\equiv& {1 \over 2} \left( {\vec{p}_i^{\, \chi} \over m_\chi}+ {\vec{p}_f^{\, \chi} \over m_\chi}-{\vec{p}_i \over m_N}-{\vec{p}_f \over m_N} \right) \nonumber \\
~& = &{1 \over 2} \left( {\vec{p}_i^{\, \chi} \over m_\chi}+ {\vec{p}_f^{\, \chi} \over m_\chi} \right)-\vec{v}_N 
\end{eqnarray}
where $\vec{p}_i^{\, \chi}$ and $\vec{p}_f^{\, \chi}$ are the incoming and outgoing WIMP velocities and $m_\chi$ the WIMP mass.
Using energy conservation for free-particle
scattering, one finds $\hat{q} \cdot \vec{v}^\perp = 0$, which allows one to eliminate one scalar from the operator 
construction.  This step depends on the relation
\be
 \vec{q} \cdot \vec{v}_N ={\vec{p}_i^{\, 2} \over 2 m_N} - {\vec{p}_f^{\, 2} \over 2m_N} 
 \label{eq:eqqdot}
 \ee
However, the operators in Eq. (\ref{eq:ops}) will be embedded 
in a nucleus where they act on all nucleons, and where these nucleons are bound in a strong potential.
The relation above has an analog in the nuclear case, where the relevant velocities are those of the center of mass
of the nucleus.  The center-of-mass motion is of no interest here as it is integrated in the
phase space computation.  As discussed in more detail in \cite{Liam2}, the interesting velocities are the $A-1$ intrinsic 
velocities that characterize the relative motions of the bound nucleons.  These velocities are intrinsic nuclear operators unconstrained by external kinematics.

Consequently, in the present work we retain the longitudinal projection of the nuclear convection
current $i \hat{q} \cdot \vec{v}_N$, generating a slightly more general nucleon-level effective theory that does not 
assume free nucleons.   Two new operators, $\CO^\prime_2$ and $\CO^\prime_{16}$
then arise, increasing
the number of operators from the 14 of DM elastic scattering to the 16 of $\mu \rightarrow e$ conversion.
(Initially $\CO_2 \equiv (v^{\perp})^2$ \cite{Liam1} was included in the DM EFT expansion, but it was
later eliminated \cite{Liam2}, being of order $1/m_N^2$.)  In addition, one operator introduced 
in DM studies is modified. The $\CO^\prime_{13}$ we have defined can be
rewritten
\be
\CO^\prime_{13} = \vec{\sigma}_L \cdot \vec{v}_N~i \hat{q} \cdot \vec{\sigma}_N - \vec{\sigma}_L \cdot \vec{\sigma}_N ~i \hat{q} \cdot \vec{v}_N
\ee
The first term on the right side corresponds to the DM EFT operator $\CO_{13}$; the addition of 
the second term, which depends on $i \hat{q} \cdot \vec{v}_N$, ensures that our $\mu \rightarrow e$ operator expansion is 
not contaminated by $[\vec{v}_N \otimes \vec{\sigma}_N]_2$.

The operators $\CO^\prime_2$ and $\CO^\prime_{16}$  do not contribute to $\mu \rightarrow e$ 
conversion if the nucleus remains in its ground state.  In this case, selection rules imposed by the assumed
good parity and $CP$ properties of the nuclear ground state 
restrict $\vec{v}_N$ to transverse magnetic nuclear multipoles, while $i \hat{q} \cdot \vec{v}_N$ is
longitudinal.   However, no similar restriction exists for inelastic scattering.  The primed operators
generate a nucleon-level interaction that can be used in either elastic or inelastic reactions.

Another difference in the present construction is that we have treated the LECs of the NRET as complex, though
all operators we constructed are Hermitian.  In the DM work the LECs are real.  This difference arises because, in $\mu \rightarrow e$ 
conversion, the theory is reduced to a nonrelativistic  form by integrating out the relativistic electron's velocity, leaving only bound-state
nonrelativistic degrees of freedom.  In the spinor of Eq. (\ref{eq:Delectron}), the electron velocity operator has been replaced by the three-momentum direction $\hat{q}$.  While the electron's velocity and $\hat{q}$ are identical, the underlying operators, the leptonic velocity
and three-momentum transfer $i \vec{q}$, are distinct
\begin{eqnarray}
 \vec{v}_L=-{1 \over i} \cev{\nabla}_r \delta(\vec{r}-\vec{r_i})+ \delta(\vec{r}-\vec{r}_i) {1 \over i} \vec{\nabla}_r ~~~~~~~~~~ \nonumber \\
 i \vec{q}= \left[ \vec{\nabla}_r,\, \delta(\vec{r}-\vec{r}_i) \right] =  - \cev{\nabla}_r \delta(\vec{r}-\vec{r_i})- \delta(\vec{r}-\vec{r}_i) \vec{\nabla}_r \nonumber
 \end{eqnarray}
Thus $i \hat{q}$ represents two operators, accounting for the ambiguity in whether the coefficient is real or pure imaginary.
Further, the component of the lepton velocity operator we retain and treat explicitly, the nonrelativistic operator $\vec{v}_\mu$, acts only on the initial state.

\section{Decay Rate Calculation}
\label{AppendixB}
The amplitude of Eq. (\ref{eq:HamAA}) can be squared to yield the transition probability, then averaged over 
nuclear spins and summed over final spins to obtain a probability.  We find
\begin{widetext}
\allowdisplaybreaks
\begin{eqnarray}
&& {1 \over 2j_N+1} \sum_{m_f  m_i } | \langle \tfrac{1}{2} {s_f}; j_N m_f |\CM | \tfrac{1}{2} {s_i}; j_N m_i \rangle |^2 = {{E_e \over 2 m_e}}~ {q_\mathrm{eff}
^2 \over q^2}~ {4 \pi \over 2j_N + 1} 
\sum_{ \tau=0,1} \sum_{\tau^\prime = 0,1} \nonumber \\
 &&\left\{ \sum_{J=0,2,...}^\infty  ~ \Bigg(  ~\langle l_0^\tau \rangle \langle l_0^{\tau^\prime } \rangle^* \langle j_N||~ M^g_{J;\tau} (q_\mathrm{eff})~ || j_N \rangle 
\langle j_N ||~ M^g_{J;\tau^\prime} (q_\mathrm{eff})~ || j_N \rangle \right.  \nonumber \\
&&~~~+2 ~\mathrm{Re} \left[ i \langle  l_0^\tau \rangle \langle l_0^{\tau^\prime (2) } \rangle^* \right] \langle j_N||~ M^g_{J;\tau} (q_\mathrm{eff})~ || j_N \rangle 
\langle j_N ||~ M^{(2)f} _{J;\tau^\prime} (q_\mathrm{eff})~ || j_N \rangle \nonumber \\
&&~~~+~\langle l_0^{\tau (2)}  \rangle \langle l_0^{\tau^\prime (2) } \rangle^* \langle j_N||~ M^{(2)f}_{J;\tau} (q_\mathrm{eff})~ || j_N \rangle 
\langle j_N ||~ M^{(2) f}_{J;\tau^\prime} (q_\mathrm{eff})~ || j_N \rangle \nonumber \\
&&~~~+ {\vec{q}_\mathrm{eff} \over m_N} \cdot \langle \vec{l}_E^\tau \rangle ~ {\vec{q}_\mathrm{eff} \over m_N} \cdot \langle \vec{l}_E^{\tau^\prime} \rangle^*~\langle j_N ||~\Phi^{\prime \prime g}_{J; \tau}(q_\mathrm{eff})~ || j_N \rangle  \langle j_N ||~\Phi^{\prime \prime g}_{J; \tau^\prime}(q_\mathrm{eff})~ || j_N \rangle \nonumber \\
&&~~~+   {2 \vec{q}_\mathrm{eff} \over m_N} \cdot \mathrm{Re} \left[ \langle  \vec{l}_E^\tau \rangle~\langle  l_0^{\tau^\prime} \rangle^* \right]~ \langle j_N ||~ \Phi^{\prime \prime g }_{J;\tau}(q_\mathrm{eff})~ || j_N \rangle \langle j_N ||~ M^g_{J;\tau^\prime} (q_\mathrm{eff}) ~|| j_N \rangle  \nonumber \\
&&~~~+{2 \vec{q}_\mathrm{eff} \over m_N} \cdot \mathrm{Re} \left[ i \langle  \vec{l}_E^\tau \rangle~\langle  l_0^{\tau^\prime (2)} \rangle^* \right]~ \langle j_N ||~ \Phi^{\prime \prime g }_{J;\tau}(q_\mathrm{eff})~ || j_N \rangle \langle j_N ||~ M^{(2) f}_{J;\tau^\prime} (q_\mathrm{eff}) ~|| j_N \rangle  \Bigg) ~~ \nonumber \\
 && +   \sum_{J=1,3,...}^\infty  \Bigg({1 \over 2}  \left( \langle \vec{l}_5^\tau \rangle \cdot \langle \vec{l}_5^{\tau^\prime}\rangle^* - \hat{q} \cdot \langle \vec{l}_5^\tau \rangle ~\hat{q} \cdot \langle \vec{l}_5^{\tau^\prime }\rangle^* \right)  \langle j_N ||~  \Sigma_{J;\tau}^{\prime g} (q_\mathrm{eff})~ || j_N \rangle  \langle j_N || ~\Sigma_{J;\tau^\prime}^{\prime g} (q_\mathrm{eff})~ || j_N \rangle \nonumber \\ 
 &&~~~+{1 \over 2}  \left( \langle \vec{l}_5^{\tau (0)} \rangle \cdot \langle \vec{l}_5^{\tau^\prime (0)}\rangle^* - \hat{q} \cdot \langle \vec{l}_5^{\tau (0)} \rangle ~\hat{q} \cdot \langle \vec{l}_5^{\tau^\prime (0) }\rangle^* \right)  \langle j_N ||~  \Sigma_{J;\tau}^{\prime (0) f} (q_\mathrm{eff})~ || j_N \rangle  \langle j_N || ~\Sigma_{J;\tau^\prime}^{\prime (0) f} (q_\mathrm{eff})~ || j_N \rangle \nonumber \\ 
 &&~~~+{1 \over 2}  \left( \langle \vec{l}_5^{\tau (2)} \rangle \cdot \langle \vec{l}_5^{\tau^\prime (2)}\rangle^* - \hat{q} \cdot \langle \vec{l}_5^{\tau (2)} \rangle ~\hat{q} \cdot \langle \vec{l}_5^{\tau^\prime (2) }\rangle^* \right)  \langle j_N ||~  \Sigma_{J;\tau}^{\prime (2) f} (q_\mathrm{eff})~ || j_N \rangle  \langle j_N || ~\Sigma_{J;\tau^\prime}^{\prime (2) f} (q_\mathrm{eff})~ || j_N \rangle \nonumber \\ 
 &&~~~ +\mathrm{Re} \left[ i \langle \vec{l}_5^{\tau (0)} \rangle \cdot \langle \vec{l}_5^{\tau^\prime}\rangle^* - i \hat{q} \cdot \langle \vec{l}_5^{\tau (0)} \rangle ~\hat{q} \cdot \langle \vec{l}_5^{\tau^\prime  }\rangle^* \right]  \langle j_N ||~  \Sigma_{J;\tau}^{\prime (0) f} (q_\mathrm{eff})~ || j_N \rangle  \langle j_N || ~\Sigma_{J;\tau^\prime}^{\prime  g} (q_\mathrm{eff})~ || j_N \rangle \nonumber \\ 
  &&~~~ +\mathrm{Re} \left[ i \langle \vec{l}_5^{\tau (2)} \rangle \cdot \langle \vec{l}_5^{\tau^\prime}\rangle^* - i \hat{q} \cdot \langle \vec{l}_5^{\tau (2)} \rangle ~\hat{q} \cdot \langle \vec{l}_5^{\tau^\prime  }\rangle^* \right]  \langle j_N ||~  \Sigma_{J;\tau}^{\prime (2) f} (q_\mathrm{eff})~ || j_N \rangle  \langle j_N || ~\Sigma_{J;\tau^\prime}^{\prime  g} (q_\mathrm{eff})~ || j_N \rangle \nonumber \\ 
 &&~~~ +\mathrm{Re} \left[  \langle \vec{l}_5^{\tau (0)} \rangle \cdot \langle \vec{l}_5^{\tau^\prime (2) }\rangle^* -  \hat{q} \cdot \langle \vec{l}_5^{\tau (0)} \rangle ~\hat{q} \cdot \langle \vec{l}_5^{\tau^\prime (2) }\rangle^* \right]  \langle j_N ||~  \Sigma_{J;\tau}^{\prime (0) f} (q_\mathrm{eff})~ || j_N \rangle  \langle j_N || ~\Sigma_{J;\tau^\prime }^{\prime  (2)  f} (q_\mathrm{eff})~ || j_N \rangle \nonumber \\ 
&&~~~+ {1 \over 2} \left( {q^2_\mathrm{eff} \over m_N^2} \langle \vec{l}_M^\tau \rangle \cdot
\langle \vec{l}_M^{\tau^\prime}\rangle^* - {\vec{q}_\mathrm{eff} \over m_N} \cdot \langle \vec{l}_M^\tau \rangle ~{\vec{q}_\mathrm{eff} \over m_N} \cdot \langle \vec{l}_M^{\tau^\prime }\rangle^* \right)  \langle j_N ||~ \Delta^g_{J;\tau} (q_\mathrm{eff})~ || j_N \rangle  \langle j_N ||~ \Delta^g_{J; \tau^\prime}(q_\mathrm{eff})~ || j_N \rangle \nonumber \\
&&~~~  +{\vec{q}_\mathrm{eff} \over m_N} \cdot  \mathrm{Re} \left[ i \langle \vec{l}_M^\tau \rangle \times
\langle \vec{l}_5^{\tau^\prime } \rangle^* \right] \langle j_N || ~ \Delta^g_{J;\tau} (q_\mathrm{eff}) ~|| j_N \rangle  \langle j_N || ~\Sigma^{\prime g}_{J; \tau^\prime}(q_\mathrm{eff})~ || j_N  \rangle \nonumber \\
&&~~~  +{\vec{q}_\mathrm{eff} \over m_N} \cdot  \mathrm{Re} \left[  \langle \vec{l}_M^\tau \rangle \times
\langle \vec{l}_5^{\tau^\prime (0) } \rangle^* \right] \langle j_N || ~ \Delta^g_{J;\tau} (q_\mathrm{eff}) ~|| j_N \rangle  \langle j_N || ~\Sigma^{\prime (0) f}_{J; \tau^\prime}(q_\mathrm{eff})~ || j_N  \rangle \nonumber \\
&&~~~ +{\vec{q}_\mathrm{eff} \over m_N} \cdot  \mathrm{Re} \left[ \langle \vec{l}_M^\tau \rangle \times
\langle \vec{l}_5^{\tau^\prime (2) } \rangle^* \right] \langle j_N || ~ \Delta^g_{J;\tau} (q_\mathrm{eff}) ~|| j_N \rangle  \langle j_N || ~\Sigma^{\prime (2) f}_{J; \tau^\prime}(q_\mathrm{eff})~ || j_N  \rangle   \Bigg) \nonumber \\
&& +   \sum_{J=2,4,...}^\infty  \Bigg(  ~ {1 \over 2} \left( {q^2_\mathrm{eff} \over m_N^2}\langle \vec{l}_E^\tau \rangle \cdot
\langle \vec{l}_E^{\tau^\prime}\rangle^* - {\vec{q}_\mathrm{eff} \over m_N} \cdot \langle \vec{l}_E^\tau \rangle~ {\vec{q}_\mathrm{eff} \over m_N} \cdot \langle \vec{l}_E^{\tau^\prime }\rangle^* \right)    \langle j_N || ~\tilde{\Phi}_{J; \tau}^{\prime g}(q_\mathrm{eff})~ || j_N \rangle   \langle j_N ||  ~ \tilde{\Phi}^{\prime g}_{J; \tau^\prime}(q_\mathrm{eff})~ || j_N \rangle \nonumber \\
&&~~~+~ {1 \over 2} \left( \langle \vec{l}^{\tau (1)}  \rangle \cdot
\langle \vec{l}^{\tau^\prime (1) } \rangle^* - \hat{q} \cdot \langle \vec{l}^{\tau (1)} \rangle~ \hat{q} \cdot \langle \vec{l}^{\tau^\prime (1) } \rangle^* \right)    \langle j_N || ~M_{J; \tau}^{(1) f} (q_\mathrm{eff})~ || j_N \rangle   \langle j_N ||  ~ M^{(1) f}_{J; \tau^\prime}(q_\mathrm{eff})~ || j_N \rangle \nonumber \\
&&~~~-  \mathrm{Re} \left[ {q_\mathrm{eff} \over m_N} i \langle \vec{l}_E^\tau \rangle \cdot
\langle \vec{l}^{\tau^\prime (1)}\rangle^* - {\vec{q}_\mathrm{eff} \over m_N} \cdot i  \langle \vec{l}_E^\tau \rangle~ \hat{q} \cdot \langle \vec{l}^{\tau^\prime (1) }\rangle^* \right]    \langle j_N || ~\tilde{\Phi}_{J; \tau}^{\prime g}(q_\mathrm{eff})~ || j_N \rangle   \langle j_N ||  ~ M^{(1) f}_{J; \tau^\prime}(q_\mathrm{eff})~ || j_N \rangle \Bigg)  \nonumber \\
&& +   \sum_{J=1,3,...}^\infty  \Bigg( \hat{q} \cdot \langle \vec{l}_5^\tau \rangle~ \hat{q} \cdot \langle \vec{l}_5^{\tau^\prime }\rangle^*  ~
\langle j_N ||~ \Sigma^{\prime \prime g}_{J; \tau}(q_\mathrm{eff})~ || j_N \rangle
 \langle j_N ||~ \Sigma^{\prime \prime g}_{J; \tau^\prime}(q_\mathrm{eff})~ || j_N \rangle  \nonumber \\
 &&~~~+\hat{q} \cdot \langle \vec{l}_5^{\tau (0)} \rangle~ \hat{q} \cdot \langle \vec{l}_5^{\tau^\prime  (0) }\rangle^*  ~
\langle j_N ||~ \Sigma^{\prime \prime (0) f }_{J; \tau}(q_\mathrm{eff})~ || j_N \rangle
 \langle j_N ||~ \Sigma^{\prime \prime (0) f}_{J; \tau^\prime}(q_\mathrm{eff})~ || j_N \rangle  \nonumber \\
 &&~~~+\hat{q} \cdot \langle \vec{l}_5^{\tau (2)} \rangle~ \hat{q} \cdot \langle \vec{l}_5^{\tau^\prime (2) }\rangle^*  ~
\langle j_N ||~ \Sigma^{\prime \prime (2) f}_{J; \tau}(q_\mathrm{eff})~ || j_N \rangle
 \langle j_N ||~ \Sigma^{\prime \prime (2) f}_{J; \tau^\prime}(q_\mathrm{eff})~ || j_N \rangle  \nonumber \\
  &&~~~+2 \mathrm{Re} \left[ \hat{q} \cdot \langle \vec{l}_5^{\tau (0)} \rangle~ \hat{q} \cdot \langle \vec{l}_5^{\tau^\prime (2) }\rangle^* \right] ~
\langle j_N ||~ \Sigma^{\prime \prime (0) f}_{J; \tau}(q_\mathrm{eff})~ || j_N \rangle
 \langle j_N ||~ \Sigma^{\prime \prime (2) f}_{J; \tau^\prime}(q_\mathrm{eff})~ || j_N \rangle  \nonumber \\
   &&~~~-2 \mathrm{Re} \left[ \hat{q} \cdot i \langle \vec{l}_5^{\tau (0)} \rangle~ \hat{q} \cdot \langle \vec{l}_5^{\tau^\prime }\rangle^* \right] ~
\langle j_N ||~ \Sigma^{\prime \prime (0) f}_{J; \tau}(q_\mathrm{eff})~ || j_N \rangle
 \langle j_N ||~ \Sigma^{\prime \prime g}_{J; \tau^\prime}(q_\mathrm{eff})~ || j_N \rangle  \nonumber \\
   &&~~~-2 \mathrm{Re} \left[ \hat{q} \cdot  i\langle \vec{l}_5^{\tau (2)} \rangle~ \hat{q} \cdot \langle \vec{l}_5^{\tau^\prime  }\rangle^* \right] ~
\langle j_N ||~ \Sigma^{\prime \prime (2) f}_{J; \tau}(q_\mathrm{eff})~ || j_N \rangle
 \langle j_N ||~ \Sigma^{\prime \prime g}_{J; \tau^\prime}(q_\mathrm{eff})~ || j_N \rangle  \Bigg)  \Bigg\}.
 \label{eq:BT}
\end{eqnarray}
\end{widetext}
We have used the shorthand for the leptonic matrix elements $\langle l  \rangle \equiv \langle \tfrac{1}{2} {s_f}  | l | \tfrac{1}{2} {s_i} \rangle$.
With the angular conventions used here \cite{Edmonds}, all nuclear matrix elements are real; we have assumed this above.   

 This expression contains several new lower-component leptonic operators that are generated from  $l_{0 f}^\tau(\hat{x}_i)$ and $\vec{l}_{5 f}^\tau(\hat{x}_i)$ of 
 Eq. (\ref{eq:newlep}), once the effect of the angular momentum ladder operator $\hat{x}$ on nuclear multipoles is evaluated:
 \begin{eqnarray}
 {l}_\lambda^{ \tau (1)} &=& -i( b_3^\tau \lambda + b_7^\tau ) {\sigma}_{L\lambda}, \nonumber \\
 l_0^{\tau (2)} &=& - b_2^\tau +i  b_7^\tau \, \sigma_{L 0}, \nonumber \\
 {l}^{\tau (0)}_{5 \lambda} &=&\Big[\lambda ( -b_{12}^\tau+b_{15}^\tau )+b_{14}^\tau \Big]   \sigma_{L \lambda}, \nonumber \\
 {l}^{\tau (2)}_{5 \lambda} &=&  \Big[ -b_{13}^\tau +b_{14}^\tau +\lambda b_{15}^\tau  \Big]   \sigma_{L \lambda},  \nonumber  \\
  {l}^{\tau (0)}_{5 \, 0} &=&-i b_8^\tau +b_{13} \sigma_{L0},  \nonumber \\
  {l}^{\tau (2)}_{5 \, 0} &=& (b_{13}^\tau -b_{14}^\tau)  \sigma_{L 0} -i b_{16}^\tau.
\end{eqnarray}
We then do the leptonic spin sums, averaging over the muon's spin and summing
over the spin states of the outgoing electron.  We note that this step should be examined for each 
specific nucleus, as the hyperfine splittings of the muon spin states due to interactions with the
nuclear magnetic or quadrupole moments may be sufficient to alter the spin sum, for nuclear targets with $j_N \ne 0$.
Evaluating the needed 25 traces yields
\begin{widetext}
\allowdisplaybreaks
\begin{eqnarray}
&&~~~~~~~~~~~~~~~~~~~~ \omega = {G_F^2 \over \pi} ~ {q_\mathrm{eff}^2 \over 1+{q \over M_T}} ~
\sum_{ \tau=0,1} \sum_{\tau^\prime = 0,1} \nonumber \\
 &&\left\{  ~ \Big[  ~\tilde{R}_{M M}^{\tau \tau^\prime}~ W_{M^g M^g}^{\tau \tau^\prime}(q_\mathrm{eff}) +2 \tilde{R}_{M M^{(2) } }^{\tau \tau^\prime}~ W_{M^g M^{(2)f} }^{\tau \tau^\prime}(q_\mathrm{eff}) + \tilde{R}_{M^{(2) } M^{(2) } }^{\tau \tau^\prime}~ W_{M^{(2)f}  M^{(2)f} }^{\tau \tau^\prime}(q_\mathrm{eff})    \right.  \nonumber \\
&&~~~+{q_\mathrm{eff}^2 \over m_N^2} \tilde{R}_{\Phi^{\prime \prime } \Phi^{\prime \prime }}^{\tau \tau^\prime}~ W_{\Phi^{\prime \prime g} \Phi^{\prime \prime g}}^{\tau \tau^\prime}(q_\mathrm{eff}) -   {2 {q}_\mathrm{eff} \over m_N} \Big( \tilde{R}_{\Phi^{\prime \prime } M}^{\tau \tau^\prime}~ W_{\Phi^{\prime \prime g} M^{g}}^{\tau \tau^\prime}(q_\mathrm{eff})  + \tilde{R}_{\Phi^{\prime \prime } M^{(2) }}^{\tau \tau^\prime}~ W_{\Phi^{\prime \prime g} M^{(2) f}}^{\tau \tau^\prime}(q_\mathrm{eff}) \Big) \Big] ~~ \nonumber \\
 && +   \Big[ ~\tilde{R}_{\Sigma^{\prime } \Sigma^{\prime }}^{\tau \tau^\prime}~ W_{\Sigma^{\prime g} \Sigma^{\prime g}}^{\tau \tau^\prime}(q_\mathrm{eff}) + \tilde{R}_{\Sigma^{\prime (0) } \Sigma^{\prime (0) }}^{\tau \tau^\prime}~ W_{\Sigma^{\prime (0) f} \Sigma^{\prime (0) f}}^{\tau \tau^\prime}(q_\mathrm{eff})   + \tilde{R}_{\Sigma^{\prime (2) } \Sigma^{\prime (2) }}^{\tau \tau^\prime}~ W_{\Sigma^{\prime (2) f} \Sigma^{\prime (2) f}}^{\tau \tau^\prime}(q_\mathrm{eff})     \nonumber \\ 
 &&~~~-2 \Big(  \tilde{R}_{\Sigma^{\prime (0)  } \Sigma^{\prime }}^{\tau \tau^\prime}~ W_{\Sigma^{\prime (0) f} \Sigma^{\prime g}}^{\tau \tau^\prime}(q_\mathrm{eff}) + \tilde{R}_{\Sigma^{\prime (2) } \Sigma^{\prime }}^{\tau \tau^\prime}~ W_{\Sigma^{\prime (2) f} \Sigma^{\prime g}}^{\tau \tau^\prime}(q_\mathrm{eff})   +  \tilde{R}_{\Sigma^{\prime (0) } \Sigma^{\prime (2) }}^{\tau \tau^\prime}~ W_{\Sigma^{\prime (0) f} \Sigma^{\prime (2) f}}^{\tau \tau^\prime}(q_\mathrm{eff}) \Big)    \nonumber \\ 
 &&~~~+{q_\mathrm{eff}^2 \over m_N^2} \tilde{R}_{\Delta \Delta}^{\tau \tau^\prime}~ W_{\Delta^{g} \Delta^{g}}^{\tau \tau^\prime}(q_\mathrm{eff}) -{2 q_\mathrm{eff} \over m_N} \Big( \tilde{R}_{\Delta \Sigma^{\prime }}^{\tau \tau^\prime}~ W_{\Delta^{g} \Sigma^{\prime g}}^{\tau \tau^\prime}(q_\mathrm{eff})+\tilde{R}_{\Delta \Sigma^{\prime (0) }}^{\tau \tau^\prime}~ W_{\Delta^{g} \Sigma^{\prime (0) f}}^{\tau \tau^\prime}(q_\mathrm{eff}) +\tilde{R}_{\Delta \Sigma^{\prime (2) }}^{\tau \tau^\prime}~ W_{\Delta^{g} \Sigma^{\prime (2) f}}^{\tau \tau^\prime}(q_\mathrm{eff}) \Big)  \Big] \nonumber \\
&& +  \Big[ {q_\mathrm{eff}^2 \over m_N^2}\tilde{R}_{\tilde{\Phi}^{\prime } \tilde{\Phi}^{\prime }}^{\tau \tau^\prime}~ W_{\tilde{\Phi}^{\prime g} \tilde{\Phi}^{\prime g}}^{\tau \tau^\prime}(q_\mathrm{eff})+  \tilde{R}_{M^{(1) } M^{(1) }}^{\tau \tau^\prime}~ W_{M^{(1) f} M^{(1) f}}^{\tau \tau^\prime}(q_\mathrm{eff})-{2q_\mathrm{eff} \over m_N} \tilde{R}_{\tilde{\Phi}^{\prime } M^{(1) }}^{\tau \tau^\prime}~ W_{\tilde{\Phi}^{\prime g} M^{(1) f}}^{\tau \tau^\prime}(q_\mathrm{eff}) \Big] \nonumber \\
&& +   \Big[ \tilde{R}_{\Sigma^{\prime \prime } \Sigma^{\prime \prime }}^{\tau \tau^\prime} W_{\Sigma^{\prime \prime g} \Sigma^{\prime \prime g}}^{\tau \tau^\prime}(q_\mathrm{eff})+ \tilde{R}_{\Sigma^{\prime \prime (0) } \Sigma^{\prime \prime (0) }}^{\tau \tau^\prime} W_{\Sigma^{\prime \prime (0) f} \Sigma^{\prime \prime (0) f}}^{\tau \tau^\prime}(q_\mathrm{eff})+  \tilde{R}_{\Sigma^{\prime \prime (2) } \Sigma^{\prime \prime (2) }}^{\tau \tau^\prime} W_{\Sigma^{\prime \prime (2) f} \Sigma^{\prime \prime (2) f}}^{\tau \tau^\prime}(q_\mathrm{eff}) \nonumber \\
&&\left.~~~+2 \Big( \tilde{R}_{\Sigma^{\prime \prime (0) } \Sigma^{\prime \prime (2) }}^{\tau \tau^\prime} W_{\Sigma^{\prime \prime (0) f} \Sigma^{\prime \prime (2) f}}^{\tau \tau^\prime}(q_\mathrm{eff})+ \tilde{R}_{\Sigma^{\prime \prime (0) } \Sigma^{\prime \prime }}^{\tau \tau^\prime} W_{\Sigma^{\prime \prime (0) f} \Sigma^{\prime \prime g}}^{\tau \tau^\prime}(q_\mathrm{eff})+  \tilde{R}_{\Sigma^{\prime \prime (2) } \Sigma^{\prime \prime }}^{\tau \tau^\prime} W_{\Sigma^{\prime \prime (2) f} \Sigma^{\prime \prime g}}^{\tau \tau^\prime}(q_\mathrm{eff}) \Big) \Big] ~  \right\}.
\label{eq:amp2}
\end{eqnarray}
\end{widetext}
In the phase space integration we include the tiny nonrelativistic correction for nuclear recoil, which depends on the mass of the daughter nucleus $M_T$.

Equation (\ref{eq:amp2}) separates the nuclear physics from the particle physics of the LECs, which is
isolated in the various leptonic coefficients $\tilde{R}$. The dimensionful
quantities $R$ are obtained by doing the traces over the 25 leptonic scalars appearing in Eq. (\ref{eq:BT}):
\allowdisplaybreaks
\begin{eqnarray}
\label{eq:Rs}
     \tilde{R}_{M M}^{\tau \tau^\prime} &=& \tilde{c}_1^\tau \tilde{c}_1^{\tau^{\prime} * } + \tilde{c}_{11}^\tau \tilde{c}_{11}^{\tau^\prime * },  \nonumber \\
 \tilde{R}_{M M^{(2) }}^{\tau \tau^\prime} &=& \mathrm{Im}[ \tilde{c}_1^\tau \tilde{b}_2^{\tau^{\prime} * } - \tilde{c}_{11}^\tau \tilde{b}_{7}^{\tau^\prime * } ], \nonumber \\
  \tilde{R}_{M^{(2) } M^{(2) } }^{\tau \tau^\prime} &=& \tilde{b}_2^\tau \tilde{b}_2^{\tau^{\prime} * } + \tilde{b}_{7}^\tau \tilde{b}_{7}^{\tau^\prime * },  \nonumber \\
 \tilde{R}_{\Phi^{\prime \prime }  \Phi^{\prime \prime }  }^{\tau \tau^\prime}&=& \tilde{c}_3^\tau \tilde{c}_3^{\tau^\prime * } + ( \tilde{c}_{12}^\tau- \tilde{c}_{15}^\tau ) ( \tilde{c}_{12}^{\tau^\prime *}-\tilde{c}_{15}^{\tau^\prime *} ),  \nonumber \\
     \tilde{R}_{\Phi^{\prime \prime } M}^{\tau \tau^\prime} &=& \mathrm{Re} [ \tilde{c}_3^\tau \tilde{c}_1^{\tau^\prime *} -  \left(\tilde{c}_{12}^\tau - \tilde{c}_{15}^\tau \right) \tilde{c}_{11}^{\tau^\prime *} ] \nonumber, \\
    \tilde{R}_{\Sigma^{\prime } \Sigma^{\prime }}^{\tau \tau^\prime} &=&\tilde{c}_4^\tau\tilde{c}_4^{\tau^{\prime} * } + \tilde{c}_{9}^\tau \tilde{c}_{9}^{\tau^\prime * }, \nonumber \\  
      \tilde{R}_{\Sigma^{\prime (0) } \Sigma^{\prime (0) }}^{\tau \tau^\prime} &=&(\tilde{b}^\tau_{12}-\tilde{b}_{15}^\tau) (\tilde{b}_{12}^{\tau^\prime *} -\tilde{b}_{15}^{\tau^\prime *}) + \tilde{b}_{14}^\tau \tilde{b}_{14}^{\tau^\prime * }, \nonumber \\
      \tilde{R}_{\Sigma^{\prime (2) } \Sigma^{\prime (2) }}^{\tau \tau^\prime} &=&(\tilde{b}^\tau_{13}-\tilde{b}_{14}^\tau) (\tilde{b}_{13}^{\tau^\prime *} -\tilde{b}_{14}^{\tau^\prime *}) + \tilde{b}_{15}^\tau \tilde{b}_{15}^{\tau^\prime * }, \nonumber \\
     \tilde{R}_{\Sigma^{\prime (0) } \Sigma^{\prime }}^{\tau \tau^\prime} &=&\mathrm{Im}[\tilde{b}_{14}^\tau \tilde{c}_4^{\tau^{\prime} * } - (\tilde{b}^\tau_{12}-\tilde{b}^\tau_{15}) \tilde{c}_{9}^{\tau^\prime * }], \nonumber \\   
      \tilde{R}_{\Sigma^{\prime (2) } \Sigma^{\prime }}^{\tau \tau^\prime} &=&\mathrm{Im}[-(\tilde{b}_{13}^\tau-\tilde{b}_{14}^\tau) \tilde{c}_4^{\tau^{\prime} * } + \tilde{b}_{15}^\tau \tilde{c}_{9}^{\tau^\prime * } ],\nonumber \\    
      \tilde{R}_{\Sigma^{\prime (0) } \Sigma^{\prime (2) }}^{\tau \tau^\prime} &=&\mathrm{Re}[ \tilde{b}_{14}^\tau ( \tilde{b}_{13}^{\tau^\prime * }-\tilde{b}_{14}^{\tau^\prime *}) +(\tilde{b}_{12}^\tau-\tilde{b}_{15}^\tau) \tilde{b}_{15}^{\tau^\prime * }], \nonumber \\       
      \tilde{R}_{\Delta \Delta}^{\tau \tau^\prime} &=&\tilde{c}^\tau_{5}  \tilde{c}_{5}^{\tau^\prime *} + \tilde{c}_{8}^\tau \tilde{c}_{8}^{\tau^\prime * }, \nonumber \\
      \tilde{R}_{\Delta \Sigma^{ \prime }}^{\tau \tau^\prime} &=&\mathrm{Re}[\tilde{c}^\tau_{5}  \tilde{c}_{4}^{\tau^\prime *} + \tilde{c}_{8}^\tau \tilde{c}_{9}^{\tau^\prime * }], \nonumber \\          
         \tilde{R}_{\Delta \Sigma^{ \prime (0) }}^{\tau \tau^\prime} &=&\mathrm{Im}[ \tilde{c}_{5}^{\tau }\tilde{b}_{14}^{\tau^\prime *} -\tilde{c}_{8}^{\tau}  (\tilde{b}^{\tau^\prime *}_{12}-\tilde{b}^{\tau^\prime *}_{15})   ], \nonumber \\                       
               \tilde{R}_{\Delta \Sigma^{ \prime (2) }}^{\tau \tau^\prime} &=&\mathrm{Im}[\tilde{c}_{8}^{\tau} \tilde{b}_{15}^{\tau^\prime *} -  \tilde{c}_{5}^{\tau} (\tilde{b}^{\tau^\prime *}_{13}-\tilde{b}^{\tau^\prime *}_{14}) ], \nonumber \\              
  \tilde{R}_{\tilde{\Phi}^{\prime } \tilde{\Phi}^{\prime }}^{\tau \tau^\prime}&=& \tilde{c}_{12}^\tau \tilde{c}_{12}^{\tau^\prime * }+\tilde{c}_{13}^\tau \tilde{c}_{13}^{\tau^\prime *},   \nonumber \\
    \tilde{R}_{M^{(1) } M^{(1) }}^{\tau \tau^\prime}&=& \tilde{b}_{3}^\tau \tilde{b}_{3}^{\tau^\prime * }+\tilde{b}_{7}^\tau \tilde{b}_{7}^{\tau^\prime *} ,  \nonumber \\
      \tilde{R}_{\tilde{\Phi}^{\prime } M^{(1) }}^{\tau \tau^\prime}&=& \mathrm{Im}[\tilde{c}_{12}^{\tau} \tilde{b}_{7}^{\tau ^\prime *} + \tilde{c}_{13}^{\tau }\tilde{b}_{3}^{\tau^\prime *}], \nonumber \\      
  \tilde{R}_{\Sigma^{\prime \prime } \Sigma^{\prime \prime }}^{\tau \tau^\prime} &=&  ( \tilde{c}_4^\tau- \tilde{c}_6^{\tau })  ( \tilde{c}_4^{\tau^\prime *}- \tilde{c}_6^{\tau^\prime *}) +\tilde{c}_{10}^\tau  \tilde{c}_{10}^{\tau^\prime *},   \nonumber \\
 \tilde{R}_{\Sigma^{\prime \prime (0) } \Sigma^{\prime \prime (0) }}^{\tau \tau^\prime} &=&  \tilde{b}_8^\tau  \tilde{b}_8^{\tau^\prime *} + \tilde{b}_{13}^\tau  \tilde{b}_{13}^{\tau^\prime *} ,  \nonumber \\
 \tilde{R}_{\Sigma^{\prime \prime (2) } \Sigma^{\prime \prime (2) }}^{\tau \tau^\prime} &=&  ( \tilde{b}_{13}^\tau- \tilde{b}_{14}^{\tau })  ( \tilde{b}_{13}^{\tau^\prime *}- \tilde{b}_{14}^{\tau^\prime *}) +\tilde{b}_{16}^\tau  \tilde{b}_{16}^{\tau^\prime *}  , \nonumber \\
 \tilde{R}_{\Sigma^{\prime \prime (0) } \Sigma^{\prime \prime (2) }}^{\tau \tau^\prime} &=&  \mathrm{Re}[\tilde{b}_8^\tau \tilde{b}_{16}^{\tau^\prime *}+\tilde{b}_{13}^\tau (\tilde{b}_{13}^{\tau^\prime *}-\tilde{b}_{14}^{\tau^\prime *})], \nonumber \\
  \tilde{R}_{\Sigma^{\prime \prime (0) } \Sigma^{\prime \prime }}^{\tau \tau^\prime} &=&  \mathrm{Im}[\tilde{b}_{13}^\tau (\tilde{c}_{4}^{\tau^\prime *}-\tilde{c}_{6}^{\tau^\prime *})  -\tilde{b}_{8}^\tau \tilde{c}_{10}^{\tau^\prime *}],  \\
   \tilde{R}_{\Sigma^{\prime \prime (2) } \Sigma^{\prime \prime }}^{\tau \tau^\prime} &=&  \mathrm{Im}[(\tilde{b}_{13}^\tau-\tilde{b}_{14}^\tau) (\tilde{c}_{4}^{\tau^\prime *}-\tilde{c}_{6}^{\tau^\prime *})  -\tilde{b}_{16}^\tau \tilde{c}_{10}^{\tau^\prime *}]. \nonumber 
\end{eqnarray}

The dimensionless hadronic tensors associated with the four square-bracketed response functions of Eq. (\ref{eq:amp2}) 
correspond to the multipole sums
\begin{widetext}
\begin{eqnarray}
\label{eq:nucresponse}
 W_{O O^\prime}^{\tau \tau^\prime}(q_\mathrm{eff})&\equiv&{4 \pi \over 2j_N + 1}  \sum_{J=0,2,...}^\infty    \langle j_N ||~ O_{J;\tau} (q_\mathrm{eff})~ || j_N \rangle 
\langle j_N ||~ O^\prime_{J;\tau^\prime} (q_\mathrm{eff})~ || j_N \rangle,~\mathrm{if~} O,O^\prime \in \{ M^g,M^{(2) f},\Phi^{\prime \prime g} \}, \nonumber \\
 W_{O O^\prime}^{\tau \tau^\prime}(q_\mathrm{eff}) &\equiv & {4 \pi \over 2j_N + 1} 
\sum_{J=1,3,...}^\infty \langle j_N ||~ O_{J; \tau}(q_\mathrm{eff})~ || j_N \rangle
 \langle j_N ||~ O^\prime_{J; \tau^\prime}(q_\mathrm{eff})~ || j_N \rangle,~ \mathrm{if~}O,O^\prime \in \{ \Sigma^{\prime g},\Sigma^{\prime (0) f},\Sigma^{\prime (2) f},\Delta^g \}, \nonumber \\
~&& \hspace{9.11cm} \mathrm{or} ~O,O^\prime \in \{ \Sigma^{\prime \prime g},\Sigma^{\prime \prime (0) f},\Sigma^{\prime \prime (2) f} \}, \nonumber \\
  W_{O O^\prime}^{\tau \tau^\prime}(q_\mathrm{eff})  &\equiv& {4 \pi \over 2j_N + 1}  \sum_{J=2,4,...}^\infty   \langle j_N || ~O_{J; \tau}^\prime (q_\mathrm{eff})~ || j_N \rangle  \langle j_N ||  ~ O^\prime_{J; \tau^\prime}(q_\mathrm{eff})~ || j_N \rangle,~\mathrm{if~} O,O^\prime \in \{ \tilde{\Phi}^{\prime~ g} ,M^{(1) f} \} .
\label{eq:amplitude}
\end{eqnarray}
\end{widetext}

The LECs for the operators $\CO_7$, $\CO_{14}$, $\CO_2$, and $\CO_{16}$
do not appear in Eq. (\ref{eq:Rs}).   $\CO_7$ and $\CO_{14}$ are generated by $\vec{v}_N \cdot \vec{\sigma}_N$: parity
and time-reversal (or CP) selection rules exclude all nuclear multipoles of this axial charge operator.  $\CO_2$ and $\CO_{16}$, generated
by $\vec{q} \cdot \vec{v}_N$, do not arise in elastic $\mu \rightarrow e$ conversion because Siegert's theorem
eliminates the longitudinal projection of the convection current for elastic transitions.  Similarly $b_5$, the LEC for $\CO_5^f$, 
contributes to a nuclear multipole operator that cannot simultaneously satisfy the parity and CP
constraints for an elastic transition.  If $\mu \rightarrow e$ conversion
is accompanied by nuclear excitation, all of these operators will contribute.\\

The interplay of the various NRET scales implicit in the rate can be made more explicit
by rewriting Eq. (\ref{eq:amp2}) after ``completing the squares."  The decay rate becomes 
\begin{widetext}
\begin{eqnarray}
&& ~~~~~~~~~~\omega= {G_F^2 \over \pi} ~ {q_\mathrm{eff}^2 \over 1+{q \over M_T}} ~{4 \pi \over 2j_N+1}
\sum_{ \tau=0,1} \sum_{\tau^\prime = 0,1} \nonumber \\
 && \bigg\{  ~  \sum_{J=0,2,...}^\infty   \Big[~ \Big|  \tilde{c}_1^\tau \langle  M^g_\tau \rangle-{q_\mathrm{eff} \over m_N} \tilde{c}_3^\tau \langle \Phi_\tau^{\prime \prime g}   \rangle+i \tilde{b}_2^\tau \langle  M^{(2) f}_\tau  \rangle \Big|^2_{\tau \tau^\prime}    \nonumber \\
 &&~~~~~~~~~~~~~~~~+\Big|  \tilde{c}_{11}^{\tau}  \langle  M^g_{\tau} \rangle+{q_\mathrm{eff} \over m_N} (\tilde{c}_{12}^{\tau} -\tilde{c}_{15}^{\tau}) \langle  \Phi_{\tau}^{\prime \prime g} \rangle-i \tilde{b}_7^{\tau} \langle  M^{(2) f}_{\tau}  \rangle \Big|^2_{\tau \tau^\prime}   \Big] \nonumber \\
 && + \sum_{J=1,3,...}^\infty   \Big[~ \Big|  \tilde{c}_4^\tau \langle  \Sigma^{\prime g}_\tau \rangle -  {q_\mathrm{eff} \over m_N} \tilde{c}_5^\tau \langle  \Delta_\tau^{  g} \rangle+i \tilde{b}_{14}^\tau \langle  \Sigma^{\prime (0) f}_\tau \rangle-i (\tilde{b}_{13}^\tau-\tilde{b}_{14}^\tau) \langle \Sigma^{\prime (2) f}_\tau \rangle \Big|^2_{\tau \tau^\prime}   \nonumber \\
 &&~~~~~~~~~~~~~~~~+\Big|  \tilde{c}_9^\tau \langle  \Sigma^{\prime g}_\tau \rangle -  {q_\mathrm{eff} \over m_N} \tilde{c}_8^\tau \langle  \Delta_\tau^{  g} \rangle-i (\tilde{b}_{12}^\tau-\tilde{b}_{15}^\tau) \langle  \Sigma^{\prime (0) f}_\tau \rangle+i \tilde{b}_{15}^\tau \langle \Sigma^{\prime (2) f}_\tau \rangle \Big|^2_{\tau \tau^\prime} \Big] \nonumber \\
  && + \sum_{J=2,4,...}^\infty   \Big[~\Big| {q_\mathrm{eff} \over m_N} \tilde{c}_{12}^\tau \langle \tilde{\Phi}_\tau^{\prime g} \rangle -i \tilde{b}_7^\tau \langle M_\tau^{(1) f} \rangle \Big|^2_{\tau \tau^\prime} +\Big| {q_\mathrm{eff} \over m_N} \tilde{c}_{13}^\tau \langle \tilde{\Phi}_\tau^{\prime g} \rangle -i \tilde{b}_3^\tau \langle M_\tau^{(1) f} \rangle \Big|^2_{\tau \tau^\prime}  \Big] \nonumber \\
 && + \sum_{J=1,3,...}^\infty   \Big[~ \Big| (\tilde{c}_4^\tau-\tilde{c}_6^\tau) \langle  \Sigma^{\prime \prime g}_\tau \rangle - i \tilde{b}_{13}^\tau \langle  \Sigma^{\prime \prime (0) f}_\tau \rangle-i (\tilde{b}_{13}^\tau-\tilde{b}_{14}^\tau) \langle \Sigma^{\prime  \prime (2) f}_\tau \rangle \Big|^2_{\tau \tau^\prime}~   \nonumber \\
 &&~~~~~~~~~~~~~~~~+\Big|  \tilde{c}_{10}^\tau \langle  \Sigma^{\prime \prime g}_\tau \rangle + i \tilde{b}_{8}^\tau \langle  \Sigma^{\prime \prime (0) f}_\tau \rangle+i \tilde{b}_{16}^\tau \langle \Sigma^{\prime \prime  (2) f}_\tau \rangle \Big|^2_{\tau \tau^\prime}~\Big]~   \bigg\},
\label{eq:amp3}
\end{eqnarray}
where $\langle O \rangle \equiv \langle j_N || O(q_\mathrm{eff}) || j_N \rangle$
and $|A_\tau|^2_{\tau \tau^\prime} \equiv A_\tau A^*_{\tau^\prime}$.

Alternatively, one can employ our averaging procedures to recast this ``master rate formula" in a form where its dependence
on small parameters is manifest, with response functions that can be evaluated analytically as functions
of $y$, for harmonic oscillator Slater determinants.  We find
\begin{eqnarray}
&& ~~~~~~~~~~\omega= {G_F^2 \over \pi} ~ {q_\mathrm{eff}^2 \over 1+{q \over M_T}} ~|\phi_{1s}^{Z_\mathrm{eff}} (\vec{0})|^2 ~{4 \pi \over 2j_N+1}
\sum_{ \tau=0,1} \sum_{\tau^\prime = 0,1} \nonumber \\
 && \bigg\{  ~  \sum_{J=0,2,...}^\infty   \Big[~ \Big|  \tilde{c}_1^\tau \langle  M_\tau \rangle-{q_\mathrm{eff} \over m_N} \tilde{c}_3^\tau \langle \Phi_\tau^{\prime \prime}   \rangle+  {\langle f \rangle \over \langle g \rangle} i  \tilde{b}_2^\tau \langle  M^{(2)}_\tau  \rangle \Big|^2_{\tau \tau^\prime}   \nonumber \\
 &&~~~~~~~~~~~~~~~~+\Big|  \tilde{c}_{11}^{\tau}  \langle  M_{\tau} \rangle+{q_\mathrm{eff} \over m_N} (\tilde{c}_{12}^{\tau} -\tilde{c}_{15}^{\tau}) \langle  \Phi_{\tau}^{\prime \prime} \rangle- {\langle f \rangle \over \langle g \rangle}  i \tilde{b}_7^{\tau} \langle  M^{(2) }_{\tau}  \rangle \Big|^2_{\tau \tau^\prime}   \Big] \nonumber \\
 && + \sum_{J=1,3,...}^\infty   \Big[~ \Big|  \tilde{c}_4^\tau \langle  \Sigma^{\prime}_\tau \rangle -  {q_\mathrm{eff} \over m_N} \tilde{c}_5^\tau \langle  \Delta_\tau \rangle+ {\langle f \rangle \over \langle g \rangle}  i \tilde{b}_{14}^\tau \langle  \Sigma^{\prime (0) }_\tau \rangle- {\langle f \rangle \over \langle g \rangle}  i (\tilde{b}_{13}^\tau-\tilde{b}_{14}^\tau) \langle \Sigma^{\prime (2)}_\tau \rangle \Big|^2_{\tau \tau^\prime}    \nonumber \\
 &&~~~~~~~~~~~~~~~~+\Big|  \tilde{c}_9^\tau \langle  \Sigma^{\prime}_\tau \rangle -  {q_\mathrm{eff} \over m_N} \tilde{c}_8^\tau \langle  \Delta_\tau \rangle- {\langle f \rangle \over \langle g \rangle}  i (\tilde{b}_{12}^\tau-\tilde{b}_{15}^\tau) \langle  \Sigma^{\prime (0) }_\tau \rangle+{\langle f \rangle \over \langle g \rangle}  i \tilde{b}_{15}^\tau \langle \Sigma^{\prime (2) }_\tau \rangle \Big|^2_{\tau \tau^\prime} \Big] \nonumber \\
  && + \sum_{J=2,4,...}^\infty   \Big[~\Big| {q_\mathrm{eff} \over m_N} \tilde{c}_{12}^\tau \langle \tilde{\Phi}_\tau^{\prime} \rangle - {\langle f \rangle \over \langle g \rangle}  i \tilde{b}_7^\tau \langle M_\tau^{(1) } \rangle \Big|^2_{\tau \tau^\prime} +\Big| {q_\mathrm{eff} \over m_N} \tilde{c}_{13}^\tau \langle \tilde{\Phi}_\tau^{\prime } \rangle - {\langle f \rangle \over \langle g \rangle}  i \tilde{b}_3^\tau \langle M_\tau^{(1) } \rangle \Big|^2_{\tau \tau^\prime}  \Big] \nonumber \\
 && + \sum_{J=1,3,...}^\infty   \Big[~ \Big| (\tilde{c}_4^\tau-\tilde{c}_6^\tau) \langle  \Sigma^{\prime \prime}_\tau \rangle - {\langle f \rangle \over \langle g \rangle}  i \tilde{b}_{13}^\tau \langle  \Sigma^{\prime \prime (0) }_\tau \rangle-{\langle f \rangle \over \langle g \rangle}  i (\tilde{b}_{13}^\tau-\tilde{b}_{14}^\tau) \langle \Sigma^{\prime  \prime (2) }_\tau \rangle \Big|^2_{\tau \tau^\prime}~   \nonumber \\
 &&~~~~~~~~~~~~~~~~+\Big|  \tilde{c}_{10}^\tau \langle  \Sigma^{\prime \prime }_\tau \rangle +{\langle f \rangle \over \langle g \rangle}  i \tilde{b}_{8}^\tau \langle  \Sigma^{\prime \prime (0) }_\tau \rangle+ {\langle f \rangle \over \langle g \rangle}  i \tilde{b}_{16}^\tau \langle \Sigma^{\prime \prime  (2) }_\tau \rangle \Big|^2_{\tau \tau^\prime}~\Big]~   \bigg\}.
\label{eq:amp3A}
\end{eqnarray}

If we specialize to the case of interactions involving scalar or vector mediators, then we can use Tables \ref{tab:LWL} and \ref{tab:LWL2}
to re-express Eq. (\ref{eq:amp3}) in terms of the 20 relativistic LECs $\tilde{d}_i$.  One finds
\begin{eqnarray}
&& ~~~~~~~~~~\omega= {G_F^2 \over \pi} ~ {q_\mathrm{eff}^2 \over 1+{q \over M_T}} ~{4 \pi \over 2j_N+1}
\sum_{ \tau=0,1} \sum_{\tau^\prime = 0,1} \nonumber \\
 && \bigg\{  ~  \sum_{J=0,2,...}^\infty   \Big[~ \Big|  (\tilde{d}_1^\tau-{q \over m_L} \tilde{d}_9^\tau) \langle  M^g_\tau-M^{(2) f}_\tau \rangle+ \tilde{d}_5^\tau \langle  M_\tau^g +M^{(2) f}_\tau  \rangle \Big|^2_{\tau \tau^\prime}   \nonumber \\
 &&~~~~~~~~~~~~~~~~+\Big| (\tilde{d}_3^\tau+{q \over m_L} \tilde{d}_{17}^\tau) \langle  M^g_\tau-M^{(2) f}_\tau \rangle \Big|^2_{\tau \tau^\prime}  +\Big| \tilde{d}_{13}^\tau \langle  M^g_\tau+M^{(2) f}_\tau \rangle \Big|^2_{\tau \tau^\prime}~    \Big]\nonumber \\
 && + \sum_{J=1,3,...}^\infty   \Big[~ \Big|  \tilde{d}_{15}^\tau \langle  \Sigma^{\prime g}_\tau+  \Sigma^{\prime (0) f}_\tau \rangle -  {q_\mathrm{eff} \over  m_N} (\tilde{d}_5^\tau+{q \over m_L}\tilde{d}_9^\tau ) \langle  \Delta_\tau^{  g} \rangle+{q \over 2 m_N} (\tilde{d}_5^\tau-2 \tilde{d}_{6}^\tau+{q \over m_L}(\tilde{d}_9^\tau-2 \tilde{d}_{10}^\tau)) \langle  \Sigma^{\prime g}_\tau \rangle \Big|^2_{\tau \tau^\prime}   \nonumber \\
 &&~~~~~~~~~~~~~~~~~~~~~~~~~~~~~~~~~~~~~~~ +\Big| {q \over m_L}  \tilde{d}_{19}^\tau \langle  \Sigma^{\prime g}_\tau +  \Sigma^{\prime (2) f}_\tau \rangle \Big|^2_{\tau \tau^\prime}+ \nonumber \\
 &&~~~~~~~~~~~~~~~~~\Big| \tilde{d}_{7}^\tau \langle  \Sigma^{\prime g}_\tau + \Sigma^{\prime (0) f}_\tau \rangle+{q \over m_L} \tilde{d}_{11}^\tau  \langle  \Sigma^{\prime g}_\tau+\Sigma_\tau^{\prime (2) f}\rangle  -  {q_\mathrm{eff} \over  m_N} \tilde{d}_{13}^\tau \langle \Delta_\tau^{  g}\rangle +{q \over 2m_N}( \tilde{d}_{13}^\tau- 2 \tilde{d}_{14}^\tau) \langle \Sigma_\tau^{\prime g} \rangle \Big|^2_{\tau \tau^\prime} \nonumber \\
  &&~~~~~~~~~~~~~~~~~~~~~~~~~~~~~~~~~~~~~~~+\Big| {q \over m_L} {q_\mathrm{eff} \over  m_N} \tilde{d}^\tau_{17} \langle  \Delta_\tau^g  \rangle +{q \over m_L} {q \over 2 m_N} (2 \tilde{d}^\tau_{18}-\tilde{d}_{17}^\tau) \langle \Sigma^{\prime g}_\tau \rangle \Big|^2_{\tau \tau^\prime}    \Big] \nonumber \\
  && + \sum_{J=2,4,...}^\infty   \Big[~\Big|  (\tilde{d}_{3}^\tau+{q \over m_L} \tilde{d}_{17}^{\tau}) \langle M_\tau^{(1) f} \rangle  \Big|^2_{\tau \tau^\prime}  +\Big| \tilde{d}_{13}^{\tau} \langle M_\tau^{(1) f} \rangle  \Big|^2_{\tau \tau^\prime} +\Big|  (\tilde{d}_{1}^\tau-\tilde{d}_{5}^\tau-{q \over m_L} \tilde{d}_{9}^{\tau}) \langle M_\tau^{(1) f} \rangle  \Big|^2_{\tau \tau^\prime}   \Big] \nonumber \\
 && + \sum_{J=1,3,...}^\infty   \Big[~ \Big| \tilde{d}_{15}^\tau  \langle  \Sigma^{\prime \prime g}_\tau - \Sigma^{\prime \prime (0) f}_\tau \rangle -{q \over  2m_N}  (\tilde{d}_{4}^\tau-{ 2q \over m_L} \tilde{d}^\tau_{20})  \langle  \Sigma^{\prime \prime g}_\tau \rangle \Big|^2_{\tau \tau^\prime}  +\Big| {q \over m_L}\tilde{d}_{19}^\tau  \langle  \Sigma^{\prime \prime (0) f}_\tau + \Sigma^{\prime \prime (2) f}_\tau \rangle +{q \over  m_N}  \tilde{d}_{16}^\tau  \langle  \Sigma^{\prime \prime g}_\tau \rangle \Big|^2_{\tau \tau^\prime} \nonumber \\
 &&~~~~~~~~~~~~~+\Big| {q \over 2 m_N} ( \tilde{d}_{2}^\tau -2\tilde{d}_{8}^\tau +{2 q \over m_L} \tilde{d}_{12}^\tau) \langle  \Sigma^{\prime \prime g}_\tau \rangle  \Big|^2_{\tau \tau^\prime}  +\Big| \tilde{d}_{7}^\tau  \langle  \Sigma^{\prime \prime g}_\tau - \Sigma^{\prime \prime (0) f}_\tau \rangle +{q \over  m_L}  \tilde{d}_{11}^\tau  \langle  \Sigma^{\prime \prime (0) f}_\tau+ \Sigma^{\prime \prime (2) f}_\tau\rangle \Big|^2_{\tau \tau^\prime} \Big]   \bigg\}.
\label{eq:amprel}
\end{eqnarray}
\end{widetext}
These results are used Sec. \ref{sec:NRETrates}.

\section{NRET \textit{Mathematica} and Python Scripts}
\label{AppendixC}
In this appendix, we describe the open-source {\it Mathematica} and Python scripts that were employed for the calculations
reported in this paper, which as noted below are available to others.  Possible uses include explorations of specific
UV theories of CLFV that can be reduced to a nucleon-level operator basis or
experimental investigations of alternative targets to explore relative LEC sensitivities.   The Mathematica script
interrogates users about desired input, and generally includes defaults as well as menus
allowing parameter investigation. \\

\begin{table*}
  \caption{ \label{tab:input}Input parameters and output quantities for the muon and electron Dirac solutions discussed in the text.}
 \begin{tabular}{|c|c|c|c|c|c|c|c|c|c|c|}
\hline
& & & & & & & & & &  \\
 ~Nucleus~ & ~$c$ (fm)~&  ~$\beta$ (fm)~ &~$\sqrt{\langle r^2 \rangle}$~(fm)~& ~$E_\mu^\mathrm{bind}$ (MeV)~  & ~$\int_0^\infty |F_{1s}|^2 dr$~   &~~Z~~& ~~~$Z_\mathrm{eff}$~~~ & ~~~~R~~~~ &  $q$ (MeV)  & $q_\mathrm{eff}$ (MeV)  \\[0.4cm]
\hline
& & & &  & & & & &  & \\
$^{12}$C & 2.215 & 0.491 & 2.505 & 0.1000 & 0.00047 & 6 & 5.7030 & 0.8587 & 105.07 & 108.40 \\
$^{16}$O & 2.534 & 0.514 & 2.739 & 0.1775 &0.00083 & 8 & 7.4210 & 0.7982 & 105.11 & 109.16 \\
$^{19}$F & 2.580 & 0.567 & 2.904 & 0.2242 & 0.00104& 9 & 8.2298 & 0.7646 & 105.12 & 109.44 \\
$^{23}$Na & 2.760 & 0.543 & 2.940 & 0.3337 &0.00154 & 11 & 9.8547 & 0.7190 & 105.07 & 110.25 \\
$^{27}$Al & 3.070 & 0.519 & 3.062 & 0.4630 & 0.00211 & 13 & 11.3086 & 0.6583 & 104.98 & 110.81 \\
$^{28}$Si & 3.140 & 0.537 & 3.146 & 0.5346 &0.00241 & 14 & 12.0009 & 0.6299 & 104.91 & 111.03 \\
$^{32}$S & 3.161 & 0.578 & 3.239 &  0.6924 & 0.00308  & 16 &  13.1839 &  0.5595 &  104.78 &  111.56 \\
$^{40}$Ca & 3.621 & 0.563 & 3.499 & 1.0585 &0.00453 & 20 & 15.6916 & 0.4830 & 104.45 & 112.28 \\
$^{48}$Ti & 3.843 & 0.588 & 3.693 & 1.2615 & 0.00527 & 22 & 16.6562 & 0.4340 & 104.28 & 112.43 \\
$^{56}$Fe & 4.111 & 0.558 & 3.800 & 1.7182 & 0.00690& 26 & 18.6028 & 0.3663 & 103.84 & 113.16 \\
$^{63}$Cu & 4.218 & 0.596 & 3.947 & 2.0884 &0.00811  & 29 & 19.8563 & 0.3210 & 103.48 & 113.50 \\
$^{184}$W & 6.510 & 0.535 & 5.421 & 9.0851 & 0.01169 & 74 & 32.2914 & 0.0831 & 96.54 & 114.95 \\[0.15cm]
 \hline
 \end{tabular}
\end{table*}

\begin{table*}
\caption{Shell-model values of the six nuclear response functions $W_O^{\tau\tau'}=W_{OO}^{\tau\tau'}(q_\mathrm{eff})$.}
\label{tab:response_values}
\centering
{\renewcommand{\arraystretch}{1.4}
\begin{tabular}{|c|c|c|c|c|c|c|c|c|c|c|c|}
\hline
Response & ~~~~C~~~~~ & ~~~~O~~~~~ & ~~~~F~~~~~ & ~~~~Na~~~~ & ~~~~Al~~~~ & ~~~~Si~~~~ & ~~~~S~~~~~ & ~~~~Ca~~~~ & ~~~~Ti~~~~ & ~~~~Fe~~~~ & ~~~~Cu~~~~\\
\hline
$W^{00}_M$ & 72 & 1.3$\cdot10^{2}$ & 1.5$\cdot10^{2}$ & 2.1$\cdot10^{2}$ & 2.6$\cdot10^{2}$ & 2.8$\cdot10^{2}$ & 3.3$\cdot10^{2}$ & 4.4$\cdot10^{2}$ & 5.2$\cdot10^{2}$ & 5.9$\cdot10^{2}$ & 6.5$\cdot10^{2}$ \\
$W^{11}_M$ & 4.7$\cdot10^{-3}$ & 2.6$\cdot10^{-3}$ & 0.28 & 0.28 & 0.28 & 4.5$\cdot 10^{-2}$ & 4.2$\cdot 10^{-2}$ & 4.2$\cdot 10^{-2}$ & 1.6 & 1.5 & 2.7 \\
$W^{00}_{\Sigma'}$ & 5.4$\cdot10^{-4}$ & 0 & 0.51 & 7.9$\cdot 10^{-2}$ & 9.1$\cdot 10^{-2}$ & 1.9$\cdot10^{-3}$ & 1.2$\cdot10^{-4}$ & 0 & 3.7$\cdot10^{-3}$ & 1.9$\cdot10^{-4}$ & 0.15\\
$W^{11}_{\Sigma'}$ & 3.8$\cdot10^{-4}$ & 0 & 4.7$\cdot 10^{-2}$ & 5.7$\cdot 10^{-2}$ & 5.4$\cdot 10^{-2}$ & 1.1$\cdot10^{-3}$ & 2.5$\cdot10^{-5}$ & 0 & 2.5$\cdot10^{-3}$ & 1.2$\cdot10^{-4}$ & 0.13\\
$W^{00}_{\Sigma''}$ & 1.6$\cdot10^{-3}$ & 0 & 0.25 & 6.1$\cdot 10^{-2}$ & 0.11 & 2.6$\cdot10^{-3}$ & 5.8$\cdot10^{-4}$ & 0 & 8.8$\cdot10^{-3}$ & 3.9$\cdot10^{-5}$ & 0.12\\
$W^{11}_{\Sigma''}$ & 1.3$\cdot10^{-3}$ & 0 & 0.27 & 4.8$\cdot 10^{-2}$ & 9.2$\cdot 10^{-2}$ & 2.2$\cdot10^{-3}$ & 3.2$\cdot10^{-4}$ & 0 & 7.1$\cdot10^{-3}$ & 7.4$\cdot10^{-7}$ & 0.12\\
$W^{00}_{\Delta}$ & 1.5$\cdot10^{-3}$ & 0 & 1.6$\cdot10^{-4}$ & 0.21 & 0.50 & 1.2$\cdot10^{-3}$ & 2.7$\cdot10^{-3}$ & 0 & 5.9$\cdot 10^{-2}$ & 1.0$\cdot10^{-3}$ & 8.0$\cdot 10^{-2}$ \\
$W^{11}_{\Delta}$ & 5.8$\cdot10^{-4}$ & 0 & 4.3$\cdot10^{-2}$ & 4.6$\cdot 10^{-2}$ & 0.21 & 1.2$\cdot10^{-3}$ & 1.2$\cdot10^{-3}$ & 0 & 4.7$\cdot 10^{-2}$ & 1.3$\cdot10^{-3}$ & 9.0$\cdot10^{-3}$ \\
$W^{00}_{\tilde{\Phi}'}$ & 0 & 0 & 0 & 5.7$\cdot10^{-3}$ & 8.8$\cdot10^{-4}$ & 0 & 2.9$\cdot10^{-4}$ & 0 & 2.8$\cdot 10^{-2}$ & 0 & 2.3$\cdot10^{-3}$ \\
$W^{11}_{\tilde{\Phi}'}$ & 0 & 0 & 0 & 5.0$\cdot10^{-4}$ & 6.6$\cdot10^{-2}$ & 0 & 8.8$\cdot10^{-5}$ & 0 & 1.4$\cdot10^{-3}$ & 0 & 4.7$\cdot 10^{-2}$\\
$W^{00}_{\Phi''}$ & 0.90 & 1.0$\cdot10^{-3}$ & 0.28 & 3.9& 11 & 10 & 10 & 0.10 & 14 & 38 & 53 \\
$W^{11}_{\Phi''}$ & 8.1$\cdot10^{-4}$ & 1.0$\cdot10^{-3}$ & 6.3$\cdot10^{-2}$ & 5.6$\cdot 10^{-2}$ & 8.3$\cdot 10^{-2}$ & 5.0$\cdot10^{-3}$ & 4.9$\cdot 10^{-2}$ & 0.10 & 3.7 & 0.56 & 1.4 \\
\hline
\end{tabular}}
\end{table*}
\noindent
{\it Atomic and nuclear physics input:}  The atomic input is derived from solving the Dirac equation in the nuclear Coulomb field, using the
nuclear charge density profile described in the text; the parameters $c$ and $\beta$ were either taken directly from
Ref. \cite{density} or fit to the alternative density profiles provided in that reference.   The density profiles were determined
from fits to elastic electron scattering and thus are consistent with the known charge radii of the targets.  Table \ref{tab:input}
gives for each target the values of $c$ and $\beta$ we employed, as well as the resulting rms charge radius, the muon binding energy, 
the muon lower-component probability, the computed $Z_\mathrm{eff}$ that relates the muon wave function (averaged over the nuclear density,
weighted by $j_0(q r)$) to the point Schr\"odinger density at the origin, the equivalent reduction factor $R$, the three-momentum transfer $q$,
and the computed $q_\mathrm{eff}$.  The values for $E_\mu^\mathrm{binding}$, $Z_\mathrm{eff}$, and $q_\mathrm{eff}$ are embedded
in the script.  The atomic physics is evaluated for the principal isotope indicated in the table, then used for all contributing isotopes.

The script includes options for 11 targets, ranging from C to Cu.  The selection includes the light- and medium-mass nuclei that were
employed in past $\mu \rightarrow e$ conversion experiments or are under consideration for future experiments, as well as a few other
common targets.  As described in the text, response function evaluation requires the one-body density matrix: knowledge
of the full one-body density matrix allows one to calculate the exact many-body matrix element of any one-body operator.  The shell model
(SM) approximation to the density matrix truncates the many-body Hilbert space to some valence space and adopts a single-particle basis
for representing the Slater determinants in that space.   We use the harmonic oscillator, a choice that allows us to preserve translational
invariance (as we use separable SM spaces).  With this choice, all multipole operator single-particle matrix elements can be evaluated analytically.  Table XII gives the resulting nuclear response functions for the 11 targets we considered.

Typically, natural targets are used in experiments, so the script accesses density matrices for every isotope with an abundance $\gtrsim$ 0.2\%, computing the corresponding
responses by weighting each density matrix by the appropriate abundance.   There is an option in the script, however, to compute the $\mu \rightarrow e$
conversion rate for a selected single isotope, in which case the abundance for that isotope is set to unity.  

The nuclear model dependence arises from the density matrix truncation (inherent in the choice of the finite SM space of interacting Slater
determinants) and from the choice of the single-particle basis employed in the determinants.  The truncation generates the dependence on the choice of single-particle basis;
absent truncation, all complete bases are equivalent.  The SM spaces used are shown in Table \ref{tab:nuclear} along
with a list of effective interactions that were used to generate the one-body density matrices in those spaces.   Effective interactions are typically constructed from iterated two-nucleon potentials derived from $NN$ phase shift
analyses, augmented by phenomenological terms whose strengths are
carefully tuned to reproduce nuclear spectra and other properties in the mass range being addressed.  In most cases, density matrices are provided for more than one 
interaction, so that users can explore uncertainties associated with alternative schemes for fitting the effective interaction. The uncertainties associated with the SM itself -- the basis truncation and the choice
of single particle basis -- are more difficult to assess.  Shell model calculations were done with the BIGSTICK code \cite{Johnson:2013bna,Calvin}.

\begin{table}
 \caption{ \label{tab:nuclear}Shell-model spaces employed and options available for the corresponding effective interactions.
 See text for descriptions of the two procedures used to determine the oscillator parameter $b$.}
 \begin{tabular}{|c|c|c|c|c|}
\hline
& & &  &  \\
 Target & Isotopes&  SM Space &$\begin{array}{c} \mathrm{Interaction} \\ \mathrm{options}  \end{array}$ &  b (fm)  \\[0.4cm]
\hline
& & & &  \\
C & 12,13 & $1p$ &\cite{CK} & 1.67/1.70  \\
O & 16,18 & $2s$-$1d$ & \cite{USDB,BW} & 1.73/1.83  \\
        & & 4$\hbar \omega$ & \cite{Johnson}  & 1.73/1.80  \\
F & 19 &  $2s$-$1d$ & \cite{USDB,BW}  & 1.76/1.88  \\
Na & 23 &  $2s$-$1d$ &  as above & 1.80/1.83 \\
Al & 27 &  $2s$-$1d$ &  as above & 1.84/1.85 \\
Si & 28-30 &  $2s$-$1d$ &  as above & 1.85/1.89  \\
S & 32-34 &  $2s$-$1d$ &  as above & 1.88/1.91  \\
Ca & 40,42,44 & $2p$-$1f$ &  \cite{KB3G,GX1A,KBP} & 1.94/2.02 \\
Ti & 46-50 & $2p$-$1f$ &  as above & 1.99/2.09 \\
Fe & 54,56-58 & $2p$-$1f$  &  as above & 2.03/2.08 \\
Cu & 63,65 & $1f_{5 \over 2}$-$2p$-$1g_{9 \over 2}$ &  \cite{GCN2850,jj44b,JUN45} & 2.07/2.12 \\[0.15cm]
 \hline
 \end{tabular}
\end{table}

The script provides a default value for the harmonic oscillator size parameter $b$, but allows the user to replace that with any
desired alternative value.\\

\noindent
{\it User options:}  The script gives the user various choices:
\begin{enumerate}
\item  Target: The 11 options listed in Table \ref{tab:nuclear} are available.
\item  Isotopic composition:  The default option is a natural target, in which case all isotopes with abundance $\gtrsim$ 0.2\%
are included, weighted by their abundances, as described above.  The user has the option to instead select a single isotope.
\item  Density matrix:  The user is prompted with a set of shell-model effective interactions.  Once a choice is made, the script
extracts the corresponding one-body density from a library, for use in evaluating the nuclear response functions.
\item  Oscillator parameter:  The first entries for $b$ in the last column of Table \ref{tab:nuclear} are the default values obtained from the formula
\[ b = \sqrt{41.467 \over 45 \bar{A}^{-1/3} -25 \bar{A}^{-2/3}}~\mathrm{fm} \]
but the user can specify another value.  (For example, the $b$ values listed second in Table \ref{tab:nuclear} reproduce the experimental rms 
charge radius when point protons are assumed.  This choice could be for consistency with the distributions used to evaluate lepton Coulomb wave functions,
as these were determined from fits to measured electron scattering form factors.)   Here $\bar{A}$ is the atomic mass computed
for the natural isotope, and thus weighted over the abundances.
\item  Response functions:  The user has the option to request that the calculated nuclear response functions be displayed
either as analytic functions in $y$ or as a graph.  If this option is exercised, the user is queried about each of the six response functions
and can chose to display any subset of them.  The user is also asked to specify the isospin, with options including isoscalar, isovector,
proton-only, and neutron-only couplings.
\item  Relativistic or nonrelativistic LECs:  The user has the option of using the 20 relativistic operators of 
Table \ref{tab:LWL} that arise for scalar and vector exchanges, rather than the basis of nonrelativistic operators.  If this option is exercised, it asks the user to 
specify the leptonic scale $m_L$  (in MeV) that is used in Table \ref{tab:LWL}.
\item  LEC input:  The user either inserts the relativistic LECs $\{i,\tilde{d}_i(0),\tilde{d}_i(1) \}$ or nonrelativistic
LECs $\{i,\tilde{c}_i(0),\tilde{c}_i(1)\}$.  Here $i$ is the operator index corresponding to the lists within this paper, and the two couplings are the
isoscalar and isovector components.  Relativistic LECs are immediately converted into their nonrelativistic equivalents, as in Table \ref{tab:LWL}.
The LECs are dimensionless, given in units of $1/v^2$.  Only those LECs of interest have to be inserted:  entry of a triad with $i=0$ terminates
the input.  LECs can be inserted as numbers or as symbols.
\end{enumerate}
The output is the $\mu \rightarrow e$ conversion rate in 1/sec.\\

\noindent
{\it Obtaining the Script and Accessing the Density matrix Library:} 
The Mathematica and Python scripts may be accessed via the public GitHub repository \cite{GitHub1}.
The included README file describes the setup of the scripts. 
The associated library of elastic one-body density matrices for relevant isotopes is available from a second public repository \cite{GitHub2}. 
The branch label for both repositories corresponding to this paper is v1.0.

\end{document}